\documentclass[aps,reprint,twocolumn,amsmath,amssymb,showpacs,floatfix]{revtex4-1}
\usepackage{graphicx,bm} \usepackage{color} 
\usepackage{ragged2e}
\usepackage{multirow}
\usepackage{mathrsfs}
\usepackage{appendix}

\begin{document}

\title{Majorana representation for topological edge states of massless Dirac fermion with non-quantized Berry phase}
\author{F. R. Pratama$^{1,2}$}
\email{pratama.fr@aist.go.jp}
\author{Takeshi Nakanishi$^{1}$}
\email{t.nakanishi@aist.go.jp}

\affiliation{$^{1}$Mathematics for Advanced Materials-OIL, AIST, 2-1-1 Katahira, Aoba, 980-8577 Sendai, Japan\\
$^{2}$Advanced Institute for Materials Research, Tohoku University, 2-1-1 Katahira, Aoba, 980-8577 Sendai, Japan }

\begin{abstract}
We study the bulk-boundary correspondences for zigzag ribbons (ZRs) of massless Dirac fermion in two-dimensional $\alpha$-${T}_3$ lattice. By tuning the hopping parameter $\alpha\in[0,1]$, the $\alpha$-${T}_3$ lattice interpolates between pseudospin $S=1/2$ (graphene) and $S=1$ (${T}_3$ or dice lattice), for $\alpha=0$ and $1$, respectively, which is followed by continuous change of the Berry phase from $\pi$ to $0$. The range of existence for edge states in the momentum space is determined by solving tight-binding equations at the boundaries of the ZRs. We find that the transitions of in-gap bands from bulk to edge states in the momentum space do not only occur at the positions of the Dirac cones but also at additional points depending on $\alpha\notin \{0,1\}$. The $\alpha$-${T}_3$ ZRs are mapped into stub Su-Schrieffer-Heeger chains by performing unitary transforms of the bulk Hamiltonian. The non-trivial topology of the bulk bands is revealed by the Majorana representation of the eigenstates, where the $\mathbb{Z}_2$ topological invariant is manifested by the winding numbers on the complex plane and the Bloch sphere. 
\end{abstract}

\date{\today}
\maketitle

\section{Introduction}~\label{sec:Intro}

\vspace{-5mm}

Bulk-boundary correspondence (BBC) is the relation between the number of robust edge states and the bulk topological invariant. This concept leads to the paradigm~\cite{hasan2010,cooper2019,cayssol2021} to classify phases of condensed-matters based on the topology of bulk bands. The pioneering work of Thouless \textit{et. al.} \cite{thouless1982} identified the Chern number $\mathscr{C}$ of the occupied Landau levels as the $\mathbb{Z}$ invariant for the quantized Hall conductivity \cite{klitzing1980} in a non-interacting, two-dimensional (2D) electron gas. In systems that preserve time-reversal symmetry, $\mathbb{Z}_2$ invariants~\cite{kane2005} differentiate between the ordinary and quantum spin Hall insulators.

The Berry phase~\cite{berry1984} in 1D Brillouin zone (BZ), also known as the Zak phase~\cite{zak1989}, provides the $\mathbb{Z}_2$ invariant for the Dirac matters with both time-reversal and inversion symmetries~\cite{xiao2010}. The quantized value of Zak phase to $\pi$ or $0$ is utilized to predict the presence or absence of edge states in the Su-Schrieffer-Heeger (SSH) chain of polyacetylene~\cite{su1979} and in various types of graphene ribbons~\cite{delplace2011,cao2017,groning2018,li2021}, insofar the number of unit cells is commensurate with the bulk lattice~\cite{rhim2018}. The spinless model for these systems is described by the generic Hamiltonian $\mathscr{H}=\boldsymbol{d}\cdot \boldsymbol{\sigma}$, where $\boldsymbol{\sigma}=(\sigma_x,\sigma_y,\sigma_z)$ is the vector of Pauli matrices. The bivalued Zak phase corresponds to the two possibilities whether the path of $\boldsymbol{d}=(\mathrm{Re}[\boldsymbol{d}],-\mathrm{Im}[\boldsymbol{d}],0)$ encloses the origin of the complex plane or not~\cite{ryu2002,mong2011}. $\mathscr{H}$ is also adapted to explain the edge states in phosphorene ribbons~\cite{ezawa2014,grujic2016tunable,van2016,hitomi2021}. Experimentally, the topology of SSH chain is studied by using arrays of cold atoms~\cite{atala2013,grusdt2014measuring,lu2016,meier2016soliton,mivehvar2017}, photonic waveguides~\cite{st2017lasing,longhi2018,jiang2018experimental,jiao2021}, and electrical circuits~\cite{goren2018}. Apart from the observations~\cite{tao2011,wang2016giant,prudkovskiy2022epitaxial} by scanning-tunneling microscopy, the properties of edge states in graphene ribbons have been investigated using artificial honeycomb lattices~\cite{polini2013artificial}, which encompass the photonic~\cite{rechtsman2013topological,plotnik2014observation,bellec2014manipulation,milicevic2017orbital,zhang2020soliton,xia2023photonic}, phononic~\cite{xi2021observation,wang2021zak}, plasmonic~\cite{han2009dirac,wang2016existence}, and polaritonic~\cite{jacqmin2014direct,st2021measuring} analogs of graphene.





The $\alpha$-$T_3$ lattice~\cite{raoux2014dia}~[Fig.~\ref{fig:aT3}(a)] interpolates between pseudospin $S=1/2$ (graphene~\cite{neto2009electronic}) and $S=1$ (${T}_3$ or dice lattice~\cite{bercioux2009massless,dora2011lattice}) by continuously varying the hopping parameter $\alpha$ from $0$ to $1$, respectively, while the energy dispersion [Fig.~\ref{fig:aT3}(b)] remains the same. On the other hand, the Berry phase smoothly changes from $\pi$ to $0$, which can be seen in the gradual transition of diamagnetic to paramagnetic orbital susceptibilities~\cite{raoux2014dia}, among others. Motivated by the unconventional physics phenomena that arise from the variable geometric phase and the existence of flat band---e.g. enhanced Klein tunneling and supercollimation~\cite{urban2011barrier,fang2016klein,illes2017klein,betancur2017super,weekes2021generalized,cunha2022tunneling}---there have been extensive studies on the properties of the $T_3$ and $\alpha$-${T}_3$ lattices, including  the optical~\cite{illes2015hall,malcolm2016frequency,carbotte2019optical,mojarro2020electron,han2022optical,iurov2022finite,oriekhov2022optical,iurov2021tailoring}, magnetic~\cite{raoux2014dia,illes2016magnetic,soni2020flat,roslyak2021effect}, pseudomagnetic~\cite{sun2022strain,filusch2022tunable,li2023topological}, magneto-optical~\cite{kovacs2017frequency,chen2019enhanced,chen2019nonlinear,balassis2020magnetoplasmons},~and transport~\cite{vigh2013diverging,louvet2015origin,wang2020integer,wang2021flat,zhou2021andreev,biswas2016magnetotransport,islam2017valley,duan2023seebeck,huang2019interplay} properties. The ${T}_3$ lattice is predicted to occur in several perovskite-based heterostructures~\cite{wang2011nearly,koksal2023high} and strained blue-phosphorene oxide~\cite{zhu2016blue}. Moreover, the Hamiltonian of the $\alpha$-${T}_3$ lattice can be employed to reproduce the absorption spectra~\cite{malcolm2015magneto} of 2D $\mathrm{Hg}_{1-x}\mathrm{Cd}_x\mathrm{Te}$ at the critical doping $x\approx 0.17$~\cite{orlita2014observation}. 

In the context of topology, the ${T}_3$ lattice shows quantum anomalous Hall effect with $|\mathscr{C}|=2$~\cite{andrijauskas2015three,dey2020unconventional} by including the Haldane term~\cite{haldane1988model} in the Hamiltonian. Similarly, the $\alpha$-${T}_3$ ribbon with spin-orbit interactions shows the quantum spin Hall effect~\cite{wang2021quantum}. However, the BBC for the $\alpha$-${T}_3$ lattice without opening non-trivial gaps is yet to be formulated, even though it is natural to investigate whether the topological edge states remain without the quantized Berry phase. Furthermore, it is shown in a recent work~\cite{hao2022zigzag} that the $T_3$ zigzag ribbon (ZR) hosts edge states. Our study shows that the edge states are topologically non-trivial despite the zero Berry phase. 



Our paper is organized as follows. In Sec. \ref{sec:sSSH}, we discuss the BBC for the stub SSH chain~\cite{real2017flat,bartlett2021illuminating,caceres2022experimental}, which is constructed by connecting an additional site to one of the sites in the SSH dimer as depicted in Fig.~\ref{fig:sSSH-chain}. The results of this section are useful for understanding and will be applied to uncover the BBC of the $\alpha$-$T_3$ ZRs. In the Majorana representation of the eigenstates~\cite{majorana1932atomi,hannay1998berry,liu2014representation}, we analytically prove that the presence or absence of edge states are topologically characterized by winding numbers on the complex plane and the Bloch sphere. Sec. \ref{sec:aT3-bulk} is devoted to a brief review of the relevant bulk properties of the $\alpha$-$T_3$ lattice. In Sec. \ref{sec:aT3-ZR}, first we describe the configurations of $\alpha$-$T_3$ ZR. For two types of ZRs, tight-binding equations (TBEs) at the boundaries are solved to determine the range of existence for edge states in momentum space. In Sec.~\ref{sec:aT3-Invariant}, we map the $\alpha$-$T_3$ ZRs into stub SSH chains by unitary transforms of the Hamiltonian. Here, momentum in the direction parallel to the $\alpha$-$T_3$ ZRs is transformed into hopping parameters of the corresponding stub SSH chains, and thus the dimension is reduced from 2D to 1D. We discuss the topological phase diagram for each ZR. Conclusion is given in Sec.~\ref{sec:Conclusion}.  

This paper serves as a technical companion to Ref. \cite{pratama24-letter} by providing detailed derivations of the results.



\section{Stub SSH chain}~\label{sec:sSSH}\label{subsec:bulkSSH}

\vspace{-10mm}

\subsection{Bulk Hamiltonian}

\begin{figure}[t]
\begin{center}
\includegraphics[width=85mm]{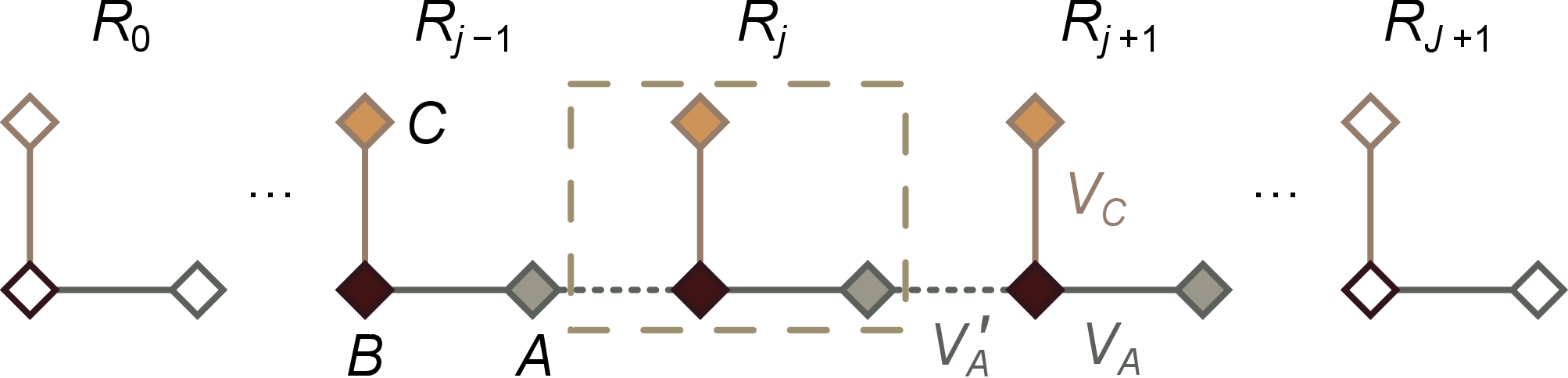}
\caption{ The stub SSH chain. The position of $j$-th unit cell (dashed rectangle) is denoted by $R_j$, for $j=1, 2,\dots, J$, where $J$ is the number of unit cells. The intracell (intercell) hopping parameter between the $A$ and $B$ sites is $V_A$ ($V_A^{\prime}$), and that between the $C$ and $B$ sites is $V_C$. The unfilled symbols at $R_0$ and $R_{J+1}$ indicate missing sites. }
\label{fig:sSSH-chain}
\end{center}
\end{figure}

Fig.~\ref{fig:sSSH-chain} illustrates the stub SSH chain. In each unit cell (dashed rectangle), the intracell hopping parameter between the $B$ and $A$ ($C$) sites is $V_A\geq 0$ ($V_C$). The intercell hopping parameter is $V_A^\prime$. For a chain consisting $J$ unit cells, the Hamiltonian $\hat{H}$ is given by 
\begin{align}
   \hat{H} = &-\sum_{j=1}^{J} \Big[ V_A  \left( \hat{a}_j^{\dagger} \hat{b}_j + \hat{b}_j^{\dagger} \hat{a}_j    \right) + V_A^{\prime} \left( \hat{a}_j^{\dagger} \hat{b}_{j+1} + \hat{b}_{j+1}^{\dagger} \hat{a}_j    \right)     \nonumber\\
   &  +V_C  \left( \hat{b}_j^{\dagger} \hat{c}_j + \hat{c}_j^{\dagger} \hat{b}_j \right) \Big],
   \label{eq:sSSH-chain}
\end{align}
where $\hat{a}_j^{\dagger}$, $\hat{b}_j^{\dagger}$, and $\hat{c}_j^{\dagger}$ ($\hat{a}_j$, $\hat{b}_j$, and $\hat{c}_j$) are the creation (annihilation) operators for the $A$, $B$, and $C$ sites, respectively, in the $j$-th unit cell. By performing the Fourier transforms for the creation and annihilation operators, the bulk Hamiltonian in the momentum space is given by $\hat{H}(k) = \sum_{k}\hat{\Psi}_k^{\dagger}  H(k) \hat{\Psi}_k$, where $\hat{\Psi}_k = \begin{pmatrix} a_k& b_k & c_k \end{pmatrix}^{T}$ is the field operator, and
\begin{align}
H(k) = - \begin{bmatrix}
     0   & F^*(k)   & 0\\
     F(k) & 0 & V_C \\
     0 & V_C & 0
\end{bmatrix}.
\label{eq:sSSH-H1}
\end{align}
Here, we define
\begin{align}
    {F}(k) = |{F}(k)| e^{-i\Phi(k)} \equiv V_A + V_A^{\prime} e^{-ika_0}.
    \label{eq:sSSH-F}
\end{align}
$a_0 \equiv R_{j+1}-R_j$ is the lattice constant~\footnote{The positions of all sites in each unit cell are regarded as identical. This treatment is equivalent to fixing a gauge in the Hamiltonian such that the intracell hopping is real~\cite{delplace2011,ezawa2014}.}. 
$\Phi(k)\in (-\pi, \pi]$ is given by~\footnote{$Z=\mathrm{Arg}(X+iY)=\mathrm{atan2}(Y,X)$, where $\mathrm{atan2}$ is two-argument arctangent function. Particularly, ${Z}=0$ for ${Y}=0,~{X}>0$ and ${Z}=\pi$ for ${Y}=0,~{X}<0$.}:
\begin{align}
    \Phi(k) = \mathrm{Arg}\left[\frac{V_A}{V_A^\prime}+\cos(ka_0)+i\sin(ka_0)\right].
    \label{eq:sSSH-Theta}
\end{align}
Without lost of generality, we only consider $V_A^\prime> 0$, and the 1D BZ is defined for $k\in [-\pi/a_0,\pi/a_0]$~\footnote{For $V_A^\prime<0$, the BZ is defined for $k\in[0,2\pi/a_0]$, because $F(k)=V_A-|V_A^\prime| e^{-ika_0} = V_A+|V_A^\prime| e^{-i(k+\pi/a_0)a_0} $}. It is noted that ${F}(-k)={F}^*(k)$ and $\Phi(-k) = -\Phi(k)$.
%
%

The energy eigenvalues of Eq.~(\ref{eq:sSSH-H1}) are given by 
\begin{subequations}
    \begin{align}
    E_s(k) &= s\sqrt{{V_A}^2+{V_A^{\prime}}^2+{V_C}^2+2V_AV_A^{\prime}\cos(ka_0)},
    \label{eq:sSSH-Es}\\
    E_0 &= 0,
    \label{eq:sSSH-E0}
\end{align}
\end{subequations}
where $E_s$ indicates the valence (conduction) band for $s=-1$ ($+1$). $E_0$ is the flat band at zero energy whose origin is understood from the Lieb theorem \cite{lieb1989two}: due to the absence of interactions between the $A$ and $C$ sites, each unit cell can be partitioned into two sublattices: one consists of the $A$ and $C$ sites, the other consists of the $B$ site. The numerical imbalance between the sublattices gives rise to $E_0$. For $V_C \neq 0$, the bulk bands do not touch at $|k| = \pi/a_0$, unlike in the SSH chain where the gap closing indicates a topological phase transition when $V_A/V_A^{\prime} = 1 $. Nevertheless, regardless the value of $V_C$, 
\begin{align}
    \Phi\left(\frac{\pi}{a_0}\right) =
    \begin{cases}
        \pi~\mathrm{for}~V_A/V_A^\prime<1,\\
        0~\mathrm{for}~V_A/V_A^\prime>1,
    \end{cases}
    \label{eq:sSSH-Phival}
\end{align}
and indeterminate for $V_A/V_A^\prime =1$. Eq.~(\ref{eq:sSSH-Phival}) will be used to calculate the topological index. 

\begin{figure}[t]
\begin{center}
\includegraphics[width=69mm]{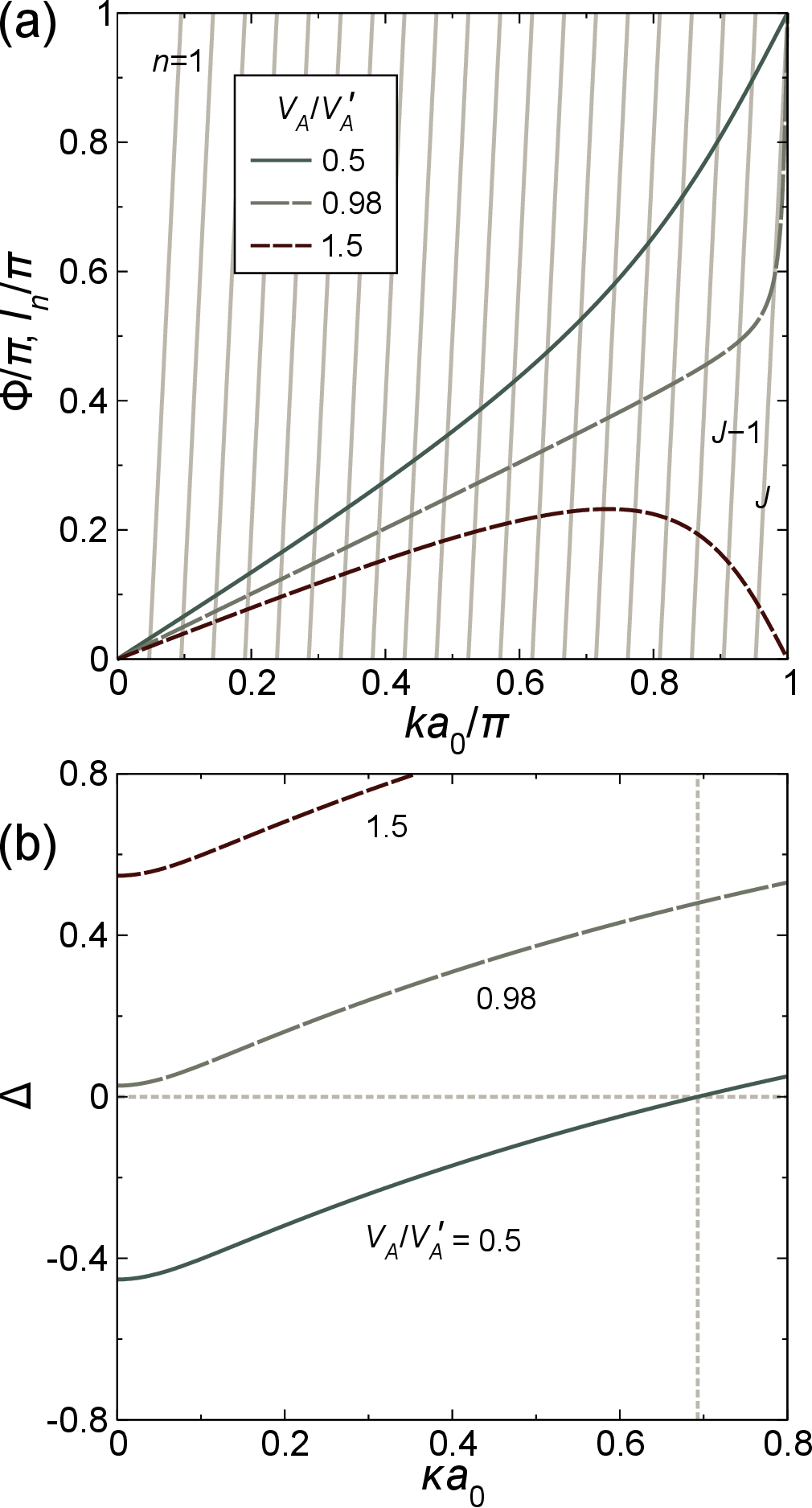}
\caption{ (a) Plot of $\Phi(k)$ for  $V_A/V_A^{\prime}=0.5,~0.98,$ and $1.5$. The straight lines are $l_n(k) =  (J+1) ka_0- n\pi$ for $n=1,~2,\dots,~J$, where $J=20$. (b) Plot of $\Delta(\kappa)$ for the same values of $V_A/V_A^{\prime}$. The vertical and horizontal dashed lines are located at $\kappa a_0 \approx 0.693 $ and $\Delta=0$, respectively.  }
\label{fig:sSSH-quant}
\end{center}
\end{figure}
%

The eigenvectors for the band $c\in\{s,0\}$ are
\begin{subequations}
    \begin{align}
|\Psi_{s}(k) \rangle = \frac{1}{\sqrt{2}}
    \begin{bmatrix}
        \cos \Theta(k) e^{i\Phi (k)}\\
        s \\
        \sin \Theta(k) 
    \end{bmatrix}
     \label{eq:sSSH-wfs}
\end{align}
for the dispersive bands, and
\begin{align}
    |\Psi_{0}(k) \rangle = 
    \begin{bmatrix}
        \sin \Theta(k) e^{i\Phi(k)}\\
        0 \\
        -\cos \Theta(k) 
    \end{bmatrix}
    \label{eq:sSSH-wf0}
\end{align}
\end{subequations}
for the flat band, where $\Theta(k) \equiv \tan^{-1} ( V_C/|F(k)| ) $. 

\subsection{Open boundary conditions}\label{subsec:boundSSH}

We derive the quantization condition of $k$ to enumerate the number of bulk states by imposing the boundary conditions on the wavefunction $\Psi_s =  \begin{pmatrix}
\Psi_s^A&\Psi_s^B&\Psi_s^C\end{pmatrix}^{T}$.

\subsubsection{Missing bulk states}

As illustrated by Fig. \ref{fig:sSSH-chain}, the missing $B$ site at $R_{J+1}$ and $A$ site at $R_0$ necessitate
\begin{align}
    \Psi_s^B(R_{J+1}) = 0,\label{eq:newBC1}\\
    \Psi_s^A(R_0)= 0\label{eq:newBC2}.
\end{align}
$\Psi_s(R_j)$ is constructed by a linear combination of the Bloch states with opposite momenta as follows~\cite{delplace2011,wakabayashi2010electronic}:
\begin{align}
    \Psi_s(R_j) \equiv \mathcal{A}_+ e^{i k R_j}|\Psi_s(k)\rangle + \mathcal{A}_- e^{-i k R_j}|\Psi_s(-k)\rangle.
 \label{eq:sSSH-wf-chain}
\end{align}
By letting $R_j = (j-J-1)a_0 $, Eq.~(\ref{eq:newBC1}) implies $\mathcal{A}_-=-\mathcal{A}_+$. Thus, Eq.~(\ref{eq:newBC2}) yields
 \begin{align}
    e^{-i(J+1)ka_0}{F}^*(k) = e^{i(J+1)ka_0}{F}(k)
    \label{eq:sSSH-q1}
\end{align}
or $\sin[(J+1)ka_0-\Phi(k)] = 0$. Equivalently, the quantization condition of $k$ for a finite $J$ is given by
\begin{align}
    (J+1)ka_0 - \Phi(k) = n\pi,~\mathrm{for}~n=1,~2,\dots,J.
    \label{eq:sSSH-qb}
\end{align}

In Fig.~\ref{fig:sSSH-quant}(a), we plot $\Phi(k)$ as a function of $k\in [0, \pi/a_0]$ for $V_A/V_A^{\prime}=0.5,~0.98,$ and $1.5$. The straight lines correspond to $l_n(k) \equiv (J+1)ka_0 - n\pi$, for $J=20$. The number of bulk states is equal to the number of solutions of Eq.~(\ref{eq:sSSH-qb}), which are given by the intersections of $\Phi(k)$ and $l_n(k)$ along $k\in (0, \pi/a_0)$. Here, $\Psi_s(R_j)$ vanishes identically at $k\in\{0,\pi/a_0\}$. There are $J$ solutions for $V_A/V_A^{\prime} = 0.98 $ and $1.5$. On the other hand, one solution is missing for $V_A/V_A^{\prime}=0.5$. The existence of $J-1$ solutions requires $\Phi(\pi/a_0)=\pi$ and $\partial_k\Phi(k)|_{\pi/a_0}<\partial_k l_J(k)|_{\pi/a_0}$, or
\begin{align}
    {V_A }/{V_A^{\prime}}  <1 - 1/(J+1).
    \label{eq:sSSH-cr}
\end{align}
We shall show that edge states emerge from the missing bulk states. The ratio $(V_A/V_A^\prime)_J\equiv 1-1/(J+1)$ is interpreted as the critical value of $V_A/V_A^\prime$ at which a bulk state for each $E<0$ and $E>0$ become edge states. In the limit $J\rightarrow\infty$, $(V_A/V_A^\prime)_\infty\sim 1$.


\subsubsection{Edge states}

By substituting $k = \pi/a_0 + i\kappa $ into Eqs. (\ref{eq:sSSH-F}) and (\ref{eq:sSSH-q1}),
\begin{align}
    e^{(J+1)\kappa a_0}\tilde{F}(-\kappa) = e^{-(J+1)\kappa a_0}\tilde{F}(\kappa),
    \label{eq:sSSH-q2}
\end{align}
where we define
\begin{align}
    \tilde{F}(\kappa) \equiv V_A - V_A^{\prime}e^{\kappa a_0}.
    \label{eq:sSSH-Fe}
\end{align}
Here, $1/\kappa>0$ is the localization length of edge states. By rearranging Eq.~(\ref{eq:sSSH-q2}), $\kappa$ satisfies
\begin{align}
    \Delta(\kappa) \equiv \frac{V_A}{V_A^{\prime}} - \frac{\sinh[J\kappa a_0]}{\sinh[(J+1) \kappa a_0]} = 0.
    \label{eq:sSSH-T}
\end{align}

In Fig. \ref{fig:sSSH-quant}(b), we plot $\Delta(\kappa)$ for $V_A/ V_A^{\prime}= 0.5,~0.98$ and $1.5$. Eq.~(\ref{eq:sSSH-T}) is satisfied for $V_A/ V_A^{\prime}= 0.5$, where $\Delta(\kappa)=0$ at $\kappa \approx 0.693/a_0 $ (vertical dashed line). Thus, the edge states exist for 
\begin{align}
       \frac{V_A}{V_A^{\prime}}   &= \frac{\sinh[J \kappa a_0]}{ \sinh[(J+1)\kappa a_0]}<1.
    \label{eq:sSSH-edge}
\end{align}

By inserting $k = \pi/a_0 + i\kappa $ into Eq. (\ref{eq:sSSH-Es}), the energies of edge states are given by
\begin{align}
   \tilde{E}_s(\kappa) =s\sqrt{{V_A}^2+{V_A^{\prime}}^2+{V_C}^2-2V_AV_A^{\prime}\cosh(\kappa a_0)}.
    \label{eq:sSSH-Ee}
\end{align}
By combining Eqs. (\ref{eq:sSSH-edge}) and (\ref{eq:sSSH-Ee}), we get
\begin{align}
    \tilde{E}_s(\kappa) \equiv s \sqrt{{V_A^{\prime}}^2\frac{\sinh^2(\kappa a _0)}{\sinh^2[(J+1)\kappa a_0]} + {V_C}^2 }.
    \label{eq:sSSH-Ee2}
\end{align}
We can see that for $J \kappa a_0\gg1$, $\tilde{E}_s(\kappa)$ becomes independent of $J$ and converges to $\tilde{E}_s \sim s |V_C|$.

\begin{figure}[t]
\begin{center}
\includegraphics[width=63mm]{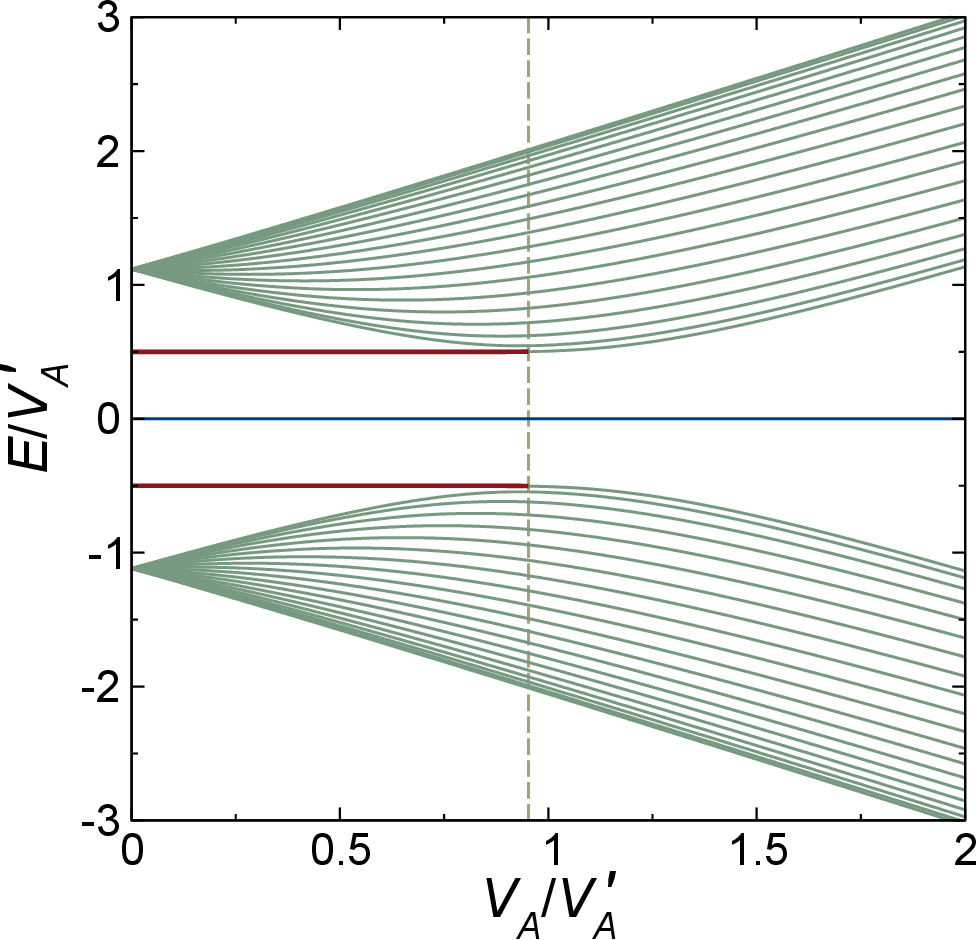}
\caption{ Energy spectra of the stub SSH chain as a function of $V_A/V_A^{\prime}$ for $V_C = 0.5 V_A^{\prime}$ and $J=20$. The bold red lines indicate the edge states. The blue line at $E=0$ is the flat band. The transition from bulk to edge states occurs at $(V_A/V_A^\prime)_{20}\approx 0.9524$ (vertical dashed line), where $|E|\sim|V_C|$. }
\label{fig:sSSH-bands}
\end{center}
\end{figure}

Fig. \ref{fig:sSSH-bands} shows the numerical calculation of energy spectra as a function of $V_A/V_A^\prime$ for $V_C =0.5 V_A^{\prime}$ and $J=20$. Since each unit cell consists of three sites, there exist $3J$ bands in total: $J$ pairs of dispersive bands and a flat band with $J$-fold degeneracy at $E=0$ (blue line). The edge states are depicted by the bold red lines. The transition from bulk to edge states is marked by the vertical dashed line at $(V_A/V_A^\prime)_{20} \approx 0.9524 $, where the in-gap bands become constant at $|E|\sim |V_C|$. Therefore, the emergence of edge states is indicated by the flattening of the in-gap bands when $V_A/V_A^\prime < 1$. Our theoretical result is consistent with a recent experiment \cite{caceres2022experimental}~that realized stub SSH chain using a photonic lattice.

\subsection{Topological invariant}\label{subsec:topoinv}

We have demonstrated that the emergence of edge states in the stub SSH chain is irrespective of $V_C$ and controlled only by $V_A/V_A^\prime$, similar to the SSH chain. Nevertheless, the presence of the coupled $C$ sites breaks the inversion symmetry \cite{bartlett2021illuminating}, and as a consequence, the Zak phase $\displaystyle\mathcal{Z}_c\equiv i\int_{\mathrm{BZ}} dk \langle \Psi_c (k) | \partial_k |\Psi_c(k)\rangle $ is no longer quantized to $\pi$ and $0$ (mod $2\pi$). By using $|\Psi_c\rangle$ in Eqs.~(\ref{eq:sSSH-wfs}) and (\ref{eq:sSSH-wf0}), $\mathcal{Z}_c$'s are analytically calculated as follows:
\begin{subequations}
    \begin{widetext}
    \begin{align}
    \mathcal{Z}_s &= -\int_{0}^{{\pi}/{a_0}} dk \cos^2\Theta(k) \frac{\partial \Phi(k)}{\partial k}= -\frac{\pi}{2}\left [  1  -  \frac{{V_A}^2 - {V_A^{\prime}}^2 + {V_C}^2}{\sqrt{(V_A - V_A^{\prime})^2 +{V_C}^2}{\sqrt{(V_A + V_A^{\prime})^2 +{V_C}^2}} }        \right],
    \label{eq:sSSH-Zs}\\
    \mathcal{Z}_0 &= -2\int_{0}^{{\pi}/{a_0}} dk \sin^2\Theta(k) \frac{\partial \Phi(k)}{\partial k}= -\pi\left [ 1 - \mathrm{sgn}\Big({V_A}^2 - {V_A^\prime}^2\Big)        \right] -2\mathcal{Z}_s.
     \label{eq:sSSH-Z0}
\end{align}
\end{widetext}
\end{subequations}


The case $V_C = 0$ corresponds to the SSH chain with an additional flat band originating from the uncoupled $C$ sites, where $\mathcal{Z}_s = -\pi$ and $0$ for $V_A/V_A^{\prime}<1$ and $V_A/V_A^{\prime}>1$, respectively, and hence $|\mathcal{Z}_s|= \Phi(\pi/a_0)$. This relation does not hold for $V_C\neq 0$. On the other hand, it was shown \cite{caceres2022experimental}~that the edge states in the stub SSH chain are robust against disorders that perturb $V_A$ and $V_A^\prime$, which suggests the topologically non-trivial character of the bulk bands. By using the Majorana representation of the eigenstates, Bartlet~\textit{et al.}~\cite{bartlett2021illuminating} found that the existence of the edge states can be predicted from the azimuthal winding number ${W}_c$ on the Bloch sphere. We complement the finding with an analytical proof that $|{W}_c|=\Phi(\pi/a_0)$.




The Majorana representation is a method to visualize the eigenstate $|\Psi\rangle$ of an $N\times N$ Hamiltonian on the Bloch sphere by a set pseudospinors for $S=1/2$. Here, $|\Psi\rangle$ is treated as a pseudospinor for $S=(N-1)/2$. According to the Schwinger theory~\cite{schwinger1952} of angular momentum, $|\Psi\rangle$ can be constructed by the bosonic creation operators $\hat{a}_{\uparrow,\downarrow}^{\dagger}$ acting on the vacuum state $|\varnothing \rangle$, as follows:
\begin{align}
    |\Psi \rangle  = \sum_{\mu = -S} ^{S} \mathcal{C}_\mu \frac{ \left(  \hat{a}_{\uparrow}^{\dagger}  \right)^{S+\mu}  }{   \sqrt{(S+\mu)!}  } \frac{ \left(  \hat{a}_{\downarrow}^{\dagger}  \right)^{S-\mu}  }{   \sqrt{(S-\mu)!}  } | \varnothing  \rangle,
    \label{eq:Maj-1}
\end{align}
where $\mathcal{C}_{\mu}$'s are the basis vector. Earlier, Majorana \cite{majorana1932atomi} discovered that $|\Psi\rangle$ is given by a product of the Pauli spinors $|\zeta_\mu\rangle = \begin{bmatrix} \cos(\eta_\mu/2) & \sin(\eta_\mu/2)e^{i\xi_\mu}  \end{bmatrix}^{T}$ as follows:
\begin{align}
    |\Psi \rangle &= \frac{1}{\mathcal{K}} \prod_{\mu=1}^{2S}\left[  \cos\left( \frac{\eta_\mu}{2}  \right) \hat{a}_{\uparrow}^{\dagger}      + \sin\left( \frac{\eta_\mu}{2}  \right)  e^{i\xi_\mu} \hat{a}_{\downarrow}^{\dagger} \right]| \varnothing  \rangle\nonumber\\
    &= \frac{1}{\mathcal{K}} \prod_{\mu=1}^{2S}|\zeta_\mu \rangle   ,
    \label{eq:Maj-2}
\end{align}
where $\mathcal{K}$ is the normalization constant. $\eta$ and $\xi$ are the polar and azimuthal angles. The trajectory of each Majorana 'star' $\boldsymbol{\zeta}_\mu \equiv \langle \zeta_\mu | \boldsymbol{\sigma} | \zeta_\mu \rangle$ thus represents the evolution of $|\Psi \rangle $. Alternatively, Eq.~(\ref{eq:Maj-2}) is expressed as
\begin{align}
    |\Psi \rangle  = \frac{\mathcal{C}_S}{\sqrt{(2S)!}}\prod_{\mu=1}^{2S}\left[  \hat{a}_{\uparrow}^{\dagger} + \lambda_\mu  \hat{a}_{\downarrow}^{\dagger}         \right ]| \varnothing  \rangle,
    \label{eq:Maj-3}
\end{align}
where $\lambda_{\mu} = \tan (\eta_\mu/2)e^{i\xi_\mu}$. Suppose that $\lambda_\mu$'s are the roots of $\prod_{\mu=1}^{2S}(\lambda - \lambda_\mu)=0$. By comparing the coefficients of Eqs. (\ref{eq:Maj-1}) and (\ref{eq:Maj-3}), $\lambda_\mu$ are given by the Majorana polynomial as follows:
\begin{align}
    \sum_{\mu=0}^{2S}(-1)^\mu\frac{ \mathcal{C}_S-\mu}{\sqrt{(2S-\mu)!\mu !}}\lambda^{2S-\mu} = 0.
    \label{eq:Maj-poly}
\end{align}

Particularly for the stub SSH chain, we insert $S=1$, where $|\Psi\rangle = \begin{pmatrix}\mathcal{C}_1 & \mathcal{C}_0 & \mathcal{C}_{-1}\end{pmatrix}^{T}$ is given by Eqs. (\ref{eq:sSSH-wfs}) or (\ref{eq:sSSH-wf0}). For the dispersive band, Eq. (\ref{eq:Maj-poly}) is reduced to
  \begin{subequations}
    \begin{align}
    {\cos\Theta(k)}e^{i\Phi(k)}{\lambda^2} -\sqrt{2}s\lambda+{\sin\Theta(k)} = 0.
    \label{eq:sSSH-Ms}
\end{align}
As for the flat band, we get
\begin{align}
    {\sin\Theta(k)}e^{i\Phi(k)}{\lambda^2} -{\cos\Theta(k)} = 0.
    \label{eq:sSSH-M0}
\end{align}
\end{subequations}


The winding number $\Omega_{c}$ around the origin of complex plane is calculated by the Cauchy integral formula. Let $\lambda_{s}^{\mp} (k)$ and $\lambda_{0}^{\mp}(k)$ be the roots of Eqs. (\ref{eq:sSSH-Ms}) and (\ref{eq:sSSH-M0}), respectively. From the coefficients of the quadratic equations, it is noted that 
\begin{subequations}
    \begin{align}
    \Lambda_{s}(k)&\equiv \prod_{\nu=\mp}\lambda_{s}^{\nu}(k)=\tan \Theta(k)e^{-i\Phi (k)}, \label{eq:lambda-dis}\\
   \Lambda_{0}(k)&\equiv \prod_{\nu=\mp}\lambda_{0}^{\nu}(k) =-\cot \Theta(k) e^{-i\Phi(k)}. \label{eq:lambda-0}
\end{align}
\end{subequations}
By regarding $\Lambda_{c}(k)$ as a curve parameterized by $k$, 
\begin{align}
    \Omega_{c}  \equiv \frac{1}{2\pi i} \int_{-{\pi}/{a_0}}^{{\pi}/{a_0}} \frac{d k}{\Lambda_{c}(k)} \frac{\partial \Lambda_{c} (k) }{ \partial k  }.
    \label{eq:sSSH-winding}
\end{align}
Here, $d\mathrm{ln}\Lambda_s(k) = d \mathrm{ln} |\tan\Theta(k)| - id\Phi(k)$ and $d\mathrm{ln}\Lambda_0(k) = d \mathrm{ln} |\cot\Theta(k)| - id\Phi(k)$. It is noted that $\Lambda_{c}(k)$ is a closed curve because $\Theta(-\pi/a_0) = \Theta(\pi/a_0)$, thus the total changes of $\mathrm{ln} |\tan\Theta(k)|$ and $\mathrm{ln} |\cot\Theta(k)|$ are zero. Therefore, $\Omega_c$ is identical for each $c$ as follows:
\begin{align}
    \Omega_{c} = -\frac{1}{2\pi}\int_{-\pi/a_0}^{\pi/a_0} dk \frac{\partial \Phi(k)}{\partial k}  = -\frac{1}{\pi} \Phi \left( \frac{\pi}{a_0} \right).
    \label{eq:sSSH-winding2}
\end{align}

\begin{figure}[t]
\begin{center}
\includegraphics[width=63mm]{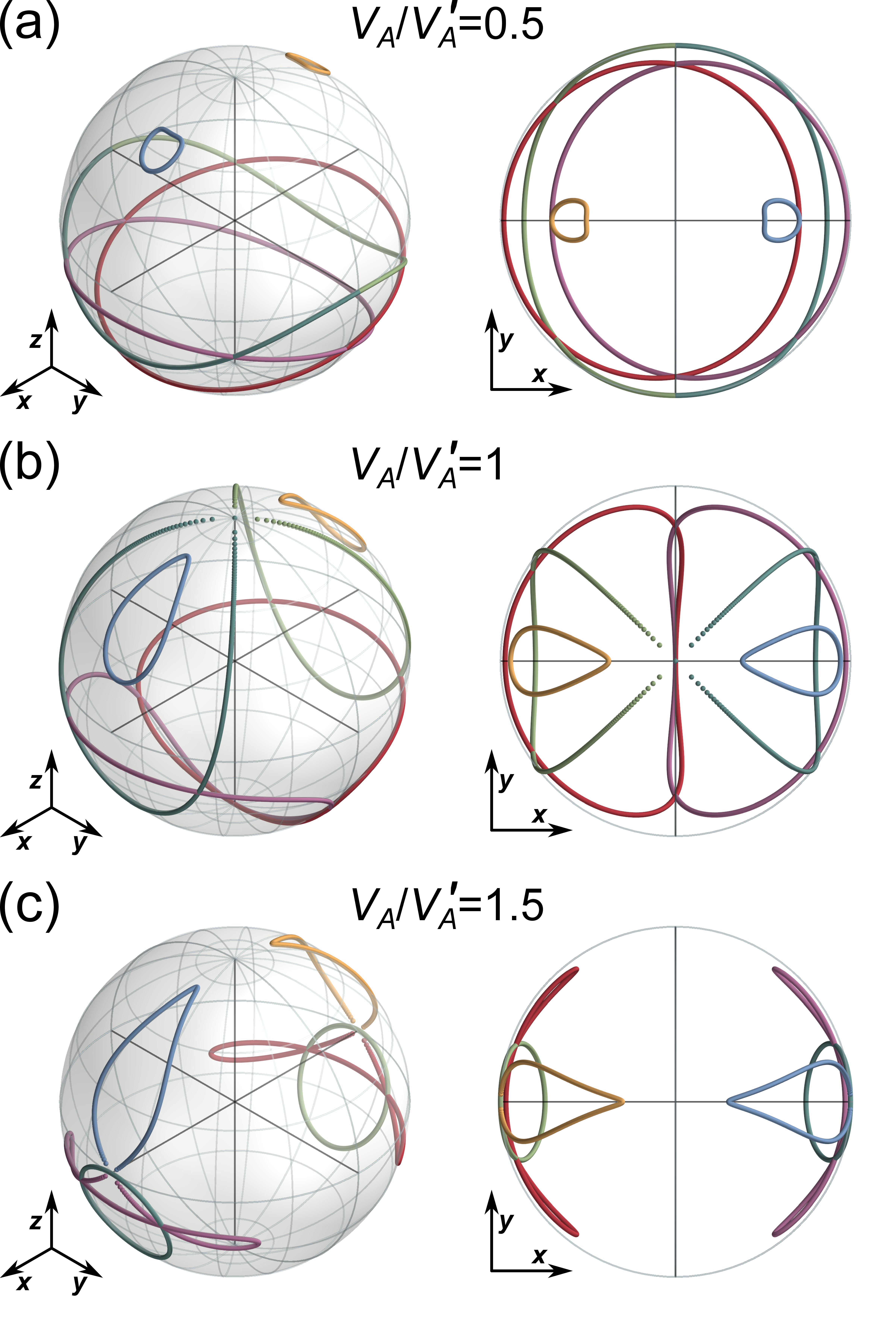}
\caption{ Trajectories of the Majorana stars on the Bloch sphere (left panels) and the $xy$ plane (right panels) for $V_C=0.5V_A^\prime$, and (a) $V_A/V_A^\prime = 0.5$, (b) $1$, (c) $1.5$. The red and yellow (blue and purple) curves correspond to the valence (conduction) band, $c=-1$ ($c=+1$). The light-green and dark-green curves correspond to the flat band, $c=0$.    }
\label{fig:sSSH-Majorana}
\end{center}
\end{figure}


The azimuthal winding number ${W}_c$ counts for the total number of times $\boldsymbol{\zeta}_{c}^\nu(k)$'s travel around the $z$ axis of the Bloch sphere. ${W}_c$ is defined by~\cite{bartlett2021illuminating}:
\begin{align}
    {W}_c \equiv \frac{1}{2\pi}   \int_{-\pi/a_0}^{\pi/a_0} dk\frac{\partial}{\partial k} \sum_{\nu=\mp}\xi_c^\nu(k).
     \label{eq:sSSH-windingZ}
\end{align}
Since $\lambda_{c}^{\nu}(k) = \tan[\eta_{c}^{\nu}(k)/2] e^{i\xi_c^\nu(k)} $ by definition,
\begin{align}
   -\sum_{\nu=\mp} \xi_{c}^\nu(k) = \Phi(k).
    \label{eq:sSSH-xi}
\end{align}
By substituting Eq. (\ref{eq:sSSH-xi}) into Eq.~(\ref{eq:sSSH-windingZ}), it is inferred that ${W}_{c} = \Omega_c$, and accordingly:
\begin{align}
 |{W}_{c}| =
 \begin{cases}
     1~\mathrm{for}~V_A/V_A^\prime<1,\\
    0~\mathrm{for}~V_A/V_A^\prime>1.
 \end{cases}
\label{eq:sSSH-windingZ3}
 \end{align}
Therefore, the trajectories of $\boldsymbol{\zeta}_{c}^{\nu}(k)$'s enclose the $z$ axis once only if $\Phi(\pi/a_0) = \pi$, as $k$ is traversing the BZ.



Fig. \ref{fig:sSSH-Majorana} shows the trajectories of $\boldsymbol{\zeta}_{c}^\nu$'s on the Bloch sphere (left panels) and their projections on the $xy$ plane (right panels) for $V_C = 0.5 V_A^{\prime}$, and (a) $V_A/V_A^{\prime} =0.5$, (b) $1$, (c) $1.5$. $\boldsymbol{\zeta}_{c}^\nu$'s associated with $c=-1$, $0$, and $1$ are depicted by the red/yellow, light-green/dark-green, and blue/purple curves, respectively. In Fig. \ref{fig:sSSH-Majorana}(a), the red, purple, and connected green loops wind around the $z$ axis, so $|{W}_{c}|=1$. In Fig. \ref{fig:sSSH-Majorana}(b), $|{W}_{c}|$ is ill-defined because the same loops intersect the poles. This is an example of topological phase transition without gap closing, which can occur provided that the topological invariant at the transition point becomes ill-defined~\cite{ezawa2013topological}. In Fig. \ref{fig:sSSH-Majorana}(c), none of the curves enclose the $z$ axis and thus $|{W}_c| = 0$. 

It is noted that Eqs. (\ref{eq:sSSH-windingZ3}) does not hold for all of the bases of $|\Psi_c\rangle$. Hereafter, the basis in Eqs.~(\ref{eq:sSSH-wfs}) or (\ref{eq:sSSH-wf0}) is referred to as $(abc)$ for simplicity. There are six permutations of basis in total: $(abc)$, $(bca)$, $(cab)$, $(cba)$, $(acb)$, and $(bac)$. In general, the $\mathbb{Z}_2$ invariant is given by the parity $P$ of the total winding numbers defined by~\cite{bartlett2021illuminating}:
\begin{align}
    P\equiv (-1)^{\sum_{c}|W_c|},
    \label{eq:parity}
\end{align}
where $P=-1$ and $+1$ indicate the non-trivial and trivial phases, respectively. We prove this statement by calculating the Majorana polynomials for all the bases, which are classified into three cases as follows:
\newline
1. For $(abc)$ and $(cba)$, $\Lambda_c\propto e^{- i\Phi}$ and $e^{ i\Phi}$, respectively. 
\newline
2. For $(bca)$ and $(acb)$, $\Lambda_s\propto e^{ i\Phi}$ and $ e^{- i\Phi}$, respectively, while a single solution of the Majorana polynomial for the flat band is $\lambda_0\propto e^{ i\Phi}$ and $e^{- i\Phi}$, respectively.
\newline
3. For $(cab)$ and $(bac)$, $\Lambda_s$ does not depend on $\Phi$, and $\lambda_0\propto e^{ i\Phi}$ and $e^{-i\Phi}$, respectively, for the single solution.
\newline
By calculating $W_c$, $\sum_{c}|W_c|=0$ for all cases when $V_A/V_A^\prime>1$. In contrast, $\sum_{c}|W_c|$ is either $ 3$ (cases 1 and 2) or $ 1$ (case 3) when $V_A/V_A^\prime<1$. Therefore,
\begin{align}
    P=
    \begin{cases}
    -1~\mathrm{for}~V_A/V_A^\prime<1,\\
    +1~\mathrm{for}~V_A/V_A^\prime>1.
    \end{cases}
    \label{eq:parity2}
\end{align}



\section{$\alpha$-${T}_3$ lattice}\label{sec:aT3-bulk}



\begin{figure}[t]
\begin{center}
\includegraphics[width=70mm]{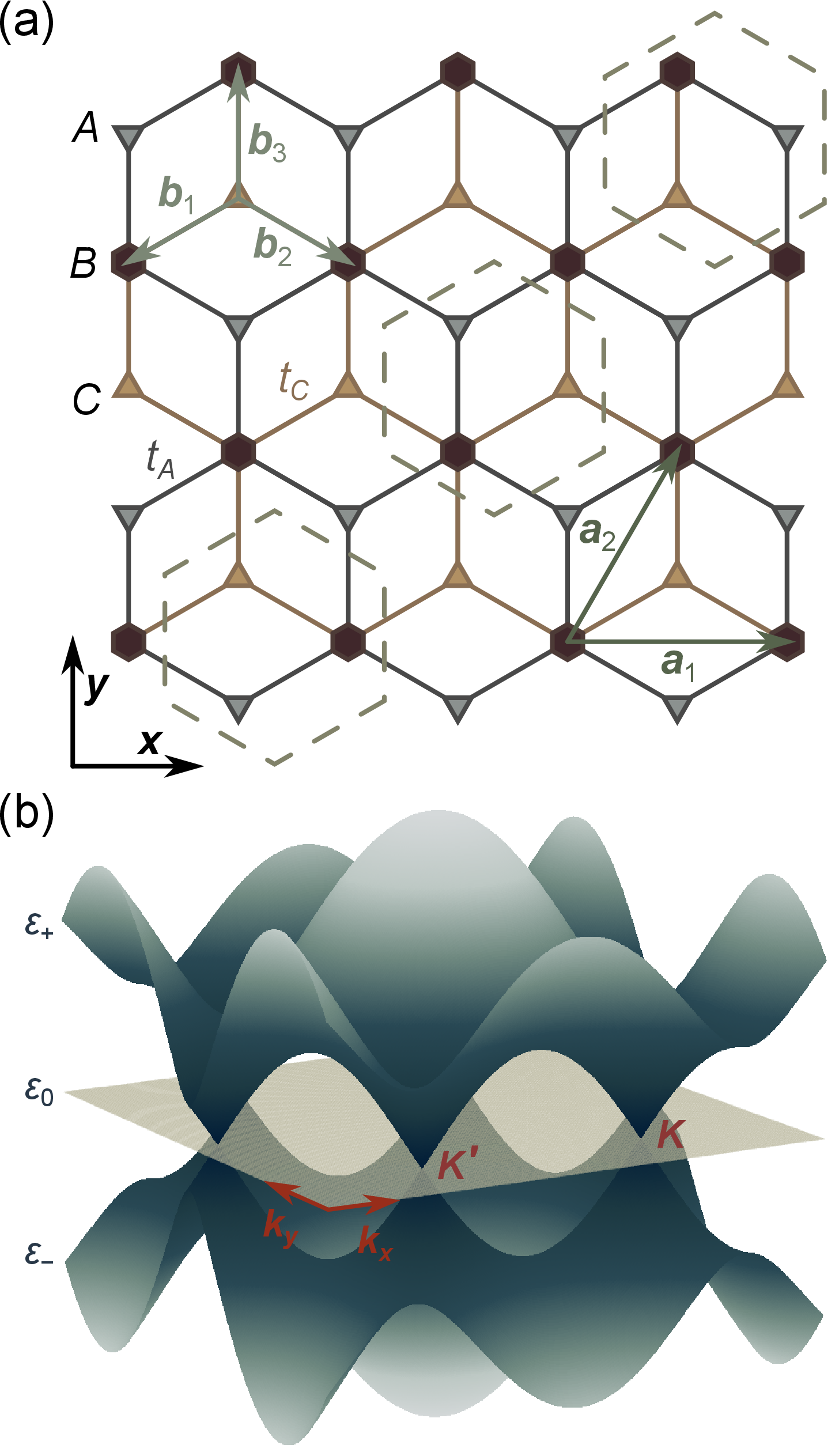}
\caption{  (a) The $\alpha $-$ {T}_3$ lattice. The hopping parameter between the $B$ and $A$ ($C$) sites is $t_A$ ($t_C$), with $t_C/t_A=\alpha\in[0,1]$. The dashed hexagons indicate three choices to define the unit cell. $\boldsymbol{b}_1$, $\boldsymbol{b}_2$, and $\boldsymbol{b}_3$ are the nearest neighbor vectors. $\boldsymbol{a}_1$ and $\boldsymbol{a}_2$ are the primitive vectors. (b) Energy dispersion of the $\alpha $-$ {T}_3$ lattice, where $\varepsilon_{-}$, $\varepsilon_{0}$, and $\varepsilon_{+}$ are the valence, flat, and conduction bands, respectively. }
\label{fig:aT3}
\end{center}
\end{figure}

Consider a honeycomb lattice composed of $A$ and $B$ sites. The $\alpha$-${T}_3$ lattice is constructed by connecting the additional $C$ sites at the center of each hexagon to the nearest $B$ sites, as illustrated by Fig.~\ref{fig:aT3}(a). The nearest-neighbour interactions between the $B$ and $A$ ($C$) sites are denoted by the hopping parameter $t_A>0$ ($t_C$), where $t_C/t_A=\alpha\in [0, 1]$.  The dashed hexagons represent three choices to define the unit cell. The vectors $\boldsymbol{b}_1 = a(-{1}/{2}, -{1}/{2\sqrt{3}} )$, $\boldsymbol{b}_2 = a({1}/{2}, -{1}/{2\sqrt{3}} )$ and $\boldsymbol{b}_3 = a(0, {1}/{\sqrt{3}} )$ connect the $B$ site to the nearest $A$ and $C$ sites, where $a$ is the length of primitive vectors $\boldsymbol{a}_1 = a(1,0) $ and $\boldsymbol{a}_2 = a({1}/{2},{\sqrt{3}}/{2}) $. 


The bulk Hamiltonian $h(\boldsymbol{k})$ in the momentum space $\boldsymbol{k} = (k_x, k_y)$ is given by 
\begin{align}
h(\boldsymbol{k}) = - \begin{bmatrix}
     0   & t_A f(\boldsymbol{k})   & 0\\
     t_A f^*(\boldsymbol{k}) & 0 & t_C f(\boldsymbol{k})\\
     0 & t_C f^*(\boldsymbol{k}) & 0
\end{bmatrix},
\label{eq:H1}
\end{align}
where we define
\begin{align}
    f(\boldsymbol{k}) \equiv \sum_{n=1}^3 e^{-i\boldsymbol{k}\cdot \boldsymbol{b}_n} = |f(\boldsymbol{k})|e^{-i\varphi(\boldsymbol{k})}.
    \label{eq:f}
\end{align}
By defining $t \equiv \sqrt{{t_A}^2 + {t_C}^2}$ and $\vartheta \equiv \tan^{-1} \alpha$, 
\begin{align}
h(\boldsymbol{k}) = -t \begin{bmatrix}
     0   & \cos\vartheta f(\boldsymbol{k})   & 0\\
     \cos\vartheta f^*(\boldsymbol{k}) & 0 & \sin\vartheta f(\boldsymbol{k})\\
     0 & \sin\vartheta f^*(\boldsymbol{k}) & 0
     \label{eq:H2}
\end{bmatrix}.
\end{align}
For $\alpha = 0$, the $\alpha$-${T}_3$ lattice reduces to graphene with a zero-energy flat band that originates from the uncoupled $C$ sites. 
\begin{figure*}[t]
\begin{center}
\includegraphics[width=160mm]{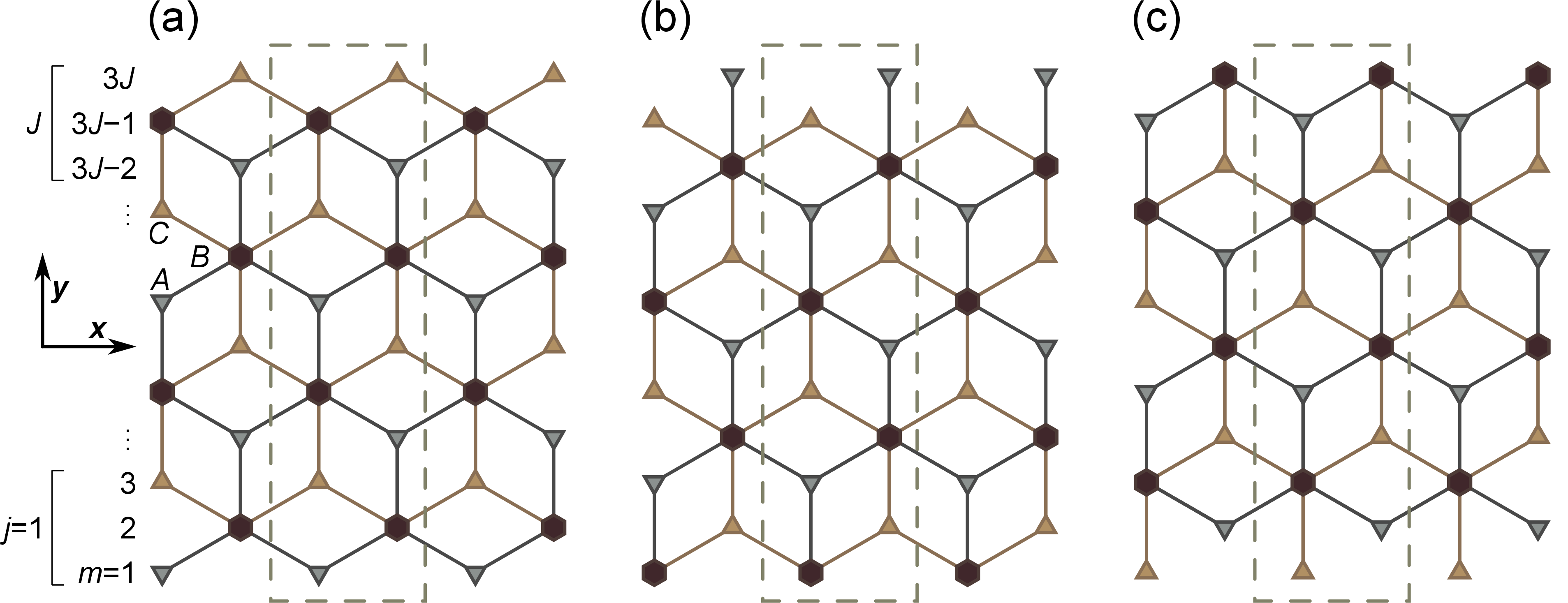}
\caption{ The commensurate $\alpha$-${T}_3$ ZRs: (a) ABC (b) BCA and (c) CAB ZRs. In each supercell (dashed rectangle), the positions of site and trimer along the $y$ axis are indicated by $m$ and $j$, respectively.  }
\label{fig:ribbon}
\end{center}
\end{figure*}

Since $h(\boldsymbol{k})$ is scaled by $t$, the energy eigenvalues of Eq.~(\ref{eq:H2}) are independent of $\alpha$ as follows:
\begin{subequations}
    \begin{align}
    \varepsilon_{s}(\boldsymbol{k}) &= st \sqrt{1+\beta^2(k_x) + 2\beta(k_x) \cos\left(\frac{\sqrt{3}}{2}k_y a\right)}, \label{eq:Energy-s}\\
    \varepsilon_0 &= 0,
    \label{eq:Energy-0}
\end{align}
\end{subequations}
where $\beta(k_x) \equiv 2 \cos (k_x a/2)\geq 0$ for $k_x\in[-\pi/a,\pi/a]$. $\varepsilon_{s}$ indicates the energy of the valence (conduction) band for $s=-1$ ($+1$). In addition, $\varepsilon_0$ appears due to the sublattice imbalance. In Fig.~\ref{fig:aT3}(b), the Dirac cones touch $\varepsilon_0$ at the $\boldsymbol{K}$ and $\boldsymbol{K^{\prime}}$ points (labeled by $\tau=+1$ and $-1$, respectively) in the corners of hexagonal BZ.

The eigenvector of $\varepsilon_s$ is given by
\begin{subequations}
    \begin{align}
|\psi_{s}(\boldsymbol{k}) \rangle = \frac{1}{\sqrt{2}}
    \begin{bmatrix}
        \cos \vartheta e^{-i\varphi (\boldsymbol{k})}\\
        s \\
        \sin \vartheta e^{i\varphi (\boldsymbol{k})}
    \end{bmatrix}.
    \label{eq:wavefunction-s}
    \end{align}
As for $\varepsilon_0$, we get
\begin{align}
    |\psi_{0}(\boldsymbol{k}) \rangle = 
    \begin{bmatrix}
        \sin \vartheta e^{-i\varphi (\boldsymbol{k})}\\
        0 \\
        -\cos \vartheta e^{i\varphi (\boldsymbol{k})}
    \end{bmatrix}.
    \label{eq:wavefunction-0}
\end{align}
\end{subequations}
The Berry phase $\displaystyle\Gamma_c \equiv i\oint d\boldsymbol{k} \cdot \langle \psi_c (\boldsymbol{k})|\nabla_{\boldsymbol{k}} | \psi_c (\boldsymbol{k}) \rangle $ can be analytically evaluated by contour integration along a path enclosing each Dirac point, as follows \cite{raoux2014dia}:
\begin{subequations}
\begin{align}
    \Gamma_{s,\tau} &= \pi\tau \cos(2\vartheta),
    \label{eq:Berry-s}\\
    \Gamma_{0,\tau} &= -2\pi\tau \cos(2\vartheta).
    \label{eq:Berry-0}
\end{align}
\end{subequations}
Thus, $\Gamma_{c} \neq \pi$ or $0$ (mod $2\pi$) for $\alpha \neq 0$ or $1$. Furthermore, $\Gamma_{c}$'s possess opposite signs for the ${K}$ and ${K^{\prime}}$ valleys except for $\alpha=0$ and $1$.

\section{ $\alpha $-$ {T}_3$ ZRs}\label{sec:aT3-ZR}

\subsection{Types of ZR}\label{subsec:type}

Fig. \ref{fig:ribbon} illustrates commensurate $\alpha$-${T}_3$ ZRs that we call (a) ABC, (b) BCA, and (c) CAB ZRs. The ZRs are respectively constructed by translating the bottom, middle, and top unit cells in Fig.~\ref{fig:aT3}(a) along the $x$ direction with the periodicity $\boldsymbol{a}_1$. The width of the ZR is characterized by a positive integer $J$, which is the number of trimers in the supercell (dashed rectangle). Starting from the bottom edge, the indices $m=1,2,\dots,3J$ and $j=1,2,\dots,J$ specify the positions of sites and trimers, respectively, in the $y$ axis. Additionally, we can construct two non-commensurate ZRs from each commensurate ZR. First, by eliminating the sites at $m=3J$, and second, by eliminating the sites at $m=3J$ and $m=3J-1$. Therefore, there are nine types of $\alpha $-${T}_3$ ZRS: three commensurate and six non-commensurate ZRs.

Here, we will discuss the BBCs for the BCA and CAB ZRs. The existence of edge states in the ABC ZR is discussed in Appendix~\ref{app:ABC}.

\subsection{Open boundary conditions for ZRs}\label{subsec:boundarycond}

In an infinite system, the TBEs with nearest-neighbor interactions at the $A$, $B$, $C$ sites are given by
    \begin{align}
    &\varepsilon\psi_s^A(\boldsymbol{r})=-t_A\sum_{n=1}^3\psi_s^B(\boldsymbol{r}-\boldsymbol{b}_n),\label{eq:tbA}\\
    &\varepsilon\psi_s^B(\boldsymbol{r})=-\sum_{n=1}^3[t_A\psi_s^A(\boldsymbol{r}+\boldsymbol{b}_n)+t_C\psi_s^C(\boldsymbol{r}-\boldsymbol{b}_n)],\label{eq:tbB}\\
    &\varepsilon\psi_s^C(\boldsymbol{r})=-t_C\sum_{n=1}^3\psi_s^B(\boldsymbol{r}+\boldsymbol{b}_n)\label{eq:tbC},
\end{align}
respectively. The missing terms of TBEs on the edges of ZRs constitute the boundary conditions. The wavefunction ${\psi}_s=\begin{pmatrix}
        {\psi}_s^A& {\psi}_s^B& {\psi}_s^C
    \end{pmatrix}^{T}$ is given by a linear combination of the Bloch states~\cite{delplace2011}:
\begin{align}
    {\psi}_s(\boldsymbol{r})  
    \equiv \mathcal{A}{e^{i\boldsymbol{k}\cdot\boldsymbol{r}}}|\psi_{s}(\boldsymbol{k})\rangle+\mathcal{A}^{\prime}{e^{i\boldsymbol{k^\prime}\cdot\boldsymbol{r}}}|\psi_{s}(\boldsymbol{k^\prime})\rangle.
    \label{eq:psi-ribbon}
\end{align}
The amplitudes  $\mathcal{A}$ and $\mathcal{A}^\prime$, and momenta $\boldsymbol{k}$ and $\boldsymbol{k^\prime}$ are determined to satisfy the boundary conditions.

%


\subsubsection{BCA ZR}

In Fig. \ref{fig:boundaries}(a), the vector $\boldsymbol{L}=(L_x,L_y)$ connects an empty $A$ site at $m=0$ to an empty $B$ site at $m=3J+1$ chosen as the origin $\boldsymbol{O}=(0,0)$. $L_x=0$ and $-a/2$ for odd and even $J$, respectively, while
\begin{align}
    L_y = \rho-\frac{a}{\sqrt{3}},~\mathrm{with}~\rho\equiv\frac{\sqrt{3}}{2}(J+1)a.
    \label{eq:Ly}
\end{align}

Let us define $\boldsymbol{L}_l\equiv\boldsymbol{L}+l\boldsymbol{a}$ and $\boldsymbol{O}_l\equiv\boldsymbol{O}+l\boldsymbol{a}$, for $l\in \mathbb{Z}$. Thus, $\boldsymbol{L}_0=\boldsymbol{L}$ and $\boldsymbol{O}_0=\boldsymbol{O}$. Consider the TBEs at $m=1$ and $m=3J$ where $\boldsymbol{r}=-\boldsymbol{L}_l-\boldsymbol{b}_2$ and $\boldsymbol{r}=-\boldsymbol{O}_l+\boldsymbol{b}_1$, respectively, as follows: 
    \begin{align}
    \varepsilon\psi_s^B(-\boldsymbol{L}_l-\boldsymbol{b}_2)=&-t_C\psi_s^C(-\boldsymbol{L}_l+\boldsymbol{b}_3)\nonumber\\
      &-t_C\psi_s^C(-\boldsymbol{L}_{l+1}+\boldsymbol{b}_3)\nonumber\\
    &-t_A\psi_s^A(-\boldsymbol{L}_{l+1}+\boldsymbol{a}_2),\label{eq:SEB1}\\
    \varepsilon\psi_s^A(-\boldsymbol{O}_l+\boldsymbol{b}_1)=
    &-t_A\psi_s^B(-\boldsymbol{O}_l-\boldsymbol{a}_2).\label{eq:SEB3}
\end{align}
The boundary conditions are thus given by
    \begin{align}
    {\psi}_{s}^{A}(-\boldsymbol{L}-l\boldsymbol{a}_1) + {\psi}_{s}^{A}(-\boldsymbol{L}-[l+1]\boldsymbol{a}_1)&\nonumber\\
    +\alpha {\psi}_{s}^{C}(-\boldsymbol{L}-l\boldsymbol{a}_1+\boldsymbol{b}_1)&=0,
    \label{eq:boun-B1} \\
        {\psi}_{s}^{B}(-l\boldsymbol{a}_1)&= 0.
    \label{eq:empty-Psi-B}
\end{align}


\subsubsection{CBA ZR}

\begin{figure}[t]
\begin{center}
\includegraphics[width=86.5mm]{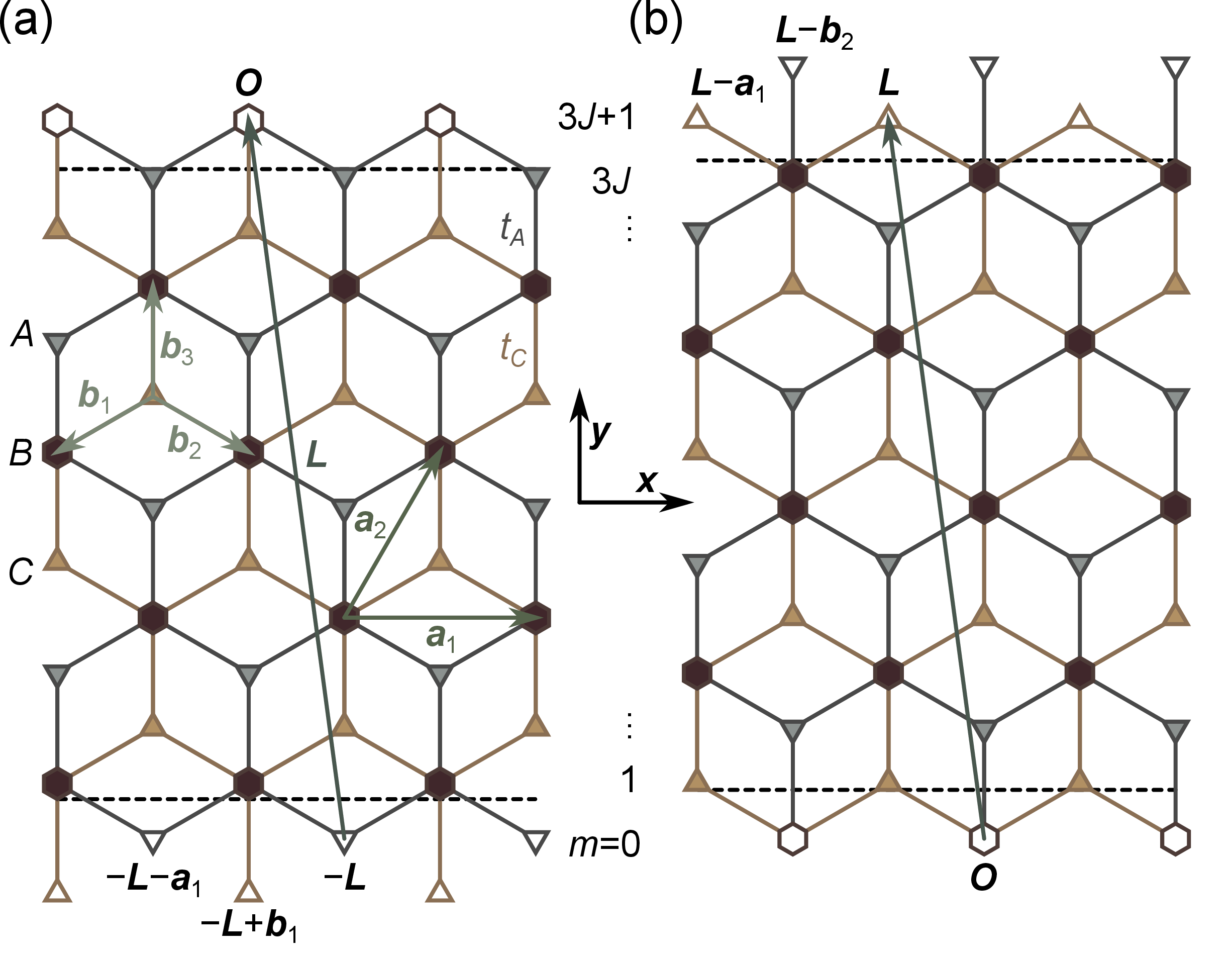}
\caption{Schematics for the boundary conditions of (a) BCA and (b) CAB ZRs. }
\label{fig:boundaries}
\end{center}
\end{figure}

Fig.~\ref{fig:boundaries}(b) shows that $\boldsymbol{O}$ is assigned to an empty $B$ site at $m=0$. The TBEs at $m=1$ and $m=3J$ are
 \begin{align}
    \varepsilon\psi_s^C(\boldsymbol{O}_l-\boldsymbol{b}_2)=
    &-t_C\psi_s^B(\boldsymbol{O}_{l-1}+\boldsymbol{a}_2),\label{eq:SEC1}\\
\varepsilon\psi_s^B(\boldsymbol{L}_l+\boldsymbol{b}_1)=
&-t_A\psi_s^A(\boldsymbol{L}_l-\boldsymbol{b}_3)\nonumber\\
&-t_A\psi_s^A(\boldsymbol{L}_{l-1}-\boldsymbol{b}_3)\nonumber\\
&-t_C\psi_s^C(\boldsymbol{L}_l-\boldsymbol{a}_2).\label{eq:SEC3}
\end{align}   
The boundary conditions given as follows:
    \begin{align}
    {\psi}_{s}^{B}(l\boldsymbol{a}_1)&= 0,
    \label{eq:empty-Psi-B2}\\
    \alpha\left\{ {\psi}_{s}^{C}(\boldsymbol{L}+l\boldsymbol{a}_1) + {\psi}_{s}^{C}(\boldsymbol{L}+[l-1]\boldsymbol{a}_1) \right\}\nonumber\\
    + {\psi}_{s}^{A}(\boldsymbol{L}+l\boldsymbol{a}_1-\boldsymbol{b}_2)&=0.
    \label{eq:boun-C1}
    \end{align}


\subsection{The existence of edge states in ZRs}\label{subsec:edgeexist}

\subsubsection{BCA ZR}

In Appendix~\ref{sapp:deriv1}, we show in detail that by combining Eqs. (\ref{eq:boun-B1}) and (\ref{eq:empty-Psi-B}), we obtain
\begin{align}
    e^{-ik_y\rho}{f}_1^* (\boldsymbol{k}) = e^{ik_y\rho}{f}_1 (\boldsymbol{k}).
    \label{eq:boun-B4}
\end{align}
Here, $ {f}_1 (\boldsymbol{k}) = | {f}_1 (\boldsymbol{k})|e^{-i\phi_1(\boldsymbol{k})} $ is defined by
\begin{align}
     {f}_1 (\boldsymbol{k}) &\equiv v_A(k_x) + v_{A1}^\prime(k_x) e^{-i\frac{\sqrt{3}}{2}k_ya},\label{eq:FB}
     \end{align}
where $v_A(k_x)$ and $v_{A1}^\prime(k_x)$ depend on $\beta(k_x)$ as follows:
\begin{align}
   v_A\equiv \beta (1 + \alpha^2),\label{eq:B-vA}\\
    v_{A1}^\prime \equiv \alpha^2 + \beta^2.
    \label{eq:B-vAp}
\end{align}
It is clear that $\phi_1(k_x,2\pi/\sqrt{3}a) = \pi $ and $0$ for $v_A/v_{A1}^\prime<1$ and $v_A/v_{A1}^\prime>1$, respectively. The quantization condition of the bulk states is therefore given by
\begin{align}
    \frac{\sqrt{3}}{2} (J + 1) k_y a-\phi_1(\boldsymbol{k}) = n\pi,~\mathrm{for}~n=1,\dots,J.  
    \label{eq:q-B}
\end{align}
The intersections of $\phi_1(\boldsymbol{k})$ and $l^\prime_n(k_y)\equiv \sqrt{3}(J+1)k_ya/2 - n\pi$ along $k_y\in(0,2\pi/\sqrt{3}a)$ gives the solutions of Eq.~(\ref{eq:q-B}). As discussed in Sec.~\ref{sec:sSSH}, a pair of bulk states become edge states when $(v_A/v_{A1}^\prime)_J < 1- 1/(J+1)$, or in terms of $\beta$:
\begin{align}
  \beta^2  - \frac{1+\alpha^2}{1-(J+1)^{-1}}\beta + \alpha^2 > 0. 
    \label{eq:beta-B}
\end{align}
For $\alpha\neq 0$, Eq.~(\ref{eq:beta-B}) possesses two solutions: $\beta_J^+<\beta\leq 2$ or $0\leq\beta<\beta_J^-$, where $\beta_J^\pm \equiv 2\cos(\chi_J^\pm a/2)$. Thus, $\chi_J^\pm$ correspond to $|k_x|$'s at which the transitions from bulk to edge states occur. The edge states exist in the range
\begin{align}
    0\leq |k_x|<\chi_J^+~\mathrm{and}~\chi_J^-<|k_x| \leq \pi/a.
    \label{eq:BCA-exist}
\end{align}
In the limit $J\rightarrow\infty$, $\chi_J^\pm$ converge to 
\begin{subequations}
\begin{align}
\chi_{\infty}^{+} &\sim ({2}/{a})\cos^{-1}\left( {1}/{2} \right ) = {2\pi}/{3a},\label{eq:B-qp} \\
\chi_{\infty}^{-} &\sim ({2}/{a})\cos^{-1}\left( {\alpha^2}/{2} \right ).\label{eq:B-qm} 
\end{align}    
\end{subequations}
Thus, apart from the Dirac point $\chi_{\infty}^+$, the bulk states undergo transition to edge states at $\chi_{\infty}^-$ for $\alpha\in(0,1)$.

Let $1/\kappa_y$ be the localization length of edge states. By substituting $k_y = 2\pi/\sqrt{3}a + i\kappa_y $ into Eqs. (\ref{eq:boun-B4}) and (\ref{eq:FB}), 
\begin{align}
    e^{ \kappa_ y\rho}\tilde{f}_1(-\kappa_y) = e^{-\kappa_ y\rho}\tilde{f}_1(\kappa_y),
    \label{eq:qe-B}
\end{align}
where we define
\begin{align}
   \tilde{f}_1(\kappa_y) \equiv v_A(k_x) - v_{A1}^\prime(k_x) e^{\frac{\sqrt{3}}{2}\kappa_ya}.
    \label{eq:FpB}
\end{align}
Therefore, $\kappa_y$ is computed by solving 
\begin{align}
    \delta_1(\kappa_y) &\equiv \frac{v_A}{v_{A1}^\prime} - \frac{\mathrm{sinh}\left( \sqrt{3}J\kappa_y a/2 \right)}{\mathrm{sinh}\left( \rho\kappa_y  \right)} = 0.
    \label{eq:T1}
\end{align}


Analytically, the energies of edge states are calculated by inserting $k_y = 2\pi/\sqrt{3}a + i\kappa_y $ into Eq. (\ref{eq:Energy-s}) as follows:
\begin{align}
    \tilde{\varepsilon}_{s}(\kappa_y) = st \sqrt{1+\beta^2 - 2\beta \mathrm{cosh}\left(\frac{\sqrt{3}}{2}\kappa_y a\right)}.
    \label{eq:EnergyEdge-s}
\end{align}
By substituting Eq.~(\ref{eq:T1}) into (\ref{eq:EnergyEdge-s}), $\tilde{\varepsilon}_s$ for $J\kappa_y a\gg 1$ is
\begin{align}
    \tilde{\varepsilon}_{s} = st \frac{\alpha\left|1-\beta^2\right|}{\sqrt{1+\alpha^2}\sqrt{\alpha^2+\beta^2}}.
    \label{eq:Ened-B}
\end{align}
Since $\beta$ depends on $k_x$, $\tilde{\varepsilon}_s$ is dispersive. In particular, we get $\tilde{\varepsilon}_s \approx 0$ and $\tilde{\varepsilon}_s= st_A$ at $|k_x|\approx 2\pi/3a$ and $|k_x|=\pi/a$, respectively. Thus, the ZR is gapless for $J\rightarrow\infty$.   


\begin{figure}
\begin{center}
32\includegraphics[width=65mm]{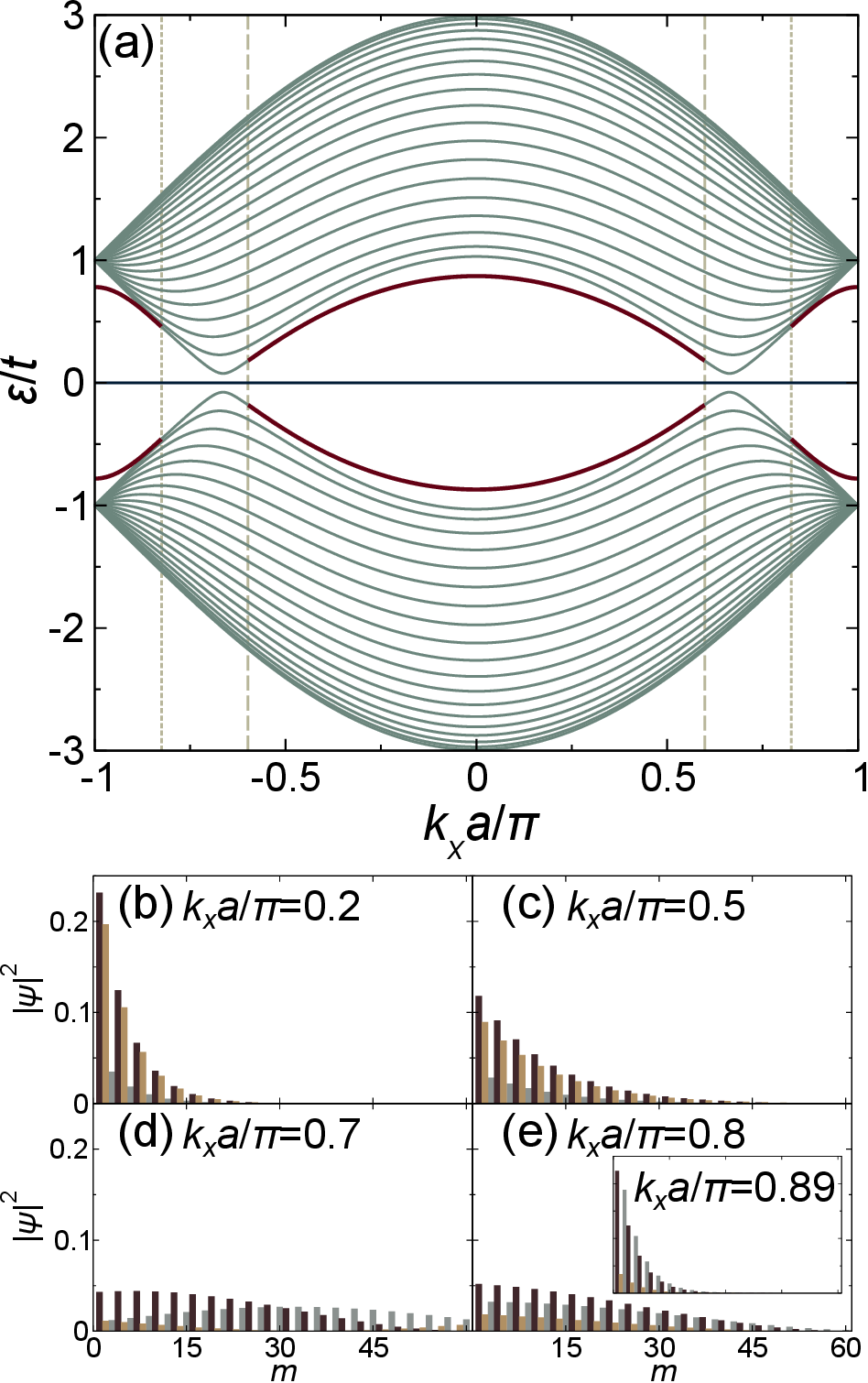}
\caption{(a) Energy spectra of the BCA ZR as a function of $k_x$ for $\alpha=0.8$ and $J=20$. The bold red lines indicate the edge states. The blue line at $\varepsilon=0$ is the flat band. The transitions from bulk to edge states occur at $|k_x| \approx 0.5985\pi/a$ and $0.8251\pi/a$, accordingly marked by the vertical dashed and dotted lines. Probability density for the in-gap states per site $m$ at (b) $k_x  = 0.2\pi/a $, (c) $ 0.5\pi/a $, (d) $ 0.7\pi/a $, and (e) $ 0.8\pi/a$ (Inset: $0.89\pi/a$).  The grey, brown, and yellow bars correspond to the $A$, $B$, and $C$ sites, respectively. }
\label{fig:Band-Psi-BCA}
\end{center}
\end{figure}


Fig. \ref{fig:Band-Psi-BCA}(a) depicts the numerical calculation of energy spectra as function of $k_x $ for $\alpha=0.8$ and $J=20$. By solving Eq.~(\ref{eq:beta-B}), we get $\chi_{20}^+ \approx 0.5985\pi/a$ and $\chi_{20}^-\approx  0.8251\pi/a$, which are indicated by the vertical dashed and dotted lines, respectively (we check that $\chi_J^+$ is undefined for $J<3$ ). The bold red lines depict the edge states, which are pinned at $|\varepsilon| = t_A\approx 0.7809t$ at $|k_x|= \pi/a$, in agreement with Eq.~(\ref{eq:Ened-B}). Due to the finite size effect, band gaps open at $|k_x|=2\pi/3a$.

In Figs.~\ref{fig:Band-Psi-BCA}(b)-(e), we plot the electronic probability density $|{\psi}|^2$ for the in-gap states per site $m$ at several $k_x$'s. It is noted that $|{\psi}|^2$'s are identical for both signs of $k_x$ and $\varepsilon$. The grey, brown, and yellow bars in the histogram correspond to the $A$, $B$, and $C$ sites, respectively. The distributions of $|{\psi}|^2$ in Figs.~\ref{fig:Band-Psi-BCA}(b) and (c) exhibit the profiles of asymmetric edge states. Here, $|{\psi}|^2$'s are concentrated around the bottom edge of the ZR and decay exponentially towards the top edge. Figs.~\ref{fig:Band-Psi-BCA}(d) and (e) indicate the bulk state because $|{\psi}|^2$'s are extended throughout the ZR. In Fig.~\ref{fig:Band-Psi-BCA}(e), $|{\psi}|^2$ is high around the bottom edge because $k_x$ is close to $\chi_{20}^-$. The inset shows the re-emergence of the edge state at a larger $k_x$.   


\subsubsection{CAB ZR}

In Appendix~\ref{sapp:deriv2}, we show in detail that Eqs. (\ref{eq:empty-Psi-B2}) and (\ref{eq:boun-C1}) lead to:
\begin{align}
     e^{ik_y\rho}{f}_2 (\boldsymbol{k}) = e^{-ik_y\rho}{f}_2^* (\boldsymbol{k}),
     \label{eq:boun-C4}
\end{align}
where $ {f}_2 (\boldsymbol{k}) = | {f}_2 (\boldsymbol{k})|e^{-i\phi_2(\boldsymbol{k})}$ is defined by
\begin{align}
    {f}_2 (\boldsymbol{k}) \equiv v_A(k_x) + v_{A2}^\prime(k_x)e^{-i\frac{\sqrt{3}}{2}k_ya},
    \label{eq:FC}
\end{align}
and $v_{A2}^\prime(k_x)$ is given as follows:
\begin{align}
    v_{A2}^\prime \equiv (1+\alpha^2\beta^2).
    \label{eq:C-vA}
\end{align}
The quantization condition of $k_y$ is given by replacing $\phi_1$ in Eq.~(\ref{eq:q-B}) with $\phi_2$. Likewise, edge states appear in the gap when $(v_A/v_{A2}^\prime)_J < 1- 1/(J+1)$, or
\begin{align}
    \alpha^2 \beta^2  - \frac{1+\alpha^2}{1-(J+1)^{-1}}\beta  + 1 > 0. 
    \label{eq:beta-C}
\end{align}
By denoting the solutions of Eq.~(\ref{eq:beta-C}) as $\beta_J^+<\beta\leq 2$ or $0\leq\beta<\beta_J^-$, the edge states for $\alpha\neq 0$ exist in the range 
\begin{align}
    0\leq |k_x|<\chi_J^+~\mathrm{and}~\chi_J^-<|k_x| \leq \pi/a.
    \label{eq:CAB-exist}
\end{align}
Note that $\chi_J^\pm$ in Eq.~(\ref{eq:CAB-exist}) are not equal to those in Eq.~(\ref{eq:BCA-exist}) except for $\alpha=1$. Here, in the limit $J\rightarrow\infty$, $\chi_J^\pm$ become
\begin{subequations}
    \begin{align}
    \chi_\infty^+ &\sim ({2}/{a})\cos^{-1}\left( {1}/{2\alpha^{2}} \right ),\label{eq:C-qp} \\
    \chi_\infty^-&\sim ({2}/{a})\cos^{-1}\left( {1}/{2} \right ) = {2\pi}/{3a}.
\label{eq:C-qm} 
\end{align}
\end{subequations}
It is noticed that $\chi_\infty^+$ is undefined for $\alpha<1/\sqrt{2} $.

\begin{figure}
\begin{center}
\includegraphics[width=65mm]{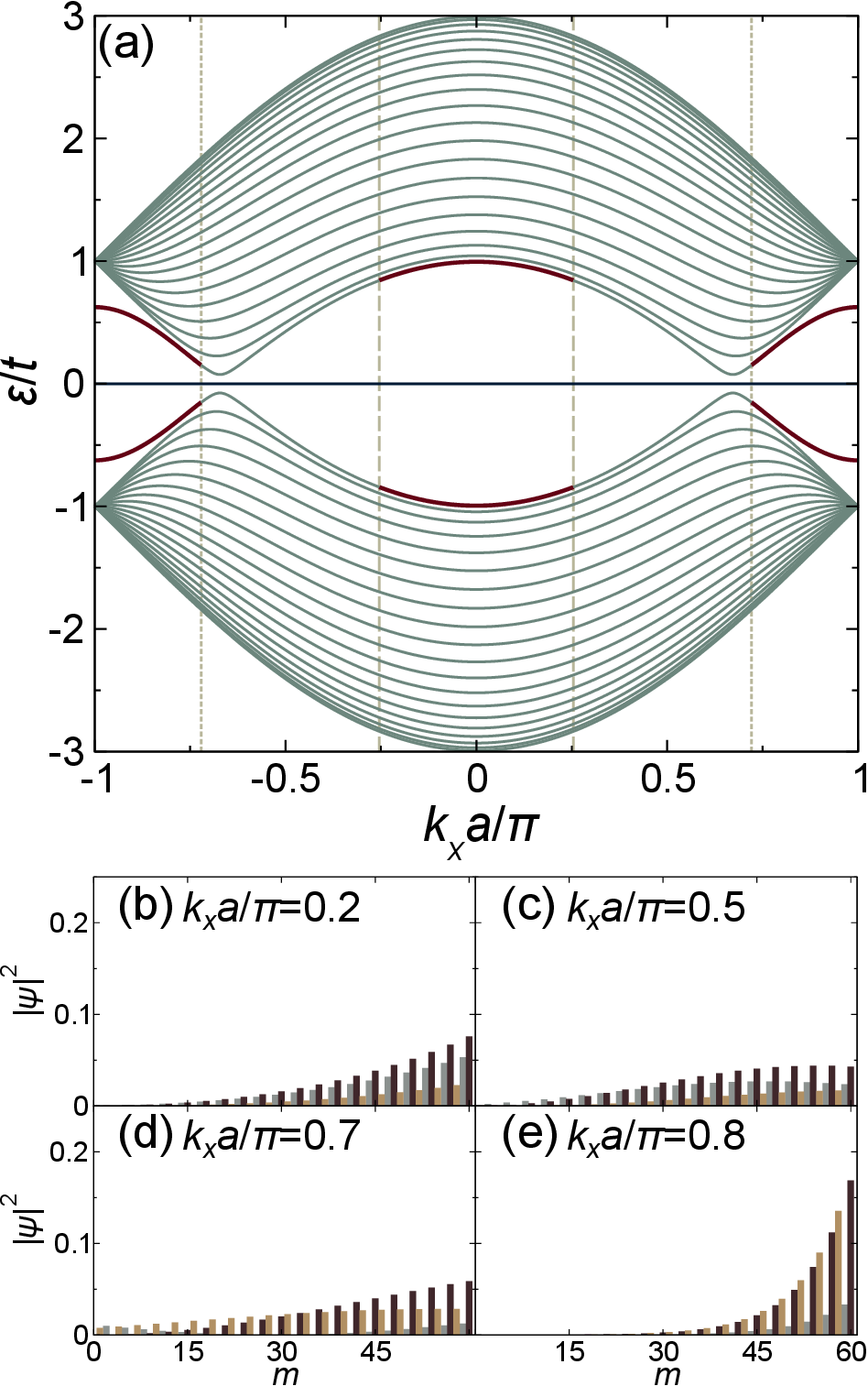}
\caption{ (a) Energy spectra ${\varepsilon}$ of the CAB ZR as a function of $k_x$ for $\alpha=0.8$ and $J=20$. The bold red lines indicate the edge states. The blue line at $\varepsilon=0$ is the flat band. The transitions from bulk to edge states occur at $|k_x| \approx 0.2542\pi/a$ and $0.7213\pi/a$, accordingly marked by the vertical dashed and dotted lines. Probability density for the in-gap states per site $m$ at (b) $k_x  = 0.2\pi/a $, (c) $ 0.5\pi/a $, (d) $ 0.7\pi/a $, and (e) $ 0.8\pi/a$.  The grey, brown, and yellow bars correspond to the $A$, $B$, and $C$ sites, respectively.}
\label{fig:Band-Psi-CAB}
\end{center}
\end{figure}

To determine $\kappa_y$, we define $\delta_2$ by replacing $v_{A1}^\prime$ in Eq.~(\ref{eq:T1}) with $v_{A2}^\prime$. $\tilde{\varepsilon}_s$ for $J\kappa_y a\gg 1$ is given by
\begin{align}
    \tilde{\varepsilon}_{s} = st \frac{\alpha\left|1-\beta^2\right|}{\sqrt{1+\alpha^2}\sqrt{1+\alpha^2\beta^2}}.
    \label{eq:Ened-C}
\end{align}
At $|k_x|\approx 2\pi/3a$ ($|k_x|=\pi/a$), $\tilde{\varepsilon}_s \approx 0$ ($\tilde{\varepsilon}_s=st_C$).


Fig. \ref{fig:Band-Psi-CAB}(a) shows the numerical calculation of energy spactra as a function of $k_x $ for $\alpha=0.8$ and $J=20$. By solving Eq.~(\ref{eq:beta-C}), $\chi_{20}^+ \approx 0.2542\pi/a$ and $\chi_{20}^- \approx  0.7213\pi/a$, respectively marked by the vertical dashed and dotted lines ($\chi_J^+$ is undefined for $J<12$). The bold red lines indicate the edge states, where $|\varepsilon| = t_C\approx 0.6247t$ at $|k_x|=\pi/a$, in accordance with Eq.~(\ref{eq:Ened-C}).

Figs.~\ref{fig:Band-Psi-CAB}(b)-(e) show $|{\psi}|^2$'s for the in-gap states per site $m$ at several $k_x$'s. Figs.~\ref{fig:Band-Psi-CAB}(b) and (e) exhibit the profiles of asymmetric edge states where $|{\psi}|^2$'s are high around the top edge of ZR and decay exponentially toward the bottom edge. In contrast, Figs.~\ref{fig:Band-Psi-CAB}(c) and (d) indicate the bulk states due to the extended $|{\psi}|^2$'s.

\section{Topological invariant of $\alpha $-${T}_3$ ZRs}~\label{sec:aT3-Invariant}

\vspace{-10mm}

\subsection{ Unitary transforms of bulk Hamiltonian}\label{subsec:unitrans}

In Section \ref{sec:aT3-ZR}, we found that for each $k_x$, the $\alpha$-${T}_3$ ZRs are isomorphic to the stub SSH chain along the $y$ direction. By unitary transforms of $h(\boldsymbol{k})$ in Eq. (\ref{eq:H1}), the BCA and CAB ZRs are mapped into stub SSH chains with distinct unit cells and hopping parameters. As a result, the $\mathbb{Z}_2$ invariant for the $\alpha$-${T}_3$ ZRs is given by $|W_c|$.   


\subsubsection{ BCA ZR }

\begin{figure}[t]
\begin{center}
\includegraphics[width=80mm]{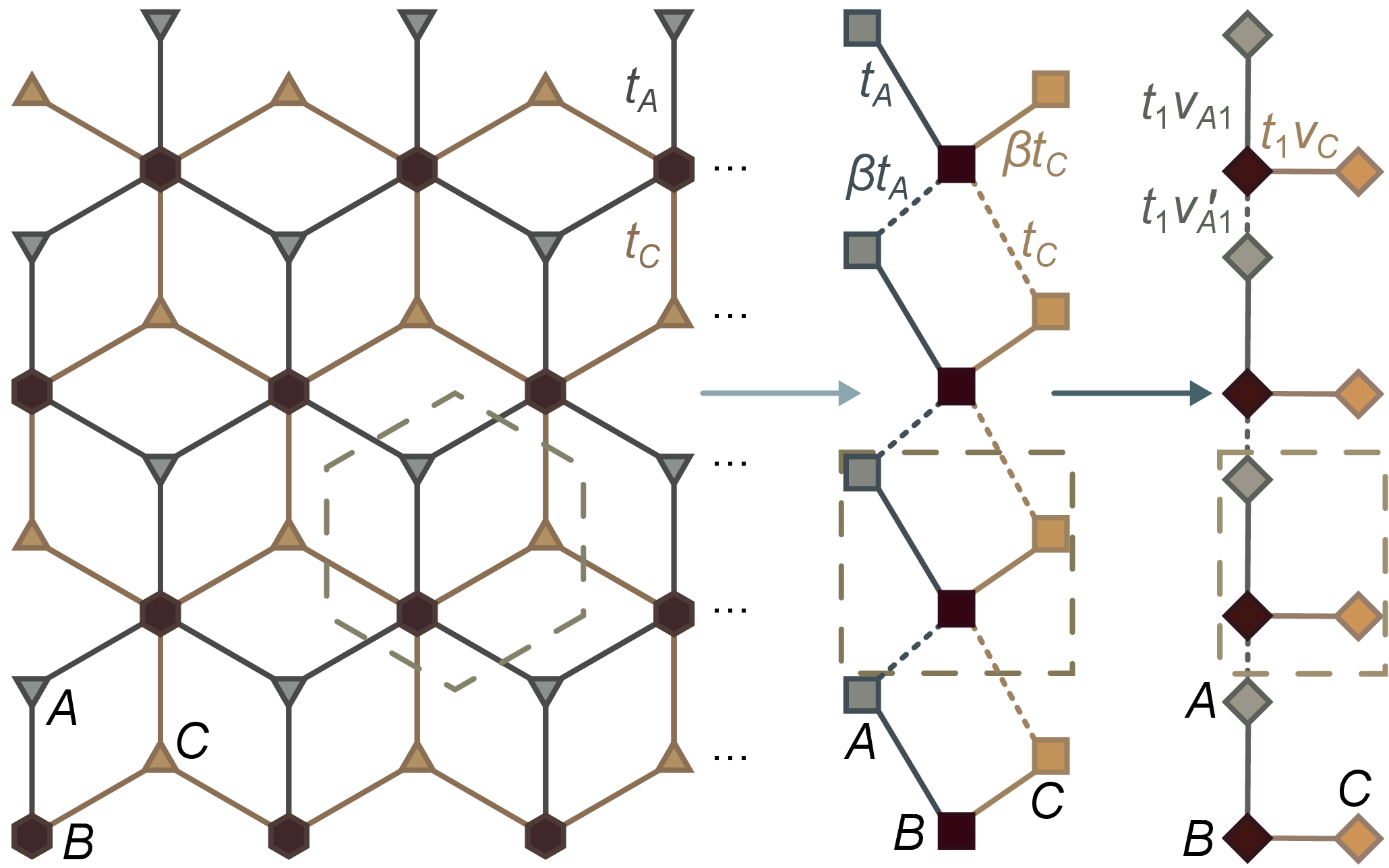}
\caption{ Schematic for the mappings of BCA ZR into rhombic and stub SSH chains.  }
\label{fig:Ut-BCA}
\end{center}
\end{figure}

\begin{figure*}[t]
\begin{center}
\includegraphics[width=176mm]{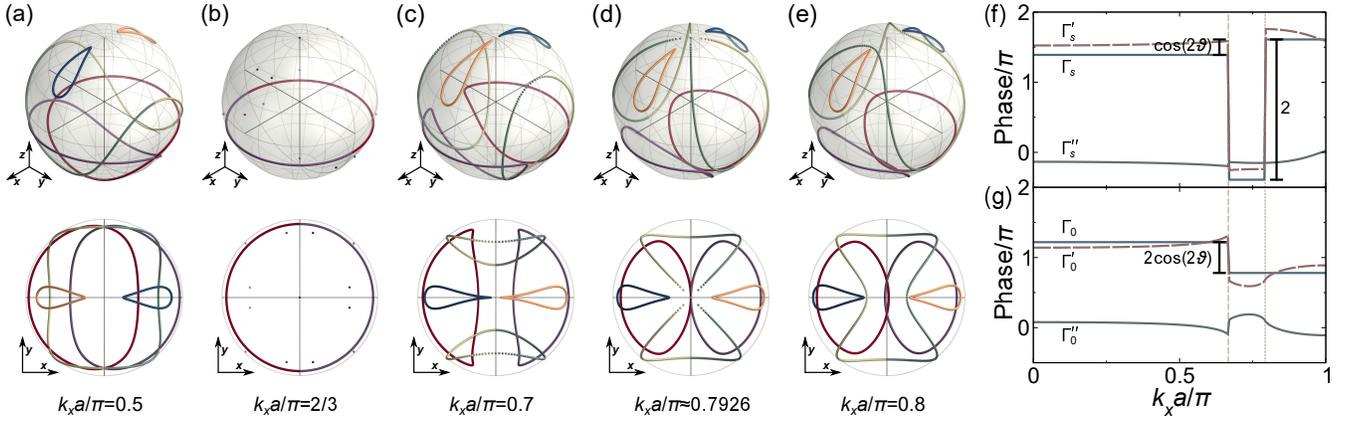}
\caption{ Trajectories of the Majorana stars [from the eigenstates of $\mathcal{H}_1(\boldsymbol{k})$, with $\alpha=0.8$] on the Bloch sphere (top panels) and the $xy$ plane (bottom panels) at (a) $k_x = 0.5 \pi/a$, (b) $\chi_{\infty}^+= 2\pi/3a $ [Eq.~(\ref{eq:B-qp})] (c) $0.7\pi/a$, (d) $\chi_\infty^- \approx 0.7926\pi/a $ [Eq.~(\ref{eq:B-qm})], and (d) $0.8\pi/a$. The red and yellow (blue and purple) curves correspond to the valence (conduction) band, $c=-1$ ($c=+1$). The light- and dark-green curves correspond to the flat band, $c=0$. Plots of $\Gamma_c$, $\Gamma_c^\prime$, and $\Gamma_c^{\prime\prime}$ for (f) $c=s=\pm 1$, (g) $c=0$. }
\label{fig:Maj-B}
\end{center}
\end{figure*}


To transform $h(\boldsymbol{k})$, we define a unitary matrix
\begin{align}
    {U}(\boldsymbol{u}_A,\boldsymbol{u}_C) \equiv 
    \begin{bmatrix}
        e^{i\boldsymbol{k}\cdot \boldsymbol{u}_A} & 0 & 0\\
        0 & 1 & 0\\
        0 & 0 & e^{i\boldsymbol{k}\cdot \boldsymbol{u}_C }
    \end{bmatrix}.
    \label{eq:U}
\end{align}
$ h_1(\boldsymbol{k})\equiv U(\boldsymbol{b}_3,-\boldsymbol{b}_1-\frac{\boldsymbol{a}_1}{2})h(\boldsymbol{k})U^\dagger(\boldsymbol{b}_3,-\boldsymbol{b}_1-\frac{\boldsymbol{a}_1}{2})$ is
\begin{align}
    h_1(\boldsymbol{k})
    =-\begin{bmatrix}
       0 & t_A p^*(\boldsymbol{k}) & 0\\
       t_A p(\boldsymbol{k}) & 0 & t_C q(\boldsymbol{k})\\
       0 & t_C q^*(\boldsymbol{k}) & 0
       \end{bmatrix},
    \label{eq:HU1}
\end{align}
where we define
\begin{align}
    p(\boldsymbol{k}) &\equiv 1 + \beta(k_x) e^{-i\frac{\sqrt{3}}{2}k_ya}, \label{eq:pp}\\
    q(\boldsymbol{k}) &\equiv \beta(k_x) +  e^{-i\frac{\sqrt{3}}{2}k_ya}.
    \label{eq:qq}
\end{align}
$h_1(\boldsymbol{k})$ is the Hamiltonian of a rhombic or diamond chain~\cite{longhi2014aharonov,mukherjee2015observation,pelegri2019topological} illustrated in the middle panel of Fig.~\ref{fig:Ut-BCA}. Here, the $A$ and $C$ sites are connected to the $B$ sites by the intracell (intercell) hopping parameters $t_A$ and $\beta t_C$ ($\beta t_A$ and $t_C$), respectively. Experimental realizations of the rhombic chain have been achieved using photonic lattices~\cite{longhi2014aharonov,mukherjee2015observation}. The present method is therefore applicable for topological characterization of edge states~\cite{pelegri2019topological} in the rhombic chains.

Now, consider a rotation matrix 
\begin{align}
    {\Upsilon} (\gamma_1)\equiv 
    \begin{bmatrix}
        \cos\gamma_1 & 0 & \sin\gamma_1\\
        0 & 1 & 0\\
        -\sin\gamma_1 & 0 & \cos\gamma_1
    \end{bmatrix}.
    \label{eq:V1}
\end{align}
By setting $\gamma_1= \tan^{-1}(\alpha/\beta)$, we transform $h_1$ into $\mathcal{H}_1(\boldsymbol{k})=\Upsilon(\gamma_1) h_1(\boldsymbol{k}) \Upsilon^\dagger(\gamma_1)$ as follows:
\begin{align}
    \mathcal{H}_1(\boldsymbol{k})
    =-t_1\begin{bmatrix}
        0 & f_1^*(\boldsymbol{k}) & 0\\
        f_1(\boldsymbol{k}) & 0 & -v_C \\
        0 & -v_C   & 0
    \end{bmatrix},
    \label{eq:HV1}
\end{align}
where $t_1\equiv t_A\cos \gamma_1/\beta$, $v_C \equiv \alpha(1-\beta^2)$, and $f_1(\boldsymbol{k})$ is defined by Eq.~(\ref{eq:FB}). $\mathcal{H}_1(\boldsymbol{k})$ constitutes the Hamiltonian of stub SSH chain illustrated in the right panel of Fig.~\ref{fig:Ut-BCA}. The mapping of $h(\boldsymbol{k})$ into $\mathcal{H}_1(\boldsymbol{k})$ explains the emergence edge states with energies $\tilde{\epsilon}_s = s|t_1 v_C|$, identical to Eq. (\ref{eq:Ened-B}), when $v_A/v_{A1}^\prime <1 $.


The eigenstates of $\mathcal{H}_1(\boldsymbol{k})$ are derived by replacing $\Phi(k)$ and $\Theta(k)$ in see Eqs.~(\ref{eq:sSSH-wfs}), (\ref{eq:sSSH-wf0}) with $\phi_1(\boldsymbol{k})$ and $\theta_1(\boldsymbol{k})\equiv \tan^{-1}(-v_C(k_x)/|f(\boldsymbol{k})|) $, respectively. Here, the BZ is $k_y\in[-2\pi/\sqrt{3}a, 2\pi/\sqrt{3}a]$. $\zeta_c^\nu$ and $|W_c|$ are calculated by using the method discussed in Sec.~\ref{sec:sSSH}. 

Fig.~\ref{fig:Maj-B} shows the trajectories of $\zeta_c^{\nu}$'s for $\alpha=0.8$ at (a) $k_x = 0.5\pi/a$, (b) $\chi_{\infty}^+=2\pi/3a$ [${K}$ point, see Eq.~(\ref{eq:B-qp})], (c) $0.7\pi/a$, (d) $\chi_{\infty}^-\approx 0.7926\pi/a$ [see Eq.~(\ref{eq:B-qm})], and (e) $0.8\pi/a$. The red/yellow, light-green/dark-green, and blue/purple curves correspond to $c=-1$, $0$, and $1$, respectively. In Figs.~\ref{fig:Maj-B}(a) and (e), $|{W}_c|=1$ because $0<k_x<\chi_\infty^+$ and $\chi_\infty^-<k_x<\pi/a$, respectively. In Fig.~\ref{fig:Maj-B}(c), $|{W}_c|=0$ because $\chi_\infty^+<k_x<\chi_\infty^-$.  ${W}_c$'s are ill-defined at $\chi_\infty^\pm$  for different reasons. In Fig.~\ref{fig:Maj-B}(b), the trajectories of $\zeta_c^{\nu}$'s become discontinuous due to  the gap closing at $|k_y|= 2\pi/\sqrt{3}a$. Here, the system is reduced to the metallic SSH chain because $\beta=1$ and consequently $v_C=0$, $v_A=v_{A1}^\prime$.  In Fig.~\ref{fig:Maj-B}(d), the red, purple, and connected green curves intersect the poles of the Bloch sphere. Accordingly, Figs.~\ref{fig:Maj-B}(b) and (d) indicate topological phase transitions with and without gap closing.

To show that the unitary transforms preserve the topological properties of the $\alpha$-$T_3$ ZR, we numerically compute the accumulated Berry phase $\Gamma_{c}$ from the adiabatic evolution of $\zeta_c^{\nu}$'s. For $S=1$, $\Gamma_{c}$ is given by~\cite{hannay1998berry,liu2014representation}:
\begin{align}
     \Gamma_c &= \Gamma_c^\prime + \Gamma_c^{\prime\prime},~\mathrm{where} 
     \label{eq:Maj-G1}\\
         \Gamma_c^\prime&\equiv-\frac{1}{2} \sum_{\nu=\mp}\int_{\mathrm{BZ}} (1-\cos\eta_{c}^{\nu})d\xi_{c}^{\nu},
      \label{eq:Maj-G2}\\
      \Gamma_c^{\prime\prime}&\equiv-\frac{1}{2}\int_{\mathrm{BZ}}\frac{\boldsymbol{\zeta}_{c}^-\times\boldsymbol{\zeta}_{c}^+}{3+\boldsymbol{\zeta}_{c}^-\cdot\boldsymbol{\zeta}_{c}^+}\cdot d(\boldsymbol{\zeta}_{c}^--\boldsymbol{\zeta}_{c}^+)\label{eq:Maj-G3}.
\end{align}
$\Gamma_c^\prime$ corresponds to the sum of solid angles subtended by $\zeta_c^{-}$ and $\zeta_c^{+}$, while $\Gamma_c^{\prime\prime}$ is interpreted as the correlation term between $\zeta_c^{-}$ and $\zeta_c^{+}$ due to their relative motions. 


Fig.~\ref{fig:Maj-B}(f) shows $\Gamma_s$, $\Gamma_s^\prime$, and $\Gamma_s^{\prime\prime}$ as a function of $k_x$, where the vertical dashed and dotted lines mark $k_x=\chi_\infty^+$ and $\chi_\infty^-$, respectively. It is noted that $\Gamma_s^\prime$, $\Gamma_s^{\prime\prime}$ and hence $\Gamma_s$ are identical for $s=\pm1$. For $0\leq k_x <\chi_\infty^+$, $\Gamma_s^\prime$ includes the large portion of the Bloch sphere bounded by the red (or purple) loop that encloses the $z$ axis [see Fig.~\ref{fig:Maj-B}(a) for instance], where $\Gamma_s\approx 1.390244\pi$. The trajectories of $\zeta_s^\nu$'s become disconnected at $\chi_\infty^+$ [Fig.~\ref{fig:Maj-B}(b)] and reconnect for $\chi_\infty^+< k_x \leq \chi_\infty^-$, where $\Gamma_s\approx-0.390244\pi$. The abrupt decrease of $\Gamma_s$ is because $\Gamma_s^\prime$ comprises the small portion of the Bloch sphere as the red (or purple) loop does not enclose the $z$ axis [Figs.~\ref{fig:Maj-B}(c) and (d) for examples]. For $\chi_\infty^-<k_x\leq \pi/a$, $\Gamma_s$ increases by $2\pi$ to $\Gamma_s\approx1.609756\pi$ as the red (or purple) loop re-encloses the $z$ axis [see Figs.~\ref{fig:Maj-B}(e)]. Since $\Gamma_s$ is defined modulo $2\pi$, $\Gamma_s$ can not distinguish between the trivial and non-trivial phases of the ZR for $\chi_\infty^+< k_x < \chi_\infty^-$ and $\chi_\infty^-<k_x\leq \pi/a$, respectively. It is noted that $\Gamma_s$ for $k_x\in(\chi_\infty^-, \pi/a]$ and $k_x\in[0,\chi_\infty^+)$ differs by $\pi\cos(2\vartheta)\approx0.219512\pi$ (mod $2\pi$), consistent with Eq.~(\ref{eq:Berry-s}) for $\vartheta=\tan^{-1}(0.8)$ and $\tau=1$.

Fig.~\ref{fig:Maj-B}(g) shows $\Gamma_0$, $\Gamma_0^\prime$, and $\Gamma_0^{\prime\prime}$ as a function of $k_x$. $\Gamma_0^\prime$ either corresponds to a portion of the Bloch sphere bounded by a single loop [as in Figs.~\ref{fig:Maj-B}(a) and (e)], or two loops [as in Figs.~\ref{fig:Maj-B}(c) and (d)]. The jump discontinuity from $\Gamma_0\approx 1.219512\pi$ to $\Gamma_0\approx 0.780488\pi$ occurs at $k_x=\chi_\infty^+$. Therefore, $\Gamma_0$ for $k_x\in(\chi_\infty^+, \pi/a]$ and $k_x\in[0,\chi_\infty^+)$ differs by $-2\pi\cos(2\vartheta)\approx-0.439024\pi$, in agreement with Eq.~(\ref{eq:Berry-0}) for $\vartheta=\tan^{-1}(0.8)$ and $\tau=1$.

\subsubsection{ CAB ZR }

\begin{figure}[t]
\begin{center}
\includegraphics[width=80mm]{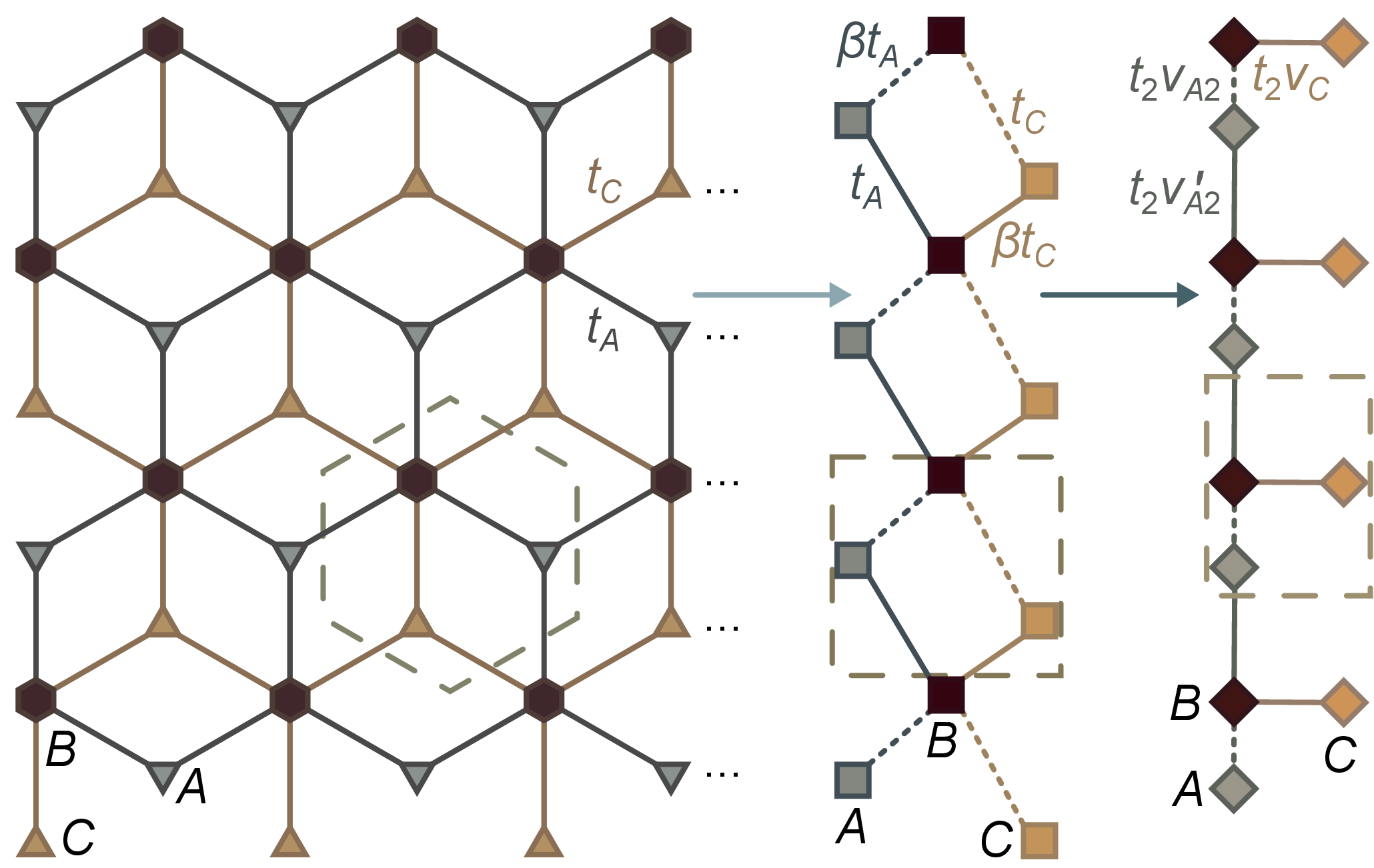}
\caption{ Schematic for the mappings of CAB ZR into rhombic and stub SSH chains.}
\label{fig:Ut-CAB}
\end{center}
\end{figure}


By using Eq.~(\ref{eq:U}), we can transform $h(\boldsymbol{k})$ into $ h_2(\boldsymbol{k})\equiv U(\boldsymbol{b}_2-\frac{\boldsymbol{a}_1}{2},-\boldsymbol{b}_3)h(\boldsymbol{k})U^\dagger(\boldsymbol{b}_2-\frac{\boldsymbol{a}_1}{2},-\boldsymbol{b}_3)$ as follows:
\begin{align}
    h_2(\boldsymbol{k})
    =-\begin{bmatrix}
       0 & t_A q(\boldsymbol{k}) & 0\\
       t_A q^*(\boldsymbol{k}) & 0 & t_C p^*(\boldsymbol{k})\\
       0 & t_C p(\boldsymbol{k}) & 0
       \end{bmatrix}.
    \label{eq:HU2}
\end{align}
The middle panel of Fig.~\ref{fig:Ut-CAB} shows the system described by Eq.~(\ref{eq:HU2}), which is a rhombic chain with a distinct choice of unit cell compared with the one in Fig.~\ref{fig:Ut-BCA}.

By using $\Upsilon$ in Eq.~(\ref{eq:V1}), $h_2(\boldsymbol{k})$ is transformed into $\mathcal{H}_2(\boldsymbol{k})=\Upsilon(\gamma_2) h_2(\boldsymbol{k}) \Upsilon^\dagger(\gamma_2)$ as follows:
\begin{align}
    \mathcal{H}_2(\boldsymbol{k})
    =-t_2\begin{bmatrix}
        0 & f_2(\boldsymbol{k}) & 0\\
        f_2^*(\boldsymbol{k}) & 0 & v_C \\
        0 & v_C   & 0
    \end{bmatrix},
    \label{eq:HV2}
\end{align}
where $\gamma_2\equiv\mathrm{tan}^{-1}(\alpha\beta)$, $t_2\equiv t_A\cos\gamma_2$, and $f_2(\boldsymbol{k})$ is defined by Eq.~(\ref{eq:FC}). The right panel of Fig.~\ref{fig:Ut-CAB} depicts the stub SSH chain represented by Eq.~(\ref{eq:HV2}). Thus, edge states with energies $\tilde{\epsilon}_s = s|t_2 v_C|$ [identical to Eq. (\ref{eq:Ened-C})] appear when for $v_A/v_{A2}^\prime <1 $. 

Similarly, we derive the eigenstates of $\mathcal{H}_2(\boldsymbol{k})$ by replacing $\Phi(k)$ and $\Theta(k)$ in Eqs.~(\ref{eq:sSSH-wfs}), (\ref{eq:sSSH-wf0}) with $\phi_2(\boldsymbol{k})$ and $\theta_2(\boldsymbol{k})=-\theta_1(\boldsymbol{k})$, respectively.

\subsection{Topological phase diagrams of ZRs}\label{subsec:topophase}

Having shown that the distinction between the trivial and nontrivial phases of the $\alpha$-$T_3$ ZRs is indicated by $|{W}_c|$, we discuss the topological phase diagrams as a function $|k_x|\in[0,\pi/a]$ and $\alpha\in[0,1]$. 

\vspace{-2mm}

\subsubsection{BCA ZR}

\vspace{-2mm}


Fig.~\ref{fig:phase}(a) depicts the topological phase diagram of the BCA ZR. The light and dark regions indicate the trivial and nontrivial phases, respectively. The pink and red lines mark the phase transition with and without gap closing. For $\alpha =0$, the edge states exist in the range $|k_x|\in[0, 2\pi/3a)$. This result is expected because removing the $C$ sites reduces the BCA ZR to bearded graphene ribbons~\cite{ryu2002,delplace2011}, i.e. ZR with dangling bonds or the Klein defects. For $\alpha =1 $ ($T_3$ ZR), the edge states exist for $|k_x|\in[0, \pi/a]$ except at $|k_x|= 2\pi/3a$, even though the Berry phase $\Gamma_s=0$. In the Majorana representation of the corresponding stub SSH chain, $|{W}_c|=1$ before and after gap closing because $\chi_\infty^+=\chi_\infty^-=2\pi/3a$. Thus, when we plot $\Gamma_s$ as a function of $k_x$, the value of $\Gamma_s$ remains the same for $|k_x|\in[0, 2\pi/3a)$ and $(2\pi/3a, \pi/a]$.

\vspace{-2mm}

\subsubsection{CAB ZR}

\vspace{-2mm}

\begin{figure}[t]
\begin{center}
\includegraphics[width=86.5mm]{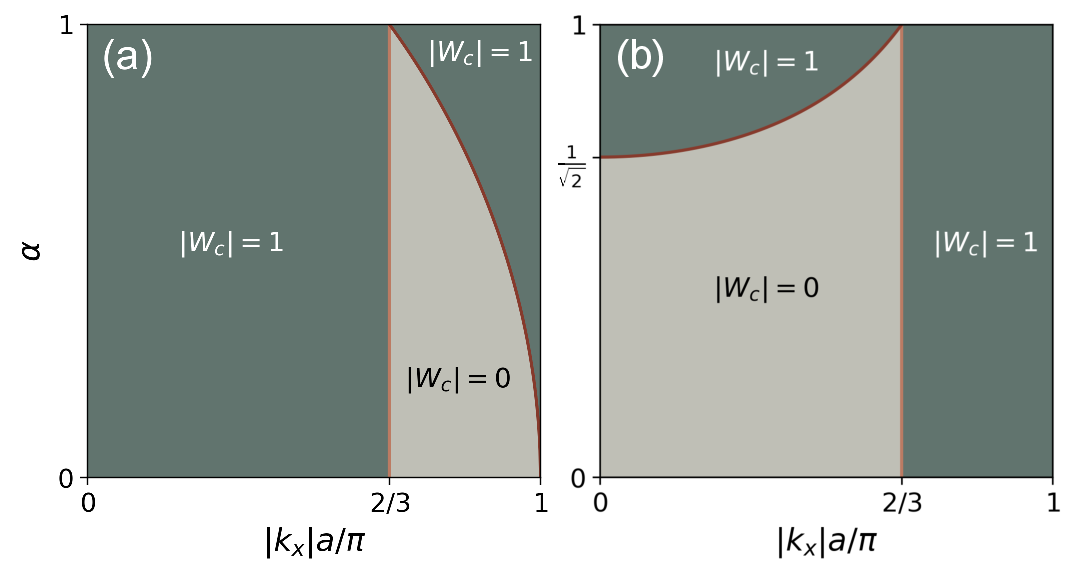}
\caption{ Topological phase diagrams of (a) BCA and (b) CAB ZRs. The light (dark) region indicates the trivial (nontrivial) phase, where edge states are absent (present). The pink (red) lines mark the phase transition with (without) gap closing.  } 
\label{fig:phase}
\end{center}
\end{figure}

Fig.~\ref{fig:phase}(b) shows the topological phase diagram of the CAB ZR. For $\alpha<1/\sqrt{2} $, the edge states appear in the range $|k_x|\in(2\pi/3a,\pi/a]$, similar to those in the graphene ZR~\cite{ryu2002,delplace2011,wakabayashi2010electronic}. 

\vspace{-2mm}

\section{Conclusion}~\label{sec:Conclusion}

\vspace{-2mm}

In the first part of this paper, we investigate the bulk-boundary correspondence for the stub SSH chain. The edge states appear when the intracell hopping parameter is smaller than the intercell one, identical to the SSH chain. By using the Majorana representation of the eigenstates on the Bloch sphere, we prove that the $\mathbb{Z}_2$ topological invariant is given by the winding number instead of the Zak phase, which is not quantized to $\pi$ or $0$ due to the broken inversion symmetry.


In the second part, we demonstrate the isomorphism between the $\alpha$-$T_3$ ZRs and the stub SSH chains from the boundary conditions of the former. The equivalence between the two systems is revealed by the unitary transforms of the bulk Hamiltonian, which maps the $\alpha$-$T_3$ ZRs into stub SSH chains. As a result, the edge states in the $\alpha$-$T_3$ ZRs possess the same topological origin and characterization as those in the stub SSH chains. Therefore, the $\alpha$-$T_3$ lattice is identified as a topologically non-trivial, massless Dirac fermion in two dimensions without a quantized Berry phase. The $T_3$ lattice particularly exemplifies a topological system with zero Berry phase.

\section*{Acknowledgment}

We thank R. Saito, N. T. Hung, S. Hayashi, and E. H. Hasdeo for helpful discussions. This work is partly supported by Grant-in-Aid for Scientific Research, JSPS KAKENHI (23K03293), and JST-CREST (JPMJCR18T1).

\renewcommand{\appendixname}{APPENDIX}
\appendix
\section{ABC ZR}\label{app:ABC}

Here, we derive the quantization condition of the bulk states and prove the existence of two dispersive edge states for each $\varepsilon<0$ and $\varepsilon>0$. The topological characterization of the edge states is reserved for future work. 

The boundary conditions are derived by expressing the nearest-neighbor interactions for the $B$ sites in terms of the next-nearest-neighbor interactions. First, we multiply Eq.~(\ref{eq:tbB}) by $\varepsilon$, then we substitute $\psi_s^A$ and $\psi_s^C$ by the right-hand sides of Eqs. (\ref{eq:tbA}) and (\ref{eq:tbC}), respectively. As a result, Eq.~(\ref{eq:tbB}) becomes
\begin{align}
    \varepsilon^2\psi_s^{B}(\boldsymbol{r}) =&~t^2\Big[ \sum_{\nu=\pm} \Big\{\psi_s^{B}(\boldsymbol{r}+\nu\boldsymbol{a}_1) + \psi_s^{B}(\boldsymbol{r}+\nu\boldsymbol{a}_2)\nonumber\\
    &+\psi_s^{B}(\boldsymbol{r}+\nu[\boldsymbol{a}_2-\boldsymbol{a}_1])\Big\} +3\psi_s^B(\boldsymbol{r})  \Big].\label{eq:sqrd1}
\end{align}

\begin{figure}[t]
\begin{center}
\includegraphics[width=86.5mm]{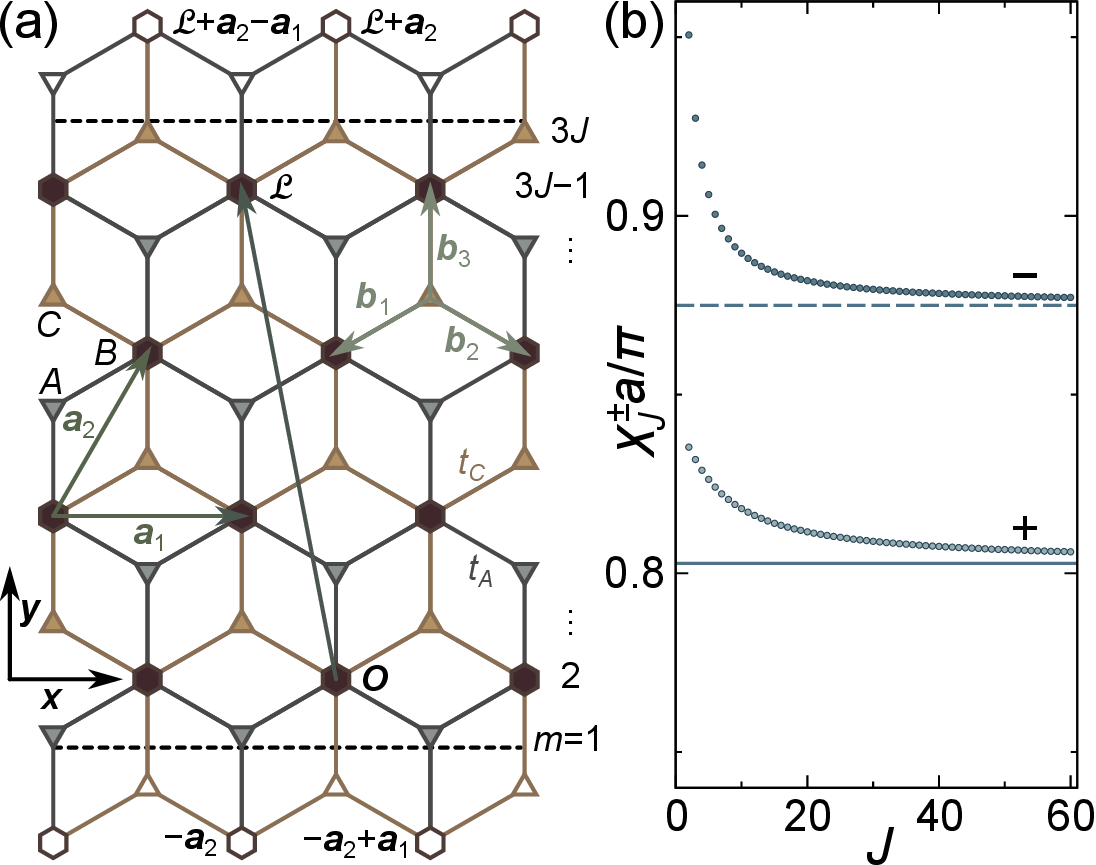}
\caption{ABC ZR: (a) Schematic for the boundary conditions. (b) Plot $\chi_{J}^\pm$ as a function of $J$ for $\alpha=0.8$. The solid and dashed lines mark $\chi_\infty^+\approx 0.802769\pi/a$ and $\chi_\infty^-\approx 0.874979\pi/a$, respectively.   }
\label{fig:ABC-Appendix}
\end{center}
\end{figure}

In Fig.~\ref{fig:ABC-Appendix}(a), we define a vector $\boldsymbol{\mathcal{L}}=(\mathcal{L}_x,\mathcal{L}_y)$, where $\mathcal{L}_x=0$ and $-a/2$ for odd and even $J$, respectively, while ${\mathcal{L}_y} = \sqrt{3}(J-1)a/2$. For each $l\in\mathbb{Z}$, the TBEs at $m=2$ and $m=3J-1$ are accordingly given as follows:
    \begin{align}
    \varepsilon\psi_s^B(\boldsymbol{O}_l)=&-t_A\sum_{n=1}^3\psi_s^A(\boldsymbol{O}_l+\boldsymbol{b}_n)-t_C\sum_{n=1}^2\psi_s^C(\boldsymbol{O}_l-\boldsymbol{b}_n),\label{eq:bot1}\\
    \varepsilon\psi_s^B(\boldsymbol{\mathcal{L}}_l)=&-t_A\sum_{n=1}^2\psi_s^A(\boldsymbol{\mathcal{L}}_l+\boldsymbol{b}_n)-t_C\sum_{n=1}^3\psi_s^C(\boldsymbol{\mathcal{L}}_l-\boldsymbol{b}_n).\label{eq:top1}
\end{align}
 We define $\boldsymbol{\mathcal{L}}_l\equiv \boldsymbol{\mathcal{L}}+l\boldsymbol{a}_1$ and $\boldsymbol{O}_l\equiv \boldsymbol{O}+l\boldsymbol{a}_1$, thus $\boldsymbol{\mathcal{L}}_0=\boldsymbol{\mathcal{L}}$ and $\boldsymbol{O}_0=\boldsymbol{O}$. By expressing $\psi_s^A$ and $\psi_s^C$ in terms of $\psi_s^B$, Eqs.~(\ref{eq:bot1}) and (\ref{eq:top1}) are respectively given by
\begin{align}
    {\varepsilon}^2\psi_s^B(\boldsymbol{O}_l)=&~{t_A}^2\psi_s^B(\boldsymbol{O}_l) + t^2\Big[2\psi_s^B(\boldsymbol{O}_l)+\psi_s^B(\boldsymbol{O}_l+\boldsymbol{a}_2)\nonumber\\
    &+\psi_s^B(\boldsymbol{O}_l+\boldsymbol{a}_2-\boldsymbol{a}_1) +\sum_{\nu=\pm}\psi_s^B(\boldsymbol{O}_l+\nu\boldsymbol{a}_1)\Big],\label{eq:bot2}\\
    {\varepsilon}^2\psi_s^B(\boldsymbol{\mathcal{L}}_l)=&~{t_C}^2\psi_s^B(\boldsymbol{\mathcal{L}}_l) + t^2\Big[2\psi_s^B(\boldsymbol{\mathcal{L}}_l)+\psi_s^B(\boldsymbol{\mathcal{L}}_l-\boldsymbol{a}_2)\nonumber\\
    &+\psi_s^B(\boldsymbol{\mathcal{L}}_l-\boldsymbol{a}_2+\boldsymbol{a}_1) +\sum_{\nu=\pm}\psi_s^B(\boldsymbol{\mathcal{L}}_l-\nu\boldsymbol{a}_1)\Big].
    \label{eq:top2}
\end{align}
The missing terms of the TBEs at $\boldsymbol{O}_l$ and $\boldsymbol{\mathcal{L}}_l$ constitute the boundary conditions as follows:
\begin{align}
    &{t_C}^2\psi_s^B(\boldsymbol{O}_l)+t^2\Big[\psi_s^B(\boldsymbol{O}_l-\boldsymbol{a}_2)
    +\psi_s^B(\boldsymbol{O}_l-\boldsymbol{a}_2+\boldsymbol{a}_1)\Big]=0,\label{eq:SA1}\\
    &{t_A}^2\psi_s^B(\boldsymbol{\mathcal{L}}_l)+t^2\Big[\psi_s^B(\boldsymbol{\mathcal{L}}_l+\boldsymbol{a}_2)
    +\psi_s^B(\boldsymbol{\mathcal{L}}_l+\boldsymbol{a}_2-\boldsymbol{a}_1)\Big]=0.\label{eq:SA2}
\end{align}
Therefore, the boundary conditions are not indeterminate as suggested by Ref. \cite{oriekhov2018electronic}. By adopting $\psi_s({\boldsymbol{r}}$) in Eq.~(\ref{eq:psi-ribbon}), and assuming $\boldsymbol{k^\prime}=(k_x,-k_y)$ because momentum is conserved in the $x$ direction, Eqs.~(\ref{eq:SA1}) and (\ref{eq:SA2}) respectively reduce to
\begin{align}
    \mathcal{A}f_C(\boldsymbol{k})+\mathcal{A}^\prime f_C(\boldsymbol{k^\prime})&=0,\label{eq:GC}\\\mathcal{A}e^{i\boldsymbol{k\cdot \mathcal{L}}}f_A(\boldsymbol{k^\prime})+\mathcal{A}^\prime e^{i\boldsymbol{k^\prime\cdot \mathcal{L}}}f_A(\boldsymbol{k})&=0,\label{eq:GA}   
\end{align}
where $f_{\lambda}(\boldsymbol{k})=|f_{\lambda} (\boldsymbol{k})|e^{-i\phi_{\lambda}(\boldsymbol{k})}$, $\lambda=A,C$ are defined by 
\begin{align}
    f_C (\boldsymbol{k})\equiv \mathrm{sin}^2\vartheta + \beta(k_x) e^{-i\sqrt{3}k_ya/2},\label{eq:AfC}\\
    f_A (\boldsymbol{k})\equiv \mathrm{cos}^2\vartheta + \beta(k_x) e^{-i\sqrt{3}k_ya/2}.\label{eq:AfA}
\end{align}
By combining Eqs.~(\ref{eq:GC}) and (\ref{eq:GA}), we get
\begin{align}
     e^{ik_y\mathcal{L}_y}f_A(\boldsymbol{k^\prime})f_C(\boldsymbol{k^\prime})+e^{-ik_y\mathcal{L}_y}f_A(\boldsymbol{k})f_C(\boldsymbol{k}) = 0.
     \label{eq:comb-A}
\end{align}
Thus, the quantization condition is derived as follows:
\begin{align}
    \frac{\sqrt{3}}{2}(J-1)k_ya + \phi_0(\boldsymbol{k}) = n\pi,~\mathrm{for}~n=1,\dots, J, 
    \label{eq:ABC-quant}
\end{align}
where we define
\begin{align}
     \phi_0(\boldsymbol{k})= \phi_A(\boldsymbol{k})+ \phi_C(\boldsymbol{k}).
     \label{eq:ABC-Phi0}
\end{align}
Note that Eq.~(\ref{eq:ABC-quant}) is not mathematically analogous to those of the BCA and CAB ZRs [see Eq.~(\ref{eq:q-B})].

\begin{figure}[t]
\begin{center}
\includegraphics[width=72mm]{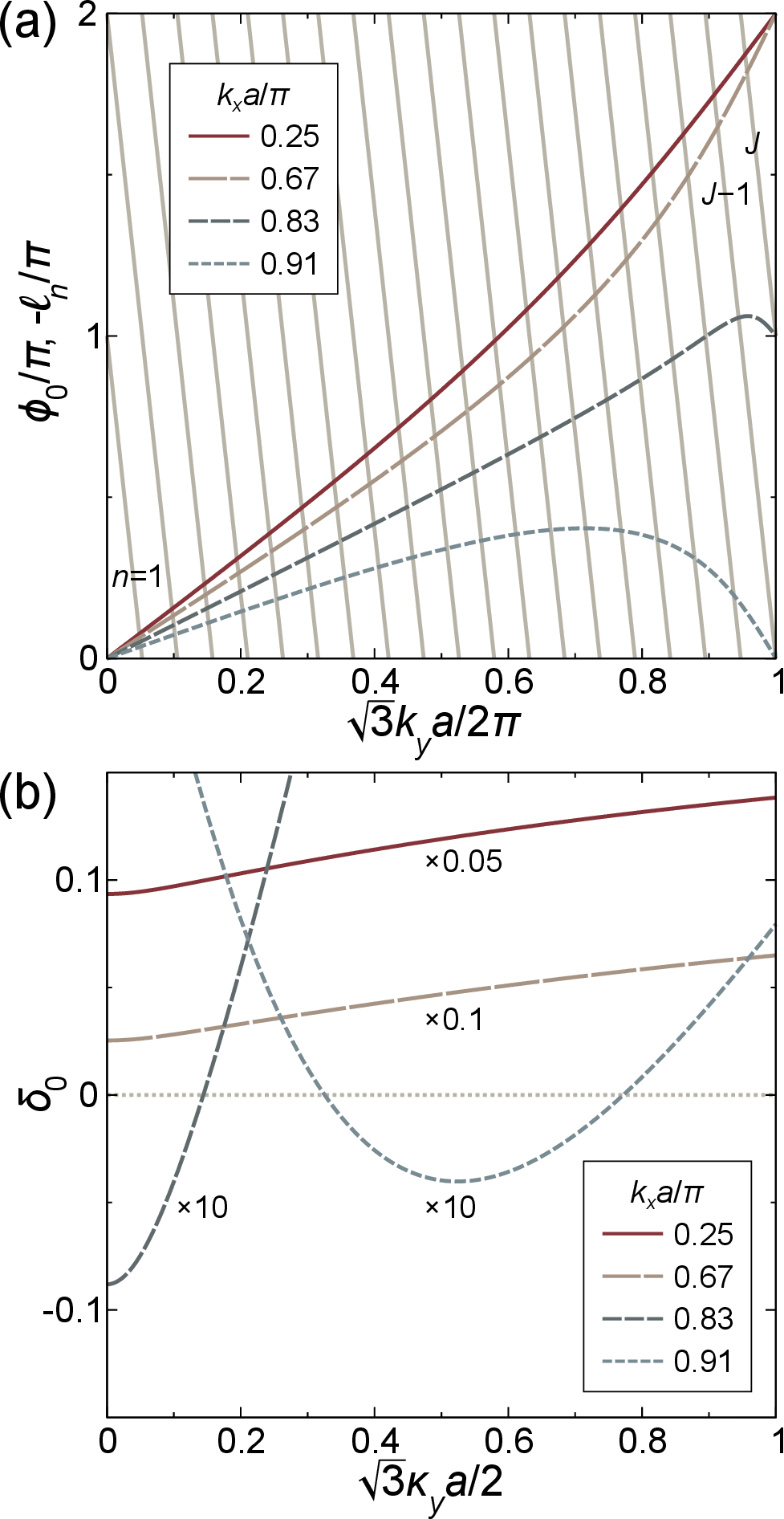}
\caption{(a) Plot of $\phi_0(\boldsymbol{k})$ as a function of $k_y$ for $k_x=0.25\pi/a$, $0.67\pi/a$, $0.83\pi/a$ and $0.91\pi/a$, where $\alpha=0.8$. The straight lines are $-\ell_n(k_y) = -\sqrt{3}(J-1)k_ya/2 + n\pi $, $n=1,\dots,J$, for $J=20$. (b) Plot of $\delta_0(\kappa_y)$ for the same $k_x$'s.   }
\label{fig:Phi-ABC-a0p8}
\end{center}
\end{figure}

\begin{figure}[t]
\begin{center}
\includegraphics[width=76mm]{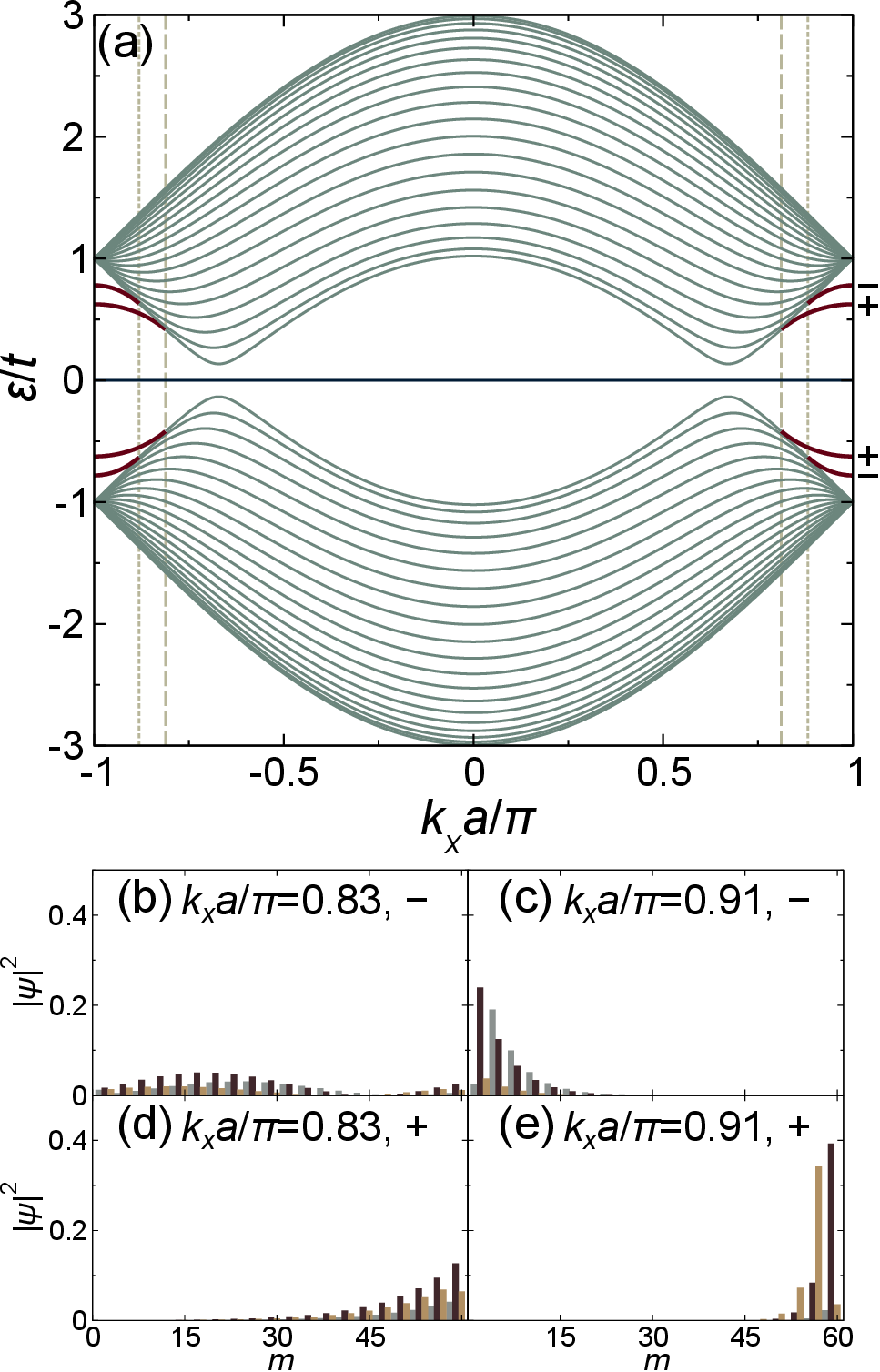}
\caption{(a) Energy spectra of the ABC ZR as a function of $k_x$ for $\alpha=0.8$ and $J=20$. The bold red lines indicate the edge states. The blue line at $\varepsilon=0$ is the flat band. The transitions from bulk to edge states for the "$+$" ["$-$"] bands occur at $|k_x| \approx 0.8116\pi/a$ [$0.8818\pi/a$], marked by the vertical dashed [dotted] lines. (b) and (c) [(d) and (e)] Probability density $|{\psi}|^2$ per site $m$ for the "$-$" ["$+$"] bands at $k_x=0.83\pi/a$ and $0.91\pi/a$, respectively. }
\label{fig:Band-Psi-ABC}
\end{center}
\end{figure}


By taking derivative of Eq.~(\ref{eq:ABC-quant}) with respect to $k_y$ and setting $k_y = 2\pi/\sqrt{3}a$, we obtain 
 \begin{align}
     \frac{J+1}{J-1}\beta^2 - \frac{J}{J-1}\beta + \frac{1}{4}\sin^2(2\vartheta)=0,~\mathrm{for}~J\geq 2. 
     \label{eq:ABC-quad1}
 \end{align}
Similarly, the solutions of Eq.~(\ref{eq:ABC-quad1}) are denoted by $\beta_{J}^\pm=2\cos(\chi_J^\pm/2)$. $\chi_J^\pm$ converge to 
\begin{subequations}
   \begin{align}
    \chi_\infty^+&\sim (2/ a)\cos^{-1}\left({\cos^2\vartheta/ 2}\right),\label{eq:chi-ABCp}\\
    \chi_\infty^-&\sim (2/ a)\cos^{-1}\left({\sin^2\vartheta/ 2}\right).\label{eq:chi-ABCm}
\end{align} 
\end{subequations}
Except for $\alpha=0$, the transitions from bulk to edge states do not occur at $|k_x|=2\pi/3a$, unlike the BCA and CAB ZRs. Fig.~\ref{fig:ABC-Appendix}(b) shows $\chi_J^\pm$ as a function of $J=2,\dots, 60$ for $\alpha=0.8$. The solid and dashed lines indicate $\chi_\infty^+\approx 0.80277\pi/a$ and $\chi_\infty^-\approx 0.87498\pi/a$, respectively.


Fig.~\ref{fig:Phi-ABC-a0p8}(a) shows $\phi_0(\boldsymbol{k})$ for several $k_x$'s and $-\ell_n(k_y)$, where $\ell_n(k_y)\equiv \sqrt{3}(J-1)k_ya/2 - n\pi$ for $J=20$. We can see that along $k_y\in(0,2\pi/\sqrt{3}a)$, $\phi_0$ for $k_x=0.83\pi/a$ [$0.91\pi/a$] does not intersect $-\ell_{J}$ [$-\ell_{J}$ and $-\ell_{J-1}$], thus two [four] bulk states are missing in the dispersive bands.   

To demonstrate that the edge states arise from the missing missing bulk states, we substitute $k_y=2\pi/\sqrt{3}a+i\kappa_y$ into Eq.~(\ref{eq:comb-A}). The existence of edge states requires 
\begin{align}
    \delta_0(\kappa_y)&\equiv\beta^2 - \frac{\sinh(\sqrt{3}J\kappa_ya/2)}{\sinh(\rho \kappa_y)}\beta+\frac{\sinh(\mathcal{L}_y \kappa_y)}{\sinh(\rho \kappa_y)}\frac{\sin^2(2\vartheta)}{4}\nonumber\\
    &=0.\label{eq:T0}
\end{align}
Fig.~\ref{fig:Phi-ABC-a0p8}(b) shows $\delta_0(\kappa_y)$ for several $k_x$'s. The number of zero-crossings for $k_x=0.83\pi/a$ [$0.91\pi/a$] is one [two], which indicates the existence of two [four] edge states. 


For $J\kappa_y a\gg 1$, Eq.~(\ref{eq:T0}) is rearranged into
\begin{align}
    e^{2\kappa_y}\beta^2 - e^{\kappa_y}\beta + \sin^2(2\vartheta)/4=0.\label{eq:T0b}
\end{align}
It is noted that Eqs.~(\ref{eq:T0b}) and (\ref{eq:ABC-quad1}) become identical in the limits $\kappa_y\rightarrow 0$ and $J\rightarrow \infty$, respectively. The solutions of Eq.~(\ref{eq:T0b}) are given by
\begin{subequations}
    \begin{align}
        \kappa_y^+ &= \mathrm{ln}(\cos^2\vartheta/\beta),\label{eq:kappap}\\
        \kappa_y^- &= \mathrm{ln}(\sin^2\vartheta/\beta)\label{eq:kappam}.
    \end{align}
\end{subequations}
By substituting $\kappa_y^\pm$ into Eq.~(\ref{eq:EnergyEdge-s}), the energies of edge states are given as follows:
\begin{subequations}
   \begin{align}
    \tilde{\varepsilon}_s^+&=st_C\sqrt{1-(\beta/\cos\vartheta)^2},\label{eq:EsAp}\\
    \tilde{\varepsilon}_s^-&=st_A\sqrt{1-(\beta/\sin\vartheta)^2}.\label{eq:EsAm}
\end{align} 
\end{subequations}
Since $\kappa_y^\pm>0$, the edge states associated with $\tilde{\varepsilon}_s^\pm$ appear in the range $\chi_\infty^\pm<|k_x|\leq \pi/a$, where $t_C\sin\vartheta < |\tilde{\varepsilon}_s^+| \leq t_C$ and $t_A\cos\vartheta < |\tilde{\varepsilon}_s^-| \leq t_A$. Note that for $\alpha=0$, the edge states with $\tilde{\varepsilon}_s^-$ vanish. The remaining edge states with $\tilde{\varepsilon}_s^+=0$ exist for $|k_x|\in (2\pi/3a,\pi/a]$ because the ABC ZR reduces to the graphene ZR by removing the $C$ sites.


Fig.~\ref{fig:Band-Psi-ABC}(a) shows the numerical calculation of energy spectra as a function of $k_x$. The vertical dashed and dotted lines accordingly mark $\chi_{20}^+\approx 0.8116\pi/a$ and $\chi_{20}^-\approx 0.8818\pi/a$. The bold red lines in the bands labeled by "$+$" and "$-$" depict the edge states with energies  $\tilde{\varepsilon}_s^+$ and $\tilde{\varepsilon}_s^-$, respectively. At $|k_x| =\pi/a$, the edge states are pinned at $\varepsilon=t_C\approx 0.6247t$ and $\varepsilon=t_A\approx 0.7809t$, consistent with Eqs. (\ref{eq:EsAp}) and (\ref{eq:EsAm}). 

Figs.~\ref{fig:Band-Psi-ABC}(b) and (c) [(d) and (e)] depict the probability density $|\psi|^2$ per site $m$ for the "$-$" ["$+$"] bands at $k_x=0.83\pi/a$ and $0.91\pi/a$, respectively. Since $0.83\pi/a<\chi_{20}^-<0.91\pi/a$, Figs.~\ref{fig:Band-Psi-ABC}(b) and (c) accordingly show the profile of bulk and edge states. On the other hand, both Figs.~\ref{fig:Band-Psi-ABC}(d) and (e) exhibit the profile of edge states because $\chi_{20}^+<0.83\pi/a<0.91\pi/a$.

\renewcommand\thesubsection{\arabic{subsection}}
\setcounter{equation}{0}
\renewcommand{\theequation}{B\arabic{equation}}
\section{DERIVATIONS OF EQS.~(\ref{eq:boun-B4}), (\ref{eq:boun-C4})}\label{app:derivation}

First, let us determine $\mathcal{A}^\prime$ and $\boldsymbol{k^\prime}$ in $\psi_s(\boldsymbol{r})$ [see Eq.~(\ref{eq:psi-ribbon})]. Eqs.~(\ref{eq:empty-Psi-B}) or (\ref{eq:empty-Psi-B2}) imply $\mathcal{A}^{\prime} = - \mathcal{A}$ and $\boldsymbol{k}\cdot \boldsymbol{a}_1 = \boldsymbol{k^{\prime}}\cdot \boldsymbol{a}_1$, due to the conservation of momentum in the $x$ direction. It can be shown that the conservation of energy $\varepsilon (\boldsymbol{k}) = \varepsilon (\boldsymbol{k^{\prime}})$ gives $\boldsymbol{k}\cdot \boldsymbol{a}_1 = \boldsymbol{k^{\prime}}\cdot \boldsymbol{a}_1=(\boldsymbol{k} + \boldsymbol{k^{\prime}})\cdot \boldsymbol{a}_2$ [see Eq.~(\ref{eq:Energy-s})]. By using the vector components of $\boldsymbol{a}_1$ and $\boldsymbol{a}_2$, we get 
\begin{align}
    \boldsymbol{k} = (k_x, k_y),~\mathrm{and}~\boldsymbol{k^{\prime}} = (k_x, -k_y).
    \label{eq:kk}
\end{align}


Next, we write Eq.~(\ref{eq:f}) as $f(\boldsymbol{k})=e^{-i\boldsymbol{k}\cdot\boldsymbol{b}_n}g_n(\boldsymbol{k})$, where 
\begin{align}
    g_1(\boldsymbol{k}) = 1 + e^{-i\boldsymbol{k}\cdot\boldsymbol{a}_1} + e^{-i\boldsymbol{k}\cdot\boldsymbol{a}_2} = q(\boldsymbol{k})e^{-i\boldsymbol{k}\cdot\boldsymbol{a}_1/2},\label{eq:g1}\\
    g_2(\boldsymbol{k}) = 1 + e^{i\boldsymbol{k}\cdot\boldsymbol{a}_1} + e^{i\boldsymbol{k}\cdot(\boldsymbol{a}_1-\boldsymbol{a}_2)} = q(\boldsymbol{k})e^{i\boldsymbol{k}\cdot\boldsymbol{a}_1/2},\label{eq:g2}\\
    g_3(\boldsymbol{k}) = 1 + e^{i\boldsymbol{k}\cdot\boldsymbol{a}_2} + e^{i\boldsymbol{k}\cdot(\boldsymbol{a}_2-\boldsymbol{a}_1)} = p^*(\boldsymbol{k}).\label{eq:g3}
\end{align}
Note that $|g_n(\boldsymbol{k})|=|f(\boldsymbol{k})|$. $p(\boldsymbol{k})$ and $q(\boldsymbol{k})$ are defined in Eqs.~(\ref{eq:pp}) and (\ref{eq:qq}), respectively.

\subsection{Derivation of Eq.~(\ref{eq:boun-B4})}\label{sapp:deriv1}

Let us express $\psi_s(\boldsymbol{r})$ in terms of $|\psi_s(\boldsymbol{k})\rangle$ [see Eq. (\ref{eq:wavefunction-s})]. Multiplying Eq.~(\ref{eq:boun-B1}) by  $|f(\boldsymbol{k})|$ yields
\begin{align}
    e^{-i\boldsymbol{k}\cdot\boldsymbol{L}}f(\boldsymbol{k})-e^{-i\boldsymbol{k^\prime}\cdot\boldsymbol{L}}f(\boldsymbol{k^\prime})+&\nonumber\\
    e^{-i\boldsymbol{k}\cdot(\boldsymbol{L}+\boldsymbol{a}_1)}f(\boldsymbol{k})-e^{-i\boldsymbol{k^\prime}\cdot(\boldsymbol{L}+\boldsymbol{a}_1)}f(\boldsymbol{k^\prime})+&\nonumber\\
    \alpha^2\{e^{-i\boldsymbol{k}\cdot(\boldsymbol{L}-\boldsymbol{b}_1)}f^*(\boldsymbol{k})-e^{-i\boldsymbol{k^\prime}\cdot(\boldsymbol{L}-\boldsymbol{b}_1)}f^*(\boldsymbol{k^\prime})\}&=0.
    \label{eq:boun-B2}
\end{align}
In the middle unit cell in Fig.~\ref{fig:aT3}(a), the $A$ and $C$ sites are connected to the $B$ site by $\boldsymbol{b}_3$ and $\boldsymbol{b}_1$, respectively. By expressing $f$ ($f^*$) in terms of $g_3$ ($g_1^*$), Eq.~(\ref{eq:boun-B2}) becomes
\begin{align}
    \{1+e^{-ik_xa}\}\{e^{-ik_y\rho}g_3(\boldsymbol{k})-e^{ik_y\rho}g_3(\boldsymbol{k^\prime}) \}+&\nonumber\\
    \alpha^2e^{-ik_xa}\{e^{-ik_y\rho}g_1^*(\boldsymbol{k}) - e^{ik_y\rho}g_1^*(\boldsymbol{k^\prime})  \}&=0.
    \label{eq:boun-B3}
\end{align}
Multiplying Eq.~(\ref{eq:boun-B3}) by $e^{ik_xa/2}$ and grouping the resulting terms gives Eq.~(\ref{eq:boun-B4}).


\subsection{Derivation of Eq.~(\ref{eq:boun-C4})}\label{sapp:deriv2}

By following the same procedure, we multiply Eq. (\ref{eq:boun-C1}) by $|f(\boldsymbol{k})|$ as follows:
\begin{align}
    e^{i\boldsymbol{k}\cdot(\boldsymbol{L}-\boldsymbol{b}_2)}f(\boldsymbol{k})-e^{i\boldsymbol{k^\prime}\cdot(\boldsymbol{L}-\boldsymbol{b}_2)}f(\boldsymbol{k^\prime})+&\nonumber\\
    \alpha^2\{e^{i\boldsymbol{k}\cdot\boldsymbol{L}}f^*(\boldsymbol{k})-e^{i\boldsymbol{k^\prime}\cdot\boldsymbol{L}}f^*(\boldsymbol{k^\prime})\}+&\nonumber\\
    \alpha^2\{e^{i\boldsymbol{k}\cdot(\boldsymbol{L}-\boldsymbol{a}_1)}f^*(\boldsymbol{k})-e^{i\boldsymbol{k^\prime}\cdot(\boldsymbol{L}-\boldsymbol{a}_1)}f^*(\boldsymbol{k^\prime})\}&=0.
    \label{eq:boun-C2}
\end{align}
In the top unit cell in Fig.~\ref{fig:aT3}(a), the $A$ ($C$) site is connected the $B$ site by $\boldsymbol{b}_2$ ($\boldsymbol{b}_3$). Hence, $f$ and $f^*$ in Eq.~(\ref{eq:boun-C2}) are expressed in terms of $g_2$ and $g_3^*$, respectively,
\begin{align}
    e^{-ik_xa}\{e^{ik_y\rho}g_2(\boldsymbol{k})-e^{-ik_y\rho}g_2(\boldsymbol{k^\prime})\}+&\nonumber\\
    \alpha^2\{1+e^{-ik_xa}\}\{e^{ik_y\rho}g_3^*(\boldsymbol{k}) - e^{-ik_y\rho}g_3^*(\boldsymbol{k^\prime})\}&=0.
    \label{eq:boun-C3}
\end{align}
Multiplying Eq.~(\ref{eq:boun-C3}) by $e^{ik_xa/2}$ and grouping the resulting terms leads to Eq.~(\ref{eq:boun-C4}).






\bibliographystyle{pratrev}
\bibliography{pratama}

\begin{thebibliography}{112}%
    \makeatletter
    \providecommand \@ifxundefined [1]{%
     \@ifx{#1\undefined}
    }%
    \providecommand \@ifnum [1]{%
     \ifnum #1\expandafter \@firstoftwo
     \else \expandafter \@secondoftwo
     \fi
    }%
    \providecommand \@ifx [1]{%
     \ifx #1\expandafter \@firstoftwo
     \else \expandafter \@secondoftwo
     \fi
    }%
    \providecommand \natexlab [1]{#1}%
    \providecommand \enquote  [1]{``#1''}%
    \providecommand \bibnamefont  [1]{#1}%
    \providecommand \bibfnamefont [1]{#1}%
    \providecommand \citenamefont [1]{#1}%
    \providecommand \href@noop [0]{\@secondoftwo}%
    \providecommand \href [0]{\begingroup \@sanitize@url \@href}%
    \providecommand \@href[1]{\@@startlink{#1}\@@href}%
    \providecommand \@@href[1]{\endgroup#1\@@endlink}%
    \providecommand \@sanitize@url [0]{\catcode `\\12\catcode `\$12\catcode
      `\&12\catcode `\#12\catcode `\^12\catcode `\_12\catcode `\%12\relax}%
    \providecommand \@@startlink[1]{}%
    \providecommand \@@endlink[0]{}%
    \providecommand \url  [0]{\begingroup\@sanitize@url \@url }%
    \providecommand \@url [1]{\endgroup\@href {#1}{\urlprefix }}%
    \providecommand \urlprefix  [0]{URL }%
    \providecommand \Eprint [0]{\href }%
    \providecommand \doibase [0]{http://dx.doi.org/}%
    \providecommand \selectlanguage [0]{\@gobble}%
    \providecommand \bibinfo  [0]{\@secondoftwo}%
    \providecommand \bibfield  [0]{\@secondoftwo}%
    \providecommand \translation [1]{[#1]}%
    \providecommand \BibitemOpen [0]{}%
    \providecommand \bibitemStop [0]{}%
    \providecommand \bibitemNoStop [0]{.\EOS\space}%
    \providecommand \EOS [0]{\spacefactor3000\relax}%
    \providecommand \BibitemShut  [1]{\csname bibitem#1\endcsname}%
    \let\auto@bib@innerbib\@empty
    \bibitem [{\citenamefont {Hasan}\ and\ \citenamefont {Kane}(2010)}]{hasan2010}%
      \BibitemOpen
      \bibfield  {author} {\bibinfo {author} {\bibfnamefont {M.~Z.}\ \bibnamefont
      {Hasan}}\ and\ \bibinfo {author} {\bibfnamefont {C.~L.}\ \bibnamefont
      {Kane}},\ }\bibfield  {title} {\enquote {\bibinfo {title} {Colloquium:
      topological insulators},}\ }\href@noop {} {\bibfield  {journal} {\bibinfo
      {journal} {Rev. Mod. Phys.}\ }\textbf {\bibinfo {volume} {82}},\ \bibinfo
      {pages} {3045} (\bibinfo {year} {2010})}\BibitemShut {NoStop}%
    \bibitem [{\citenamefont {Cooper}\ \emph {et~al.}(2019)\citenamefont {Cooper},
      \citenamefont {Dalibard},\ and\ \citenamefont {Spielman}}]{cooper2019}%
      \BibitemOpen
      \bibfield  {author} {\bibinfo {author} {\bibfnamefont {N.}~\bibnamefont
      {Cooper}}, \bibinfo {author} {\bibfnamefont {J.}~\bibnamefont {Dalibard}}, \
      and\ \bibinfo {author} {\bibfnamefont {I.}~\bibnamefont {Spielman}},\
      }\bibfield  {title} {\enquote {\bibinfo {title} {Topological bands for
      ultracold atoms},}\ }\href@noop {} {\bibfield  {journal} {\bibinfo  {journal}
      {Rev. Mod. Phys.}\ }\textbf {\bibinfo {volume} {91}},\ \bibinfo {pages}
      {015005} (\bibinfo {year} {2019})}\BibitemShut {NoStop}%
    \bibitem [{\citenamefont {Cayssol}\ and\ \citenamefont
      {Fuchs}(2021)}]{cayssol2021}%
      \BibitemOpen
      \bibfield  {author} {\bibinfo {author} {\bibfnamefont {J.}~\bibnamefont
      {Cayssol}}\ and\ \bibinfo {author} {\bibfnamefont {J.-N.}\ \bibnamefont
      {Fuchs}},\ }\bibfield  {title} {\enquote {\bibinfo {title} {Topological and
      geometrical aspects of band theory},}\ }\href@noop {} {\bibfield  {journal}
      {\bibinfo  {journal} {J. Phys. Mater.}\ }\textbf {\bibinfo {volume} {4}},\
      \bibinfo {pages} {034007} (\bibinfo {year} {2021})}\BibitemShut {NoStop}%
    \bibitem [{\citenamefont {Thouless}\ \emph {et~al.}(1982)\citenamefont
      {Thouless}, \citenamefont {Kohmoto}, \citenamefont {Nightingale},\ and\
      \citenamefont {den Nijs}}]{thouless1982}%
      \BibitemOpen
      \bibfield  {author} {\bibinfo {author} {\bibfnamefont {D.~J.}\ \bibnamefont
      {Thouless}}, \bibinfo {author} {\bibfnamefont {M.}~\bibnamefont {Kohmoto}},
      \bibinfo {author} {\bibfnamefont {M.~P.}\ \bibnamefont {Nightingale}}, \ and\
      \bibinfo {author} {\bibfnamefont {M.}~\bibnamefont {den Nijs}},\ }\bibfield
      {title} {\enquote {\bibinfo {title} {Quantized {H}all conductance in a
      two-dimensional periodic potential},}\ }\href@noop {} {\bibfield  {journal}
      {\bibinfo  {journal} {Phy. Rev. Lett.}\ }\textbf {\bibinfo {volume} {49}},\
      \bibinfo {pages} {405} (\bibinfo {year} {1982})}\BibitemShut {NoStop}%
    \bibitem [{\citenamefont {Klitzing}\ \emph {et~al.}(1980)\citenamefont
      {Klitzing}, \citenamefont {Dorda},\ and\ \citenamefont
      {Pepper}}]{klitzing1980}%
      \BibitemOpen
      \bibfield  {author} {\bibinfo {author} {\bibfnamefont {K.~v.}\ \bibnamefont
      {Klitzing}}, \bibinfo {author} {\bibfnamefont {G.}~\bibnamefont {Dorda}}, \
      and\ \bibinfo {author} {\bibfnamefont {M.}~\bibnamefont {Pepper}},\
      }\bibfield  {title} {\enquote {\bibinfo {title} {New method for high-accuracy
      determination of the fine-structure constant based on quantized {H}all
      resistance},}\ }\href@noop {} {\bibfield  {journal} {\bibinfo  {journal}
      {Phy. Rev. Lett.}\ }\textbf {\bibinfo {volume} {45}},\ \bibinfo {pages} {494}
      (\bibinfo {year} {1980})}\BibitemShut {NoStop}%
    \bibitem [{\citenamefont {Kane}\ and\ \citenamefont {Mele}(2005)}]{kane2005}%
      \BibitemOpen
      \bibfield  {author} {\bibinfo {author} {\bibfnamefont {C.~L.}\ \bibnamefont
      {Kane}}\ and\ \bibinfo {author} {\bibfnamefont {E.~J.}\ \bibnamefont
      {Mele}},\ }\bibfield  {title} {\enquote {\bibinfo {title} {Quantum spin
      {H}all effect in graphene},}\ }\href@noop {} {\bibfield  {journal} {\bibinfo
      {journal} {Phys. Rev. Lett.}\ }\textbf {\bibinfo {volume} {95}},\ \bibinfo
      {pages} {226801} (\bibinfo {year} {2005})}\BibitemShut {NoStop}%
    \bibitem [{\citenamefont {Berry}(1984)}]{berry1984}%
      \BibitemOpen
      \bibfield  {author} {\bibinfo {author} {\bibfnamefont {M.~V.}\ \bibnamefont
      {Berry}},\ }\bibfield  {title} {\enquote {\bibinfo {title} {Quantal phase
      factors accompanying adiabatic changes},}\ }\href@noop {} {\bibfield
      {journal} {\bibinfo  {journal} {Proc. R. Soc. A: Math. Phys. Eng. Sci.}\
      }\textbf {\bibinfo {volume} {392}},\ \bibinfo {pages} {45} (\bibinfo {year}
      {1984})}\BibitemShut {NoStop}%
    \bibitem [{\citenamefont {Zak}(1989)}]{zak1989}%
      \BibitemOpen
      \bibfield  {author} {\bibinfo {author} {\bibfnamefont {J.}~\bibnamefont
      {Zak}},\ }\bibfield  {title} {\enquote {\bibinfo {title} {{B}erry’s phase
      for energy bands in solids},}\ }\href@noop {} {\bibfield  {journal} {\bibinfo
       {journal} {Phys. Rev. Lett.}\ }\textbf {\bibinfo {volume} {62}},\ \bibinfo
      {pages} {2747} (\bibinfo {year} {1989})}\BibitemShut {NoStop}%
    \bibitem [{\citenamefont {Xiao}\ \emph {et~al.}(2010)\citenamefont {Xiao},
      \citenamefont {Chang},\ and\ \citenamefont {Niu}}]{xiao2010}%
      \BibitemOpen
      \bibfield  {author} {\bibinfo {author} {\bibfnamefont {D.}~\bibnamefont
      {Xiao}}, \bibinfo {author} {\bibfnamefont {M.-C.}\ \bibnamefont {Chang}}, \
      and\ \bibinfo {author} {\bibfnamefont {Q.}~\bibnamefont {Niu}},\ }\bibfield
      {title} {\enquote {\bibinfo {title} {{B}erry phase effects on electronic
      properties},}\ }\href@noop {} {\bibfield  {journal} {\bibinfo  {journal}
      {Rev. Mod. Phys.}\ }\textbf {\bibinfo {volume} {82}},\ \bibinfo {pages}
      {1959} (\bibinfo {year} {2010})}\BibitemShut {NoStop}%
    \bibitem [{\citenamefont {Su}\ \emph {et~al.}(1979)\citenamefont {Su},
      \citenamefont {Schrieffer},\ and\ \citenamefont {Heeger}}]{su1979}%
      \BibitemOpen
      \bibfield  {author} {\bibinfo {author} {\bibfnamefont {W.-P.}\ \bibnamefont
      {Su}}, \bibinfo {author} {\bibfnamefont {J.~R.}\ \bibnamefont {Schrieffer}},
      \ and\ \bibinfo {author} {\bibfnamefont {A.~J.}\ \bibnamefont {Heeger}},\
      }\bibfield  {title} {\enquote {\bibinfo {title} {Solitons in
      polyacetylene},}\ }\href@noop {} {\bibfield  {journal} {\bibinfo  {journal}
      {Phys. Rev. Lett.}\ }\textbf {\bibinfo {volume} {42}},\ \bibinfo {pages}
      {1698} (\bibinfo {year} {1979})}\BibitemShut {NoStop}%
    \bibitem [{\citenamefont {Delplace}\ \emph {et~al.}(2011)\citenamefont
      {Delplace}, \citenamefont {Ullmo},\ and\ \citenamefont
      {Montambaux}}]{delplace2011}%
      \BibitemOpen
      \bibfield  {author} {\bibinfo {author} {\bibfnamefont {P.}~\bibnamefont
      {Delplace}}, \bibinfo {author} {\bibfnamefont {D.}~\bibnamefont {Ullmo}}, \
      and\ \bibinfo {author} {\bibfnamefont {G.}~\bibnamefont {Montambaux}},\
      }\bibfield  {title} {\enquote {\bibinfo {title} {{Z}ak phase and the
      existence of edge states in graphene},}\ }\href@noop {} {\bibfield  {journal}
      {\bibinfo  {journal} {Phys. Rev. B}\ }\textbf {\bibinfo {volume} {84}},\
      \bibinfo {pages} {195452} (\bibinfo {year} {2011})}\BibitemShut {NoStop}%
    \bibitem [{\citenamefont {Cao}\ \emph {et~al.}(2017)\citenamefont {Cao},
      \citenamefont {Zhao},\ and\ \citenamefont {Louie}}]{cao2017}%
      \BibitemOpen
      \bibfield  {author} {\bibinfo {author} {\bibfnamefont {T.}~\bibnamefont
      {Cao}}, \bibinfo {author} {\bibfnamefont {F.}~\bibnamefont {Zhao}}, \ and\
      \bibinfo {author} {\bibfnamefont {S.~G.}\ \bibnamefont {Louie}},\ }\bibfield
      {title} {\enquote {\bibinfo {title} {Topological phases in graphene
      nanoribbons: junction states, spin centers, and quantum spin chains},}\
      }\href@noop {} {\bibfield  {journal} {\bibinfo  {journal} {Phys. Rev. Lett.}\
      }\textbf {\bibinfo {volume} {119}},\ \bibinfo {pages} {076401} (\bibinfo
      {year} {2017})}\BibitemShut {NoStop}%
    \bibitem [{\citenamefont {Gr{\"o}ning}\ \emph {et~al.}(2018)\citenamefont
      {Gr{\"o}ning}, \citenamefont {Wang}, \citenamefont {Yao}, \citenamefont
      {Pignedoli}, \citenamefont {Borin~Barin}, \citenamefont {Daniels},
      \citenamefont {Cupo}, \citenamefont {Meunier}, \citenamefont {Feng},
      \citenamefont {Narita} \emph {et~al.}}]{groning2018}%
      \BibitemOpen
      \bibfield  {author} {\bibinfo {author} {\bibfnamefont {O.}~\bibnamefont
      {Gr{\"o}ning}}, \bibinfo {author} {\bibfnamefont {S.}~\bibnamefont {Wang}},
      \bibinfo {author} {\bibfnamefont {X.}~\bibnamefont {Yao}}, \bibinfo {author}
      {\bibfnamefont {C.~A.}\ \bibnamefont {Pignedoli}}, \bibinfo {author}
      {\bibfnamefont {G.}~\bibnamefont {Borin~Barin}}, \bibinfo {author}
      {\bibfnamefont {C.}~\bibnamefont {Daniels}}, \bibinfo {author} {\bibfnamefont
      {A.}~\bibnamefont {Cupo}}, \bibinfo {author} {\bibfnamefont {V.}~\bibnamefont
      {Meunier}}, \bibinfo {author} {\bibfnamefont {X.}~\bibnamefont {Feng}},
      \bibinfo {author} {\bibfnamefont {A.}~\bibnamefont {Narita}},  \emph
      {et~al.},\ }\bibfield  {title} {\enquote {\bibinfo {title} {Engineering of
      robust topological quantum phases in graphene nanoribbons},}\ }\href@noop {}
      {\bibfield  {journal} {\bibinfo  {journal} {Nature}\ }\textbf {\bibinfo
      {volume} {560}},\ \bibinfo {pages} {209} (\bibinfo {year}
      {2018})}\BibitemShut {NoStop}%
    \bibitem [{\citenamefont {Li}\ \emph {et~al.}(2021)\citenamefont {Li},
      \citenamefont {Sanz}, \citenamefont {Merino-D{\'\i}ez}, \citenamefont
      {Vilas-Varela}, \citenamefont {Garcia-Lekue}, \citenamefont {Corso},
      \citenamefont {de~Oteyza}, \citenamefont {Frederiksen}, \citenamefont
      {Pe{\~n}a},\ and\ \citenamefont {Pascual}}]{li2021}%
      \BibitemOpen
      \bibfield  {author} {\bibinfo {author} {\bibfnamefont {J.}~\bibnamefont
      {Li}}, \bibinfo {author} {\bibfnamefont {S.}~\bibnamefont {Sanz}}, \bibinfo
      {author} {\bibfnamefont {N.}~\bibnamefont {Merino-D{\'\i}ez}}, \bibinfo
      {author} {\bibfnamefont {M.}~\bibnamefont {Vilas-Varela}}, \bibinfo {author}
      {\bibfnamefont {A.}~\bibnamefont {Garcia-Lekue}}, \bibinfo {author}
      {\bibfnamefont {M.}~\bibnamefont {Corso}}, \bibinfo {author} {\bibfnamefont
      {D.~G.}\ \bibnamefont {de~Oteyza}}, \bibinfo {author} {\bibfnamefont
      {T.}~\bibnamefont {Frederiksen}}, \bibinfo {author} {\bibfnamefont
      {D.}~\bibnamefont {Pe{\~n}a}}, \ and\ \bibinfo {author} {\bibfnamefont
      {J.~I.}\ \bibnamefont {Pascual}},\ }\bibfield  {title} {\enquote {\bibinfo
      {title} {Topological phase transition in chiral graphene nanoribbons: from
      edge bands to end states},}\ }\href@noop {} {\bibfield  {journal} {\bibinfo
      {journal} {Nat. Commun.}\ }\textbf {\bibinfo {volume} {12}},\ \bibinfo
      {pages} {5538} (\bibinfo {year} {2021})}\BibitemShut {NoStop}%
    \bibitem [{\citenamefont {Rhim}\ \emph {et~al.}(2018)\citenamefont {Rhim},
      \citenamefont {Bardarson},\ and\ \citenamefont {Slager}}]{rhim2018}%
      \BibitemOpen
      \bibfield  {author} {\bibinfo {author} {\bibfnamefont {J.-W.}\ \bibnamefont
      {Rhim}}, \bibinfo {author} {\bibfnamefont {J.~H.}\ \bibnamefont {Bardarson}},
      \ and\ \bibinfo {author} {\bibfnamefont {R.-J.}\ \bibnamefont {Slager}},\
      }\bibfield  {title} {\enquote {\bibinfo {title} {Unified bulk-boundary
      correspondence for band insulators},}\ }\href@noop {} {\bibfield  {journal}
      {\bibinfo  {journal} {Phys. Rev. B}\ }\textbf {\bibinfo {volume} {97}},\
      \bibinfo {pages} {115143} (\bibinfo {year} {2018})}\BibitemShut {NoStop}%
    \bibitem [{\citenamefont {Ryu}\ and\ \citenamefont {Hatsugai}(2002)}]{ryu2002}%
      \BibitemOpen
      \bibfield  {author} {\bibinfo {author} {\bibfnamefont {S.}~\bibnamefont
      {Ryu}}\ and\ \bibinfo {author} {\bibfnamefont {Y.}~\bibnamefont {Hatsugai}},\
      }\bibfield  {title} {\enquote {\bibinfo {title} {Topological origin of
      zero-energy edge states in particle-hole symmetric systems},}\ }\href@noop {}
      {\bibfield  {journal} {\bibinfo  {journal} {Phys. Rev. Lett.}\ }\textbf
      {\bibinfo {volume} {89}},\ \bibinfo {pages} {077002} (\bibinfo {year}
      {2002})}\BibitemShut {NoStop}%
    \bibitem [{\citenamefont {Mong}\ and\ \citenamefont
      {Shivamoggi}(2011)}]{mong2011}%
      \BibitemOpen
      \bibfield  {author} {\bibinfo {author} {\bibfnamefont {R.~S.}\ \bibnamefont
      {Mong}}\ and\ \bibinfo {author} {\bibfnamefont {V.}~\bibnamefont
      {Shivamoggi}},\ }\bibfield  {title} {\enquote {\bibinfo {title} {Edge states
      and the bulk-boundary correspondence in {D}irac {H}amiltonians},}\
      }\href@noop {} {\bibfield  {journal} {\bibinfo  {journal} {Phys. Rev. B}\
      }\textbf {\bibinfo {volume} {83}},\ \bibinfo {pages} {125109} (\bibinfo
      {year} {2011})}\BibitemShut {NoStop}%
    \bibitem [{\citenamefont {Ezawa}(2014)}]{ezawa2014}%
      \BibitemOpen
      \bibfield  {author} {\bibinfo {author} {\bibfnamefont {M.}~\bibnamefont
      {Ezawa}},\ }\bibfield  {title} {\enquote {\bibinfo {title} {Topological
      origin of quasi-flat edge band in phosphorene},}\ }\href@noop {} {\bibfield
      {journal} {\bibinfo  {journal} {New J. Phys.}\ }\textbf {\bibinfo {volume}
      {16}},\ \bibinfo {pages} {115004} (\bibinfo {year} {2014})}\BibitemShut
      {NoStop}%
    \bibitem [{\citenamefont {Gruji{\'c}}\ \emph {et~al.}(2016)\citenamefont
      {Gruji{\'c}}, \citenamefont {Ezawa}, \citenamefont {Tadi{\'c}},\ and\
      \citenamefont {Peeters}}]{grujic2016tunable}%
      \BibitemOpen
      \bibfield  {author} {\bibinfo {author} {\bibfnamefont {M.~M.}\ \bibnamefont
      {Gruji{\'c}}}, \bibinfo {author} {\bibfnamefont {M.}~\bibnamefont {Ezawa}},
      \bibinfo {author} {\bibfnamefont {M.~{\v{Z}}.}\ \bibnamefont {Tadi{\'c}}}, \
      and\ \bibinfo {author} {\bibfnamefont {F.~M.}\ \bibnamefont {Peeters}},\
      }\bibfield  {title} {\enquote {\bibinfo {title} {Tunable skewed edges in
      puckered structures},}\ }\href@noop {} {\bibfield  {journal} {\bibinfo
      {journal} {Phys. Rev. B}\ }\textbf {\bibinfo {volume} {93}},\ \bibinfo
      {pages} {245413} (\bibinfo {year} {2016})}\BibitemShut {NoStop}%
    \bibitem [{\citenamefont {van Miert}\ \emph {et~al.}(2016)\citenamefont {van
      Miert}, \citenamefont {Ortix},\ and\ \citenamefont {Smith}}]{van2016}%
      \BibitemOpen
      \bibfield  {author} {\bibinfo {author} {\bibfnamefont {G.}~\bibnamefont {van
      Miert}}, \bibinfo {author} {\bibfnamefont {C.}~\bibnamefont {Ortix}}, \ and\
      \bibinfo {author} {\bibfnamefont {C.~M.}\ \bibnamefont {Smith}},\ }\bibfield
      {title} {\enquote {\bibinfo {title} {Topological origin of edge states in
      two-dimensional inversion-symmetric insulators and semimetals},}\ }\href@noop
      {} {\bibfield  {journal} {\bibinfo  {journal} {2D Mater.}\ }\textbf {\bibinfo
      {volume} {4}},\ \bibinfo {pages} {015023} (\bibinfo {year}
      {2016})}\BibitemShut {NoStop}%
    \bibitem [{\citenamefont {Hitomi}\ \emph {et~al.}(2021)\citenamefont {Hitomi},
      \citenamefont {Kawakami},\ and\ \citenamefont {Koshino}}]{hitomi2021}%
      \BibitemOpen
      \bibfield  {author} {\bibinfo {author} {\bibfnamefont {M.}~\bibnamefont
      {Hitomi}}, \bibinfo {author} {\bibfnamefont {T.}~\bibnamefont {Kawakami}}, \
      and\ \bibinfo {author} {\bibfnamefont {M.}~\bibnamefont {Koshino}},\
      }\bibfield  {title} {\enquote {\bibinfo {title} {Multiorbital edge and corner
      states in black phosphorene},}\ }\href@noop {} {\bibfield  {journal}
      {\bibinfo  {journal} {Phys. Rev. B}\ }\textbf {\bibinfo {volume} {104}},\
      \bibinfo {pages} {125302} (\bibinfo {year} {2021})}\BibitemShut {NoStop}%
    \bibitem [{\citenamefont {Atala}\ \emph {et~al.}(2013)\citenamefont {Atala},
      \citenamefont {Aidelsburger}, \citenamefont {Barreiro}, \citenamefont
      {Abanin}, \citenamefont {Kitagawa}, \citenamefont {Demler},\ and\
      \citenamefont {Bloch}}]{atala2013}%
      \BibitemOpen
      \bibfield  {author} {\bibinfo {author} {\bibfnamefont {M.}~\bibnamefont
      {Atala}}, \bibinfo {author} {\bibfnamefont {M.}~\bibnamefont {Aidelsburger}},
      \bibinfo {author} {\bibfnamefont {J.~T.}\ \bibnamefont {Barreiro}}, \bibinfo
      {author} {\bibfnamefont {D.}~\bibnamefont {Abanin}}, \bibinfo {author}
      {\bibfnamefont {T.}~\bibnamefont {Kitagawa}}, \bibinfo {author}
      {\bibfnamefont {E.}~\bibnamefont {Demler}}, \ and\ \bibinfo {author}
      {\bibfnamefont {I.}~\bibnamefont {Bloch}},\ }\bibfield  {title} {\enquote
      {\bibinfo {title} {Direct measurement of the {Z}ak phase in topological
      {B}loch bands},}\ }\href@noop {} {\bibfield  {journal} {\bibinfo  {journal}
      {Nat. Phys.}\ }\textbf {\bibinfo {volume} {9}},\ \bibinfo {pages} {795}
      (\bibinfo {year} {2013})}\BibitemShut {NoStop}%
    \bibitem [{\citenamefont {Grusdt}\ \emph {et~al.}(2014)\citenamefont {Grusdt},
      \citenamefont {Abanin},\ and\ \citenamefont {Demler}}]{grusdt2014measuring}%
      \BibitemOpen
      \bibfield  {author} {\bibinfo {author} {\bibfnamefont {F.}~\bibnamefont
      {Grusdt}}, \bibinfo {author} {\bibfnamefont {D.}~\bibnamefont {Abanin}}, \
      and\ \bibinfo {author} {\bibfnamefont {E.}~\bibnamefont {Demler}},\
      }\bibfield  {title} {\enquote {\bibinfo {title} {Measuring {$\mathbb{{Z}}_2$}
      topological invariants in optical lattices using interferometry},}\
      }\href@noop {} {\bibfield  {journal} {\bibinfo  {journal} {Phys. Rev. A}\
      }\textbf {\bibinfo {volume} {89}},\ \bibinfo {pages} {043621} (\bibinfo
      {year} {2014})}\BibitemShut {NoStop}%
    \bibitem [{\citenamefont {Lu}\ \emph {et~al.}(2016)\citenamefont {Lu},
      \citenamefont {Schemmer}, \citenamefont {Aycock}, \citenamefont {Genkina},
      \citenamefont {Sugawa},\ and\ \citenamefont {Spielman}}]{lu2016}%
      \BibitemOpen
      \bibfield  {author} {\bibinfo {author} {\bibfnamefont {H.-I.}\ \bibnamefont
      {Lu}}, \bibinfo {author} {\bibfnamefont {M.}~\bibnamefont {Schemmer}},
      \bibinfo {author} {\bibfnamefont {L.~M.}\ \bibnamefont {Aycock}}, \bibinfo
      {author} {\bibfnamefont {D.}~\bibnamefont {Genkina}}, \bibinfo {author}
      {\bibfnamefont {S.}~\bibnamefont {Sugawa}}, \ and\ \bibinfo {author}
      {\bibfnamefont {I.~B.}\ \bibnamefont {Spielman}},\ }\bibfield  {title}
      {\enquote {\bibinfo {title} {Geometrical pumping with a bose-einstein
      condensate},}\ }\href@noop {} {\bibfield  {journal} {\bibinfo  {journal}
      {Phys. Rev. Lett.}\ }\textbf {\bibinfo {volume} {116}},\ \bibinfo {pages}
      {200402} (\bibinfo {year} {2016})}\BibitemShut {NoStop}%
    \bibitem [{\citenamefont {Meier}\ \emph {et~al.}(2016)\citenamefont {Meier},
      \citenamefont {An},\ and\ \citenamefont {Gadway}}]{meier2016soliton}%
      \BibitemOpen
      \bibfield  {author} {\bibinfo {author} {\bibfnamefont {E.~J.}\ \bibnamefont
      {Meier}}, \bibinfo {author} {\bibfnamefont {F.~A.}\ \bibnamefont {An}}, \
      and\ \bibinfo {author} {\bibfnamefont {B.}~\bibnamefont {Gadway}},\
      }\bibfield  {title} {\enquote {\bibinfo {title} {Observation of the
      topological soliton state in the {S}u-{S}chrieffer-{H}eeger model},}\
      }\href@noop {} {\bibfield  {journal} {\bibinfo  {journal} {Nat. Commun.}\
      }\textbf {\bibinfo {volume} {7}},\ \bibinfo {pages} {13986} (\bibinfo {year}
      {2016})}\BibitemShut {NoStop}%
    \bibitem [{\citenamefont {Mivehvar}\ \emph {et~al.}(2017)\citenamefont
      {Mivehvar}, \citenamefont {Ritsch},\ and\ \citenamefont
      {Piazza}}]{mivehvar2017}%
      \BibitemOpen
      \bibfield  {author} {\bibinfo {author} {\bibfnamefont {F.}~\bibnamefont
      {Mivehvar}}, \bibinfo {author} {\bibfnamefont {H.}~\bibnamefont {Ritsch}}, \
      and\ \bibinfo {author} {\bibfnamefont {F.}~\bibnamefont {Piazza}},\
      }\bibfield  {title} {\enquote {\bibinfo {title} {Superradiant topological
      peierls insulator inside an optical cavity},}\ }\href@noop {} {\bibfield
      {journal} {\bibinfo  {journal} {Phys. Rev. Lett.}\ }\textbf {\bibinfo
      {volume} {118}},\ \bibinfo {pages} {073602} (\bibinfo {year}
      {2017})}\BibitemShut {NoStop}%
    \bibitem [{\citenamefont {St-Jean}\ \emph {et~al.}(2017)\citenamefont
      {St-Jean}, \citenamefont {Goblot}, \citenamefont {Galopin}, \citenamefont
      {Lema{\^\i}tre}, \citenamefont {Ozawa}, \citenamefont {Le~Gratiet},
      \citenamefont {Sagnes}, \citenamefont {Bloch},\ and\ \citenamefont
      {Amo}}]{st2017lasing}%
      \BibitemOpen
      \bibfield  {author} {\bibinfo {author} {\bibfnamefont {P.}~\bibnamefont
      {St-Jean}}, \bibinfo {author} {\bibfnamefont {V.}~\bibnamefont {Goblot}},
      \bibinfo {author} {\bibfnamefont {E.}~\bibnamefont {Galopin}}, \bibinfo
      {author} {\bibfnamefont {A.}~\bibnamefont {Lema{\^\i}tre}}, \bibinfo {author}
      {\bibfnamefont {T.}~\bibnamefont {Ozawa}}, \bibinfo {author} {\bibfnamefont
      {L.}~\bibnamefont {Le~Gratiet}}, \bibinfo {author} {\bibfnamefont
      {I.}~\bibnamefont {Sagnes}}, \bibinfo {author} {\bibfnamefont
      {J.}~\bibnamefont {Bloch}}, \ and\ \bibinfo {author} {\bibfnamefont
      {A.}~\bibnamefont {Amo}},\ }\bibfield  {title} {\enquote {\bibinfo {title}
      {Lasing in topological edge states of a one-dimensional lattice},}\
      }\href@noop {} {\bibfield  {journal} {\bibinfo  {journal} {Nat. Photonics}\
      }\textbf {\bibinfo {volume} {11}},\ \bibinfo {pages} {651} (\bibinfo {year}
      {2017})}\BibitemShut {NoStop}%
    \bibitem [{\citenamefont {Longhi}(2018)}]{longhi2018}%
      \BibitemOpen
      \bibfield  {author} {\bibinfo {author} {\bibfnamefont {S.}~\bibnamefont
      {Longhi}},\ }\bibfield  {title} {\enquote {\bibinfo {title} {Probing
      one-dimensional topological phases in waveguide lattices with broken chiral
      symmetry},}\ }\href@noop {} {\bibfield  {journal} {\bibinfo  {journal} {Opt.
      Lett.}\ }\textbf {\bibinfo {volume} {43}},\ \bibinfo {pages} {4639} (\bibinfo
      {year} {2018})}\BibitemShut {NoStop}%
    \bibitem [{\citenamefont {Jiang}\ \emph {et~al.}(2018)\citenamefont {Jiang},
      \citenamefont {Guo}, \citenamefont {Ding}, \citenamefont {Sun}, \citenamefont
      {Li}, \citenamefont {Jiang},\ and\ \citenamefont
      {Chen}}]{jiang2018experimental}%
      \BibitemOpen
      \bibfield  {author} {\bibinfo {author} {\bibfnamefont {J.}~\bibnamefont
      {Jiang}}, \bibinfo {author} {\bibfnamefont {Z.}~\bibnamefont {Guo}}, \bibinfo
      {author} {\bibfnamefont {Y.}~\bibnamefont {Ding}}, \bibinfo {author}
      {\bibfnamefont {Y.}~\bibnamefont {Sun}}, \bibinfo {author} {\bibfnamefont
      {Y.}~\bibnamefont {Li}}, \bibinfo {author} {\bibfnamefont {H.}~\bibnamefont
      {Jiang}}, \ and\ \bibinfo {author} {\bibfnamefont {H.}~\bibnamefont {Chen}},\
      }\bibfield  {title} {\enquote {\bibinfo {title} {Experimental demonstration
      of the robust edge states in a split-ring-resonator chain},}\ }\href@noop {}
      {\bibfield  {journal} {\bibinfo  {journal} {Opt. Express}\ }\textbf {\bibinfo
      {volume} {26}},\ \bibinfo {pages} {12891} (\bibinfo {year}
      {2018})}\BibitemShut {NoStop}%
    \bibitem [{\citenamefont {Jiao}\ \emph {et~al.}(2021)\citenamefont {Jiao},
      \citenamefont {Longhi}, \citenamefont {Wang}, \citenamefont {Gao},
      \citenamefont {Zhou}, \citenamefont {Wang}, \citenamefont {Fu}, \citenamefont
      {Wang}, \citenamefont {Ren}, \citenamefont {Qiao} \emph {et~al.}}]{jiao2021}%
      \BibitemOpen
      \bibfield  {author} {\bibinfo {author} {\bibfnamefont {Z.-Q.}\ \bibnamefont
      {Jiao}}, \bibinfo {author} {\bibfnamefont {S.}~\bibnamefont {Longhi}},
      \bibinfo {author} {\bibfnamefont {X.-W.}\ \bibnamefont {Wang}}, \bibinfo
      {author} {\bibfnamefont {J.}~\bibnamefont {Gao}}, \bibinfo {author}
      {\bibfnamefont {W.-H.}\ \bibnamefont {Zhou}}, \bibinfo {author}
      {\bibfnamefont {Y.}~\bibnamefont {Wang}}, \bibinfo {author} {\bibfnamefont
      {Y.-X.}\ \bibnamefont {Fu}}, \bibinfo {author} {\bibfnamefont
      {L.}~\bibnamefont {Wang}}, \bibinfo {author} {\bibfnamefont {R.-J.}\
      \bibnamefont {Ren}}, \bibinfo {author} {\bibfnamefont {L.-F.}\ \bibnamefont
      {Qiao}},  \emph {et~al.},\ }\bibfield  {title} {\enquote {\bibinfo {title}
      {Experimentally detecting quantized {Z}ak phases without chiral symmetry in
      photonic lattices},}\ }\href@noop {} {\bibfield  {journal} {\bibinfo
      {journal} {Phys. Rev. Lett.}\ }\textbf {\bibinfo {volume} {127}},\ \bibinfo
      {pages} {147401} (\bibinfo {year} {2021})}\BibitemShut {NoStop}%
    \bibitem [{\citenamefont {Goren}\ \emph {et~al.}(2018)\citenamefont {Goren},
      \citenamefont {Plekhanov}, \citenamefont {Appas},\ and\ \citenamefont
      {Le~Hur}}]{goren2018}%
      \BibitemOpen
      \bibfield  {author} {\bibinfo {author} {\bibfnamefont {T.}~\bibnamefont
      {Goren}}, \bibinfo {author} {\bibfnamefont {K.}~\bibnamefont {Plekhanov}},
      \bibinfo {author} {\bibfnamefont {F.}~\bibnamefont {Appas}}, \ and\ \bibinfo
      {author} {\bibfnamefont {K.}~\bibnamefont {Le~Hur}},\ }\bibfield  {title}
      {\enquote {\bibinfo {title} {Topological {Z}ak phase in strongly coupled
      {L}{C} circuits},}\ }\href@noop {} {\bibfield  {journal} {\bibinfo  {journal}
      {Phys. Rev. B}\ }\textbf {\bibinfo {volume} {97}},\ \bibinfo {pages} {041106}
      (\bibinfo {year} {2018})}\BibitemShut {NoStop}%
    \bibitem [{\citenamefont {Tao}\ \emph {et~al.}(2011)\citenamefont {Tao},
      \citenamefont {Jiao}, \citenamefont {Yazyev}, \citenamefont {Chen},
      \citenamefont {Feng}, \citenamefont {Zhang}, \citenamefont {Capaz},
      \citenamefont {Tour}, \citenamefont {Zettl}, \citenamefont {Louie} \emph
      {et~al.}}]{tao2011}%
      \BibitemOpen
      \bibfield  {author} {\bibinfo {author} {\bibfnamefont {C.}~\bibnamefont
      {Tao}}, \bibinfo {author} {\bibfnamefont {L.}~\bibnamefont {Jiao}}, \bibinfo
      {author} {\bibfnamefont {O.~V.}\ \bibnamefont {Yazyev}}, \bibinfo {author}
      {\bibfnamefont {Y.-C.}\ \bibnamefont {Chen}}, \bibinfo {author}
      {\bibfnamefont {J.}~\bibnamefont {Feng}}, \bibinfo {author} {\bibfnamefont
      {X.}~\bibnamefont {Zhang}}, \bibinfo {author} {\bibfnamefont {R.~B.}\
      \bibnamefont {Capaz}}, \bibinfo {author} {\bibfnamefont {J.~M.}\ \bibnamefont
      {Tour}}, \bibinfo {author} {\bibfnamefont {A.}~\bibnamefont {Zettl}},
      \bibinfo {author} {\bibfnamefont {S.~G.}\ \bibnamefont {Louie}},  \emph
      {et~al.},\ }\bibfield  {title} {\enquote {\bibinfo {title} {Spatially
      resolving edge states of chiral graphene nanoribbons},}\ }\href@noop {}
      {\bibfield  {journal} {\bibinfo  {journal} {Nat. Phys.}\ }\textbf {\bibinfo
      {volume} {7}},\ \bibinfo {pages} {616} (\bibinfo {year} {2011})}\BibitemShut
      {NoStop}%
    \bibitem [{\citenamefont {Wang}\ \emph
      {et~al.}(2016{\natexlab{a}})\citenamefont {Wang}, \citenamefont {Talirz},
      \citenamefont {Pignedoli}, \citenamefont {Feng}, \citenamefont {M{\"u}llen},
      \citenamefont {Fasel},\ and\ \citenamefont {Ruffieux}}]{wang2016giant}%
      \BibitemOpen
      \bibfield  {author} {\bibinfo {author} {\bibfnamefont {S.}~\bibnamefont
      {Wang}}, \bibinfo {author} {\bibfnamefont {L.}~\bibnamefont {Talirz}},
      \bibinfo {author} {\bibfnamefont {C.~A.}\ \bibnamefont {Pignedoli}}, \bibinfo
      {author} {\bibfnamefont {X.}~\bibnamefont {Feng}}, \bibinfo {author}
      {\bibfnamefont {K.}~\bibnamefont {M{\"u}llen}}, \bibinfo {author}
      {\bibfnamefont {R.}~\bibnamefont {Fasel}}, \ and\ \bibinfo {author}
      {\bibfnamefont {P.}~\bibnamefont {Ruffieux}},\ }\bibfield  {title} {\enquote
      {\bibinfo {title} {Giant edge state splitting at atomically precise graphene
      zigzag edges},}\ }\href@noop {} {\bibfield  {journal} {\bibinfo  {journal}
      {Nat. Commun.}\ }\textbf {\bibinfo {volume} {7}},\ \bibinfo {pages} {11507}
      (\bibinfo {year} {2016}{\natexlab{a}})}\BibitemShut {NoStop}%
    \bibitem [{\citenamefont {Prudkovskiy}\ \emph {et~al.}(2022)\citenamefont
      {Prudkovskiy}, \citenamefont {Hu}, \citenamefont {Zhang}, \citenamefont {Hu},
      \citenamefont {Ji}, \citenamefont {Nunn}, \citenamefont {Zhao}, \citenamefont
      {Shi}, \citenamefont {Tejeda}, \citenamefont {Wander} \emph
      {et~al.}}]{prudkovskiy2022epitaxial}%
      \BibitemOpen
      \bibfield  {author} {\bibinfo {author} {\bibfnamefont {V.~S.}\ \bibnamefont
      {Prudkovskiy}}, \bibinfo {author} {\bibfnamefont {Y.}~\bibnamefont {Hu}},
      \bibinfo {author} {\bibfnamefont {K.}~\bibnamefont {Zhang}}, \bibinfo
      {author} {\bibfnamefont {Y.}~\bibnamefont {Hu}}, \bibinfo {author}
      {\bibfnamefont {P.}~\bibnamefont {Ji}}, \bibinfo {author} {\bibfnamefont
      {G.}~\bibnamefont {Nunn}}, \bibinfo {author} {\bibfnamefont {J.}~\bibnamefont
      {Zhao}}, \bibinfo {author} {\bibfnamefont {C.}~\bibnamefont {Shi}}, \bibinfo
      {author} {\bibfnamefont {A.}~\bibnamefont {Tejeda}}, \bibinfo {author}
      {\bibfnamefont {D.}~\bibnamefont {Wander}},  \emph {et~al.},\ }\bibfield
      {title} {\enquote {\bibinfo {title} {An epitaxial graphene platform for
      zero-energy edge state nanoelectronics},}\ }\href@noop {} {\bibfield
      {journal} {\bibinfo  {journal} {Nat. Commun.}\ }\textbf {\bibinfo {volume}
      {13}},\ \bibinfo {pages} {7814} (\bibinfo {year} {2022})}\BibitemShut
      {NoStop}%
    \bibitem [{\citenamefont {Polini}\ \emph {et~al.}(2013)\citenamefont {Polini},
      \citenamefont {Guinea}, \citenamefont {Lewenstein}, \citenamefont
      {Manoharan},\ and\ \citenamefont {Pellegrini}}]{polini2013artificial}%
      \BibitemOpen
      \bibfield  {author} {\bibinfo {author} {\bibfnamefont {M.}~\bibnamefont
      {Polini}}, \bibinfo {author} {\bibfnamefont {F.}~\bibnamefont {Guinea}},
      \bibinfo {author} {\bibfnamefont {M.}~\bibnamefont {Lewenstein}}, \bibinfo
      {author} {\bibfnamefont {H.~C.}\ \bibnamefont {Manoharan}}, \ and\ \bibinfo
      {author} {\bibfnamefont {V.}~\bibnamefont {Pellegrini}},\ }\bibfield  {title}
      {\enquote {\bibinfo {title} {Artificial honeycomb lattices for electrons,
      atoms and photons},}\ }\href@noop {} {\bibfield  {journal} {\bibinfo
      {journal} {Nat. Nanotechnol.}\ }\textbf {\bibinfo {volume} {8}},\ \bibinfo
      {pages} {625} (\bibinfo {year} {2013})}\BibitemShut {NoStop}%
    \bibitem [{\citenamefont {Rechtsman}\ \emph {et~al.}(2013)\citenamefont
      {Rechtsman}, \citenamefont {Plotnik}, \citenamefont {Zeuner}, \citenamefont
      {Song}, \citenamefont {Chen}, \citenamefont {Szameit},\ and\ \citenamefont
      {Segev}}]{rechtsman2013topological}%
      \BibitemOpen
      \bibfield  {author} {\bibinfo {author} {\bibfnamefont {M.~C.}\ \bibnamefont
      {Rechtsman}}, \bibinfo {author} {\bibfnamefont {Y.}~\bibnamefont {Plotnik}},
      \bibinfo {author} {\bibfnamefont {J.~M.}\ \bibnamefont {Zeuner}}, \bibinfo
      {author} {\bibfnamefont {D.}~\bibnamefont {Song}}, \bibinfo {author}
      {\bibfnamefont {Z.}~\bibnamefont {Chen}}, \bibinfo {author} {\bibfnamefont
      {A.}~\bibnamefont {Szameit}}, \ and\ \bibinfo {author} {\bibfnamefont
      {M.}~\bibnamefont {Segev}},\ }\bibfield  {title} {\enquote {\bibinfo {title}
      {Topological creation and destruction of edge states in photonic graphene},}\
      }\href@noop {} {\bibfield  {journal} {\bibinfo  {journal} {Phys. Rev. Lett.}\
      }\textbf {\bibinfo {volume} {111}},\ \bibinfo {pages} {103901} (\bibinfo
      {year} {2013})}\BibitemShut {NoStop}%
    \bibitem [{\citenamefont {Plotnik}\ \emph {et~al.}(2014)\citenamefont
      {Plotnik}, \citenamefont {Rechtsman}, \citenamefont {Song}, \citenamefont
      {Heinrich}, \citenamefont {Zeuner}, \citenamefont {Nolte}, \citenamefont
      {Lumer}, \citenamefont {Malkova}, \citenamefont {Xu}, \citenamefont {Szameit}
      \emph {et~al.}}]{plotnik2014observation}%
      \BibitemOpen
      \bibfield  {author} {\bibinfo {author} {\bibfnamefont {Y.}~\bibnamefont
      {Plotnik}}, \bibinfo {author} {\bibfnamefont {M.~C.}\ \bibnamefont
      {Rechtsman}}, \bibinfo {author} {\bibfnamefont {D.}~\bibnamefont {Song}},
      \bibinfo {author} {\bibfnamefont {M.}~\bibnamefont {Heinrich}}, \bibinfo
      {author} {\bibfnamefont {J.~M.}\ \bibnamefont {Zeuner}}, \bibinfo {author}
      {\bibfnamefont {S.}~\bibnamefont {Nolte}}, \bibinfo {author} {\bibfnamefont
      {Y.}~\bibnamefont {Lumer}}, \bibinfo {author} {\bibfnamefont
      {N.}~\bibnamefont {Malkova}}, \bibinfo {author} {\bibfnamefont
      {J.}~\bibnamefont {Xu}}, \bibinfo {author} {\bibfnamefont {A.}~\bibnamefont
      {Szameit}},  \emph {et~al.},\ }\bibfield  {title} {\enquote {\bibinfo {title}
      {Observation of unconventional edge states in ‘photonic graphene’},}\
      }\href@noop {} {\bibfield  {journal} {\bibinfo  {journal} {Nat. Mater.}\
      }\textbf {\bibinfo {volume} {13}},\ \bibinfo {pages} {57} (\bibinfo {year}
      {2014})}\BibitemShut {NoStop}%
    \bibitem [{\citenamefont {Bellec}\ \emph {et~al.}(2014)\citenamefont {Bellec},
      \citenamefont {Kuhl}, \citenamefont {Montambaux},\ and\ \citenamefont
      {Mortessagne}}]{bellec2014manipulation}%
      \BibitemOpen
      \bibfield  {author} {\bibinfo {author} {\bibfnamefont {M.}~\bibnamefont
      {Bellec}}, \bibinfo {author} {\bibfnamefont {U.}~\bibnamefont {Kuhl}},
      \bibinfo {author} {\bibfnamefont {G.}~\bibnamefont {Montambaux}}, \ and\
      \bibinfo {author} {\bibfnamefont {F.}~\bibnamefont {Mortessagne}},\
      }\bibfield  {title} {\enquote {\bibinfo {title} {Manipulation of edge states
      in microwave artificial graphene},}\ }\href@noop {} {\bibfield  {journal}
      {\bibinfo  {journal} {New J. Phys.}\ }\textbf {\bibinfo {volume} {16}},\
      \bibinfo {pages} {113023} (\bibinfo {year} {2014})}\BibitemShut {NoStop}%
    \bibitem [{\citenamefont {Mili{\'c}evi{\'c}}\ \emph {et~al.}(2017)\citenamefont
      {Mili{\'c}evi{\'c}}, \citenamefont {Ozawa}, \citenamefont {Montambaux},
      \citenamefont {Carusotto}, \citenamefont {Galopin}, \citenamefont
      {Lema{\^\i}tre}, \citenamefont {Le~Gratiet}, \citenamefont {Sagnes},
      \citenamefont {Bloch},\ and\ \citenamefont {Amo}}]{milicevic2017orbital}%
      \BibitemOpen
      \bibfield  {author} {\bibinfo {author} {\bibfnamefont {M.}~\bibnamefont
      {Mili{\'c}evi{\'c}}}, \bibinfo {author} {\bibfnamefont {T.}~\bibnamefont
      {Ozawa}}, \bibinfo {author} {\bibfnamefont {G.}~\bibnamefont {Montambaux}},
      \bibinfo {author} {\bibfnamefont {I.}~\bibnamefont {Carusotto}}, \bibinfo
      {author} {\bibfnamefont {E.}~\bibnamefont {Galopin}}, \bibinfo {author}
      {\bibfnamefont {A.}~\bibnamefont {Lema{\^\i}tre}}, \bibinfo {author}
      {\bibfnamefont {L.}~\bibnamefont {Le~Gratiet}}, \bibinfo {author}
      {\bibfnamefont {I.}~\bibnamefont {Sagnes}}, \bibinfo {author} {\bibfnamefont
      {J.}~\bibnamefont {Bloch}}, \ and\ \bibinfo {author} {\bibfnamefont
      {A.}~\bibnamefont {Amo}},\ }\bibfield  {title} {\enquote {\bibinfo {title}
      {Orbital edge states in a photonic honeycomb lattice},}\ }\href@noop {}
      {\bibfield  {journal} {\bibinfo  {journal} {Phys. Rev. Lett.}\ }\textbf
      {\bibinfo {volume} {118}},\ \bibinfo {pages} {107403} (\bibinfo {year}
      {2017})}\BibitemShut {NoStop}%
    \bibitem [{\citenamefont {Zhang}\ \emph {et~al.}(2020)\citenamefont {Zhang},
      \citenamefont {Wang}, \citenamefont {Zhang}, \citenamefont {Kartashov},
      \citenamefont {Li}, \citenamefont {Zhong}, \citenamefont {Guan},
      \citenamefont {Gao}, \citenamefont {Li}, \citenamefont {Zhang} \emph
      {et~al.}}]{zhang2020soliton}%
      \BibitemOpen
      \bibfield  {author} {\bibinfo {author} {\bibfnamefont {Z.}~\bibnamefont
      {Zhang}}, \bibinfo {author} {\bibfnamefont {R.}~\bibnamefont {Wang}},
      \bibinfo {author} {\bibfnamefont {Y.}~\bibnamefont {Zhang}}, \bibinfo
      {author} {\bibfnamefont {Y.~V.}\ \bibnamefont {Kartashov}}, \bibinfo {author}
      {\bibfnamefont {F.}~\bibnamefont {Li}}, \bibinfo {author} {\bibfnamefont
      {H.}~\bibnamefont {Zhong}}, \bibinfo {author} {\bibfnamefont
      {H.}~\bibnamefont {Guan}}, \bibinfo {author} {\bibfnamefont {K.}~\bibnamefont
      {Gao}}, \bibinfo {author} {\bibfnamefont {F.}~\bibnamefont {Li}}, \bibinfo
      {author} {\bibfnamefont {Y.}~\bibnamefont {Zhang}},  \emph {et~al.},\
      }\bibfield  {title} {\enquote {\bibinfo {title} {Observation of edge solitons
      in photonic graphene},}\ }\href@noop {} {\bibfield  {journal} {\bibinfo
      {journal} {Nat. Commun.}\ }\textbf {\bibinfo {volume} {11}},\ \bibinfo
      {pages} {1902} (\bibinfo {year} {2020})}\BibitemShut {NoStop}%
    \bibitem [{\citenamefont {Xia}\ \emph {et~al.}(2023)\citenamefont {Xia},
      \citenamefont {Liang}, \citenamefont {Tang}, \citenamefont {Song},
      \citenamefont {Xu},\ and\ \citenamefont {Chen}}]{xia2023photonic}%
      \BibitemOpen
      \bibfield  {author} {\bibinfo {author} {\bibfnamefont {S.}~\bibnamefont
      {Xia}}, \bibinfo {author} {\bibfnamefont {Y.}~\bibnamefont {Liang}}, \bibinfo
      {author} {\bibfnamefont {L.}~\bibnamefont {Tang}}, \bibinfo {author}
      {\bibfnamefont {D.}~\bibnamefont {Song}}, \bibinfo {author} {\bibfnamefont
      {J.}~\bibnamefont {Xu}}, \ and\ \bibinfo {author} {\bibfnamefont
      {Z.}~\bibnamefont {Chen}},\ }\bibfield  {title} {\enquote {\bibinfo {title}
      {Photonic realization of a generic type of graphene edge states exhibiting
      topological flat band},}\ }\href@noop {} {\bibfield  {journal} {\bibinfo
      {journal} {Phys. Rev. Lett.}\ }\textbf {\bibinfo {volume} {131}},\ \bibinfo
      {pages} {013804} (\bibinfo {year} {2023})}\BibitemShut {NoStop}%
    \bibitem [{\citenamefont {Xi}\ \emph {et~al.}(2021)\citenamefont {Xi},
      \citenamefont {Ma}, \citenamefont {Wan}, \citenamefont {Dong},\ and\
      \citenamefont {Sun}}]{xi2021observation}%
      \BibitemOpen
      \bibfield  {author} {\bibinfo {author} {\bibfnamefont {X.}~\bibnamefont
      {Xi}}, \bibinfo {author} {\bibfnamefont {J.}~\bibnamefont {Ma}}, \bibinfo
      {author} {\bibfnamefont {S.}~\bibnamefont {Wan}}, \bibinfo {author}
      {\bibfnamefont {C.-H.}\ \bibnamefont {Dong}}, \ and\ \bibinfo {author}
      {\bibfnamefont {X.}~\bibnamefont {Sun}},\ }\bibfield  {title} {\enquote
      {\bibinfo {title} {Observation of chiral edge states in gapped nanomechanical
      graphene},}\ }\href@noop {} {\bibfield  {journal} {\bibinfo  {journal} {Sci.
      Adv.}\ }\textbf {\bibinfo {volume} {7}},\ \bibinfo {pages} {eabe1398}
      (\bibinfo {year} {2021})}\BibitemShut {NoStop}%
    \bibitem [{\citenamefont {Wang}\ \emph
      {et~al.}(2021{\natexlab{a}})\citenamefont {Wang}, \citenamefont {Zhang},
      \citenamefont {Gu}, \citenamefont {Liao}, \citenamefont {Cheng},\ and\
      \citenamefont {Liu}}]{wang2021zak}%
      \BibitemOpen
      \bibfield  {author} {\bibinfo {author} {\bibfnamefont {G.}~\bibnamefont
      {Wang}}, \bibinfo {author} {\bibfnamefont {Z.}~\bibnamefont {Zhang}},
      \bibinfo {author} {\bibfnamefont {Y.}~\bibnamefont {Gu}}, \bibinfo {author}
      {\bibfnamefont {D.}~\bibnamefont {Liao}}, \bibinfo {author} {\bibfnamefont
      {Y.}~\bibnamefont {Cheng}}, \ and\ \bibinfo {author} {\bibfnamefont
      {X.}~\bibnamefont {Liu}},\ }\bibfield  {title} {\enquote {\bibinfo {title}
      {{Z}ak-phase-inspired acoustic topological edge states on the honeycomb
      lattice},}\ }\href@noop {} {\bibfield  {journal} {\bibinfo  {journal} {Phys.
      Rev. B}\ }\textbf {\bibinfo {volume} {103}},\ \bibinfo {pages} {094102}
      (\bibinfo {year} {2021}{\natexlab{a}})}\BibitemShut {NoStop}%
    \bibitem [{\citenamefont {Han}\ \emph {et~al.}(2009)\citenamefont {Han},
      \citenamefont {Lai}, \citenamefont {Zi}, \citenamefont {Zhang},\ and\
      \citenamefont {Chan}}]{han2009dirac}%
      \BibitemOpen
      \bibfield  {author} {\bibinfo {author} {\bibfnamefont {D.}~\bibnamefont
      {Han}}, \bibinfo {author} {\bibfnamefont {Y.}~\bibnamefont {Lai}}, \bibinfo
      {author} {\bibfnamefont {J.}~\bibnamefont {Zi}}, \bibinfo {author}
      {\bibfnamefont {Z.-Q.}\ \bibnamefont {Zhang}}, \ and\ \bibinfo {author}
      {\bibfnamefont {C.~T.}\ \bibnamefont {Chan}},\ }\bibfield  {title} {\enquote
      {\bibinfo {title} {{D}irac spectra and edge states in honeycomb plasmonic
      lattices},}\ }\href@noop {} {\bibfield  {journal} {\bibinfo  {journal} {Phys.
      Rev. Lett.}\ }\textbf {\bibinfo {volume} {102}},\ \bibinfo {pages} {123904}
      (\bibinfo {year} {2009})}\BibitemShut {NoStop}%
    \bibitem [{\citenamefont {Wang}\ \emph
      {et~al.}(2016{\natexlab{b}})\citenamefont {Wang}, \citenamefont {Zhang},
      \citenamefont {Xiao}, \citenamefont {Han}, \citenamefont {Chan},\ and\
      \citenamefont {Wen}}]{wang2016existence}%
      \BibitemOpen
      \bibfield  {author} {\bibinfo {author} {\bibfnamefont {L.}~\bibnamefont
      {Wang}}, \bibinfo {author} {\bibfnamefont {R.-Y.}\ \bibnamefont {Zhang}},
      \bibinfo {author} {\bibfnamefont {M.}~\bibnamefont {Xiao}}, \bibinfo {author}
      {\bibfnamefont {D.}~\bibnamefont {Han}}, \bibinfo {author} {\bibfnamefont
      {C.~T.}\ \bibnamefont {Chan}}, \ and\ \bibinfo {author} {\bibfnamefont
      {W.}~\bibnamefont {Wen}},\ }\bibfield  {title} {\enquote {\bibinfo {title}
      {The existence of topological edge states in honeycomb plasmonic lattices},}\
      }\href@noop {} {\bibfield  {journal} {\bibinfo  {journal} {New J. Phys.}\
      }\textbf {\bibinfo {volume} {18}},\ \bibinfo {pages} {103029} (\bibinfo
      {year} {2016}{\natexlab{b}})}\BibitemShut {NoStop}%
    \bibitem [{\citenamefont {Jacqmin}\ \emph {et~al.}(2014)\citenamefont
      {Jacqmin}, \citenamefont {Carusotto}, \citenamefont {Sagnes}, \citenamefont
      {Abbarchi}, \citenamefont {Solnyshkov}, \citenamefont {Malpuech},
      \citenamefont {Galopin}, \citenamefont {Lema{\^\i}tre}, \citenamefont
      {Bloch},\ and\ \citenamefont {Amo}}]{jacqmin2014direct}%
      \BibitemOpen
      \bibfield  {author} {\bibinfo {author} {\bibfnamefont {T.}~\bibnamefont
      {Jacqmin}}, \bibinfo {author} {\bibfnamefont {I.}~\bibnamefont {Carusotto}},
      \bibinfo {author} {\bibfnamefont {I.}~\bibnamefont {Sagnes}}, \bibinfo
      {author} {\bibfnamefont {M.}~\bibnamefont {Abbarchi}}, \bibinfo {author}
      {\bibfnamefont {D.}~\bibnamefont {Solnyshkov}}, \bibinfo {author}
      {\bibfnamefont {G.}~\bibnamefont {Malpuech}}, \bibinfo {author}
      {\bibfnamefont {E.}~\bibnamefont {Galopin}}, \bibinfo {author} {\bibfnamefont
      {A.}~\bibnamefont {Lema{\^\i}tre}}, \bibinfo {author} {\bibfnamefont
      {J.}~\bibnamefont {Bloch}}, \ and\ \bibinfo {author} {\bibfnamefont
      {A.}~\bibnamefont {Amo}},\ }\bibfield  {title} {\enquote {\bibinfo {title}
      {Direct observation of {D}irac cones and a flatband in a honeycomb lattice
      for polaritons},}\ }\href@noop {} {\bibfield  {journal} {\bibinfo  {journal}
      {Phys. Rev. Lett.}\ }\textbf {\bibinfo {volume} {112}},\ \bibinfo {pages}
      {116402} (\bibinfo {year} {2014})}\BibitemShut {NoStop}%
    \bibitem [{\citenamefont {St-Jean}\ \emph {et~al.}(2021)\citenamefont
      {St-Jean}, \citenamefont {Dauphin}, \citenamefont {Massignan}, \citenamefont
      {Real}, \citenamefont {Jamadi}, \citenamefont {Milicevic}, \citenamefont
      {Lemaitre}, \citenamefont {Harouri}, \citenamefont {Le~Gratiet},
      \citenamefont {Sagnes} \emph {et~al.}}]{st2021measuring}%
      \BibitemOpen
      \bibfield  {author} {\bibinfo {author} {\bibfnamefont {P.}~\bibnamefont
      {St-Jean}}, \bibinfo {author} {\bibfnamefont {A.}~\bibnamefont {Dauphin}},
      \bibinfo {author} {\bibfnamefont {P.}~\bibnamefont {Massignan}}, \bibinfo
      {author} {\bibfnamefont {B.}~\bibnamefont {Real}}, \bibinfo {author}
      {\bibfnamefont {O.}~\bibnamefont {Jamadi}}, \bibinfo {author} {\bibfnamefont
      {M.}~\bibnamefont {Milicevic}}, \bibinfo {author} {\bibfnamefont
      {A.}~\bibnamefont {Lemaitre}}, \bibinfo {author} {\bibfnamefont
      {A.}~\bibnamefont {Harouri}}, \bibinfo {author} {\bibfnamefont
      {L.}~\bibnamefont {Le~Gratiet}}, \bibinfo {author} {\bibfnamefont
      {I.}~\bibnamefont {Sagnes}},  \emph {et~al.},\ }\bibfield  {title} {\enquote
      {\bibinfo {title} {Measuring topological invariants in a polaritonic analog
      of graphene},}\ }\href@noop {} {\bibfield  {journal} {\bibinfo  {journal}
      {Phys. Rev. Lett.}\ }\textbf {\bibinfo {volume} {126}},\ \bibinfo {pages}
      {127403} (\bibinfo {year} {2021})}\BibitemShut {NoStop}%
    \bibitem [{\citenamefont {Raoux}\ \emph {et~al.}(2014)\citenamefont {Raoux},
      \citenamefont {Morigi}, \citenamefont {Fuchs}, \citenamefont {Pi{\'e}chon},\
      and\ \citenamefont {Montambaux}}]{raoux2014dia}%
      \BibitemOpen
      \bibfield  {author} {\bibinfo {author} {\bibfnamefont {A.}~\bibnamefont
      {Raoux}}, \bibinfo {author} {\bibfnamefont {M.}~\bibnamefont {Morigi}},
      \bibinfo {author} {\bibfnamefont {J.-N.}\ \bibnamefont {Fuchs}}, \bibinfo
      {author} {\bibfnamefont {F.}~\bibnamefont {Pi{\'e}chon}}, \ and\ \bibinfo
      {author} {\bibfnamefont {G.}~\bibnamefont {Montambaux}},\ }\bibfield  {title}
      {\enquote {\bibinfo {title} {From dia-to paramagnetic orbital susceptibility
      of massless fermions},}\ }\href@noop {} {\bibfield  {journal} {\bibinfo
      {journal} {Phys. Rev. Lett.}\ }\textbf {\bibinfo {volume} {112}},\ \bibinfo
      {pages} {026402} (\bibinfo {year} {2014})}\BibitemShut {NoStop}%
    \bibitem [{\citenamefont {Neto}\ \emph {et~al.}(2009)\citenamefont {Neto},
      \citenamefont {Guinea}, \citenamefont {Peres}, \citenamefont {Novoselov},\
      and\ \citenamefont {Geim}}]{neto2009electronic}%
      \BibitemOpen
      \bibfield  {author} {\bibinfo {author} {\bibfnamefont {A.~C.}\ \bibnamefont
      {Neto}}, \bibinfo {author} {\bibfnamefont {F.}~\bibnamefont {Guinea}},
      \bibinfo {author} {\bibfnamefont {N.~M.}\ \bibnamefont {Peres}}, \bibinfo
      {author} {\bibfnamefont {K.~S.}\ \bibnamefont {Novoselov}}, \ and\ \bibinfo
      {author} {\bibfnamefont {A.~K.}\ \bibnamefont {Geim}},\ }\bibfield  {title}
      {\enquote {\bibinfo {title} {The electronic properties of graphene},}\
      }\href@noop {} {\bibfield  {journal} {\bibinfo  {journal} {Rev. Mod. Phys.}\
      }\textbf {\bibinfo {volume} {81}},\ \bibinfo {pages} {109} (\bibinfo {year}
      {2009})}\BibitemShut {NoStop}%
    \bibitem [{\citenamefont {Bercioux}\ \emph {et~al.}(2009)\citenamefont
      {Bercioux}, \citenamefont {Urban}, \citenamefont {Grabert},\ and\
      \citenamefont {H{\"a}usler}}]{bercioux2009massless}%
      \BibitemOpen
      \bibfield  {author} {\bibinfo {author} {\bibfnamefont {D.}~\bibnamefont
      {Bercioux}}, \bibinfo {author} {\bibfnamefont {D.}~\bibnamefont {Urban}},
      \bibinfo {author} {\bibfnamefont {H.}~\bibnamefont {Grabert}}, \ and\
      \bibinfo {author} {\bibfnamefont {W.}~\bibnamefont {H{\"a}usler}},\
      }\bibfield  {title} {\enquote {\bibinfo {title} {Massless {D}irac-{W}eyl
      fermions in $\alpha$-${T}_3$ optical lattice},}\ }\href@noop {} {\bibfield
      {journal} {\bibinfo  {journal} {Phys. Rev. A}\ }\textbf {\bibinfo {volume}
      {80}},\ \bibinfo {pages} {063603} (\bibinfo {year} {2009})}\BibitemShut
      {NoStop}%
    \bibitem [{\citenamefont {D{\'o}ra}\ \emph {et~al.}(2011)\citenamefont
      {D{\'o}ra}, \citenamefont {Kailasvuori},\ and\ \citenamefont
      {Moessner}}]{dora2011lattice}%
      \BibitemOpen
      \bibfield  {author} {\bibinfo {author} {\bibfnamefont {B.}~\bibnamefont
      {D{\'o}ra}}, \bibinfo {author} {\bibfnamefont {J.}~\bibnamefont
      {Kailasvuori}}, \ and\ \bibinfo {author} {\bibfnamefont {R.}~\bibnamefont
      {Moessner}},\ }\bibfield  {title} {\enquote {\bibinfo {title} {Lattice
      generalization of the {D}irac equation to general spin and the role of the
      flat band},}\ }\href@noop {} {\bibfield  {journal} {\bibinfo  {journal}
      {Phys. Rev. B}\ }\textbf {\bibinfo {volume} {84}},\ \bibinfo {pages} {195422}
      (\bibinfo {year} {2011})}\BibitemShut {NoStop}%
    \bibitem [{\citenamefont {Urban}\ \emph {et~al.}(2011)\citenamefont {Urban},
      \citenamefont {Bercioux}, \citenamefont {Wimmer},\ and\ \citenamefont
      {H{\"a}usler}}]{urban2011barrier}%
      \BibitemOpen
      \bibfield  {author} {\bibinfo {author} {\bibfnamefont {D.~F.}\ \bibnamefont
      {Urban}}, \bibinfo {author} {\bibfnamefont {D.}~\bibnamefont {Bercioux}},
      \bibinfo {author} {\bibfnamefont {M.}~\bibnamefont {Wimmer}}, \ and\ \bibinfo
      {author} {\bibfnamefont {W.}~\bibnamefont {H{\"a}usler}},\ }\bibfield
      {title} {\enquote {\bibinfo {title} {Barrier transmission of {D}irac-like
      pseudospin-one particles},}\ }\href@noop {} {\bibfield  {journal} {\bibinfo
      {journal} {Phys. Rev. B}\ }\textbf {\bibinfo {volume} {84}},\ \bibinfo
      {pages} {115136} (\bibinfo {year} {2011})}\BibitemShut {NoStop}%
    \bibitem [{\citenamefont {Fang}\ \emph {et~al.}(2016)\citenamefont {Fang},
      \citenamefont {Zhang}, \citenamefont {Louie},\ and\ \citenamefont
      {Chan}}]{fang2016klein}%
      \BibitemOpen
      \bibfield  {author} {\bibinfo {author} {\bibfnamefont {A.}~\bibnamefont
      {Fang}}, \bibinfo {author} {\bibfnamefont {Z.}~\bibnamefont {Zhang}},
      \bibinfo {author} {\bibfnamefont {S.~G.}\ \bibnamefont {Louie}}, \ and\
      \bibinfo {author} {\bibfnamefont {C.~T.}\ \bibnamefont {Chan}},\ }\bibfield
      {title} {\enquote {\bibinfo {title} {{K}lein tunneling and supercollimation
      of pseudospin-1 electromagnetic waves},}\ }\href@noop {} {\bibfield
      {journal} {\bibinfo  {journal} {Phys. Rev. B}\ }\textbf {\bibinfo {volume}
      {93}},\ \bibinfo {pages} {035422} (\bibinfo {year} {2016})}\BibitemShut
      {NoStop}%
    \bibitem [{\citenamefont {Illes}\ and\ \citenamefont
      {Nicol}(2017)}]{illes2017klein}%
      \BibitemOpen
      \bibfield  {author} {\bibinfo {author} {\bibfnamefont {E.}~\bibnamefont
      {Illes}}\ and\ \bibinfo {author} {\bibfnamefont {E.}~\bibnamefont {Nicol}},\
      }\bibfield  {title} {\enquote {\bibinfo {title} {{K}lein tunneling in the
      $\alpha$-${T}_3$ model},}\ }\href@noop {} {\bibfield  {journal} {\bibinfo
      {journal} {Phys. Rev. B}\ }\textbf {\bibinfo {volume} {95}},\ \bibinfo
      {pages} {235432} (\bibinfo {year} {2017})}\BibitemShut {NoStop}%
    \bibitem [{\citenamefont {Betancur-Ocampo}\ \emph {et~al.}(2017)\citenamefont
      {Betancur-Ocampo}, \citenamefont {Cordourier-Maruri}, \citenamefont {Gupta},\
      and\ \citenamefont {de~Coss}}]{betancur2017super}%
      \BibitemOpen
      \bibfield  {author} {\bibinfo {author} {\bibfnamefont {Y.}~\bibnamefont
      {Betancur-Ocampo}}, \bibinfo {author} {\bibfnamefont {G.}~\bibnamefont
      {Cordourier-Maruri}}, \bibinfo {author} {\bibfnamefont {V.}~\bibnamefont
      {Gupta}}, \ and\ \bibinfo {author} {\bibfnamefont {R.}~\bibnamefont
      {de~Coss}},\ }\bibfield  {title} {\enquote {\bibinfo {title} {Super-{K}lein
      tunneling of massive pseudospin-one particles},}\ }\href@noop {} {\bibfield
      {journal} {\bibinfo  {journal} {Phys. Rev. B}\ }\textbf {\bibinfo {volume}
      {96}},\ \bibinfo {pages} {024304} (\bibinfo {year} {2017})}\BibitemShut
      {NoStop}%
    \bibitem [{\citenamefont {Weekes}\ \emph {et~al.}(2021)\citenamefont {Weekes},
      \citenamefont {Iurov}, \citenamefont {Zhemchuzhna}, \citenamefont {Gumbs},\
      and\ \citenamefont {Huang}}]{weekes2021generalized}%
      \BibitemOpen
      \bibfield  {author} {\bibinfo {author} {\bibfnamefont {N.}~\bibnamefont
      {Weekes}}, \bibinfo {author} {\bibfnamefont {A.}~\bibnamefont {Iurov}},
      \bibinfo {author} {\bibfnamefont {L.}~\bibnamefont {Zhemchuzhna}}, \bibinfo
      {author} {\bibfnamefont {G.}~\bibnamefont {Gumbs}}, \ and\ \bibinfo {author}
      {\bibfnamefont {D.}~\bibnamefont {Huang}},\ }\bibfield  {title} {\enquote
      {\bibinfo {title} {Generalized wkb theory for electron tunneling in gapped
      $\alpha$-${T}_3$ lattices},}\ }\href@noop {} {\bibfield  {journal} {\bibinfo
      {journal} {Phys. Rev. B}\ }\textbf {\bibinfo {volume} {103}},\ \bibinfo
      {pages} {165429} (\bibinfo {year} {2021})}\BibitemShut {NoStop}%
    \bibitem [{\citenamefont {Cunha}\ \emph {et~al.}(2022)\citenamefont {Cunha},
      \citenamefont {da~Costa}, \citenamefont {Pereira~Jr}, \citenamefont
      {Costa~Filho}, \citenamefont {Van~Duppen},\ and\ \citenamefont
      {Peeters}}]{cunha2022tunneling}%
      \BibitemOpen
      \bibfield  {author} {\bibinfo {author} {\bibfnamefont {S.}~\bibnamefont
      {Cunha}}, \bibinfo {author} {\bibfnamefont {D.}~\bibnamefont {da~Costa}},
      \bibinfo {author} {\bibfnamefont {J.~M.}\ \bibnamefont {Pereira~Jr}},
      \bibinfo {author} {\bibfnamefont {R.}~\bibnamefont {Costa~Filho}}, \bibinfo
      {author} {\bibfnamefont {B.}~\bibnamefont {Van~Duppen}}, \ and\ \bibinfo
      {author} {\bibfnamefont {F.}~\bibnamefont {Peeters}},\ }\bibfield  {title}
      {\enquote {\bibinfo {title} {Tunneling properties in $\alpha$-${T}_3$
      lattices: Effects of symmetry-breaking terms},}\ }\href@noop {} {\bibfield
      {journal} {\bibinfo  {journal} {Phys. Rev. B}\ }\textbf {\bibinfo {volume}
      {105}},\ \bibinfo {pages} {165402} (\bibinfo {year} {2022})}\BibitemShut
      {NoStop}%
    \bibitem [{\citenamefont {Illes}\ \emph {et~al.}(2015)\citenamefont {Illes},
      \citenamefont {Carbotte},\ and\ \citenamefont {Nicol}}]{illes2015hall}%
      \BibitemOpen
      \bibfield  {author} {\bibinfo {author} {\bibfnamefont {E.}~\bibnamefont
      {Illes}}, \bibinfo {author} {\bibfnamefont {J.}~\bibnamefont {Carbotte}}, \
      and\ \bibinfo {author} {\bibfnamefont {E.}~\bibnamefont {Nicol}},\ }\bibfield
       {title} {\enquote {\bibinfo {title} {{H}all quantization and optical
      conductivity evolution with variable {B}erry phase in the $\alpha$-${T}_3$
      model},}\ }\href@noop {} {\bibfield  {journal} {\bibinfo  {journal} {Phys.
      Rev. B}\ }\textbf {\bibinfo {volume} {92}},\ \bibinfo {pages} {245410}
      (\bibinfo {year} {2015})}\BibitemShut {NoStop}%
    \bibitem [{\citenamefont {Malcolm}\ and\ \citenamefont
      {Nicol}(2016)}]{malcolm2016frequency}%
      \BibitemOpen
      \bibfield  {author} {\bibinfo {author} {\bibfnamefont {J.}~\bibnamefont
      {Malcolm}}\ and\ \bibinfo {author} {\bibfnamefont {E.}~\bibnamefont
      {Nicol}},\ }\bibfield  {title} {\enquote {\bibinfo {title}
      {Frequency-dependent polarizability, plasmons, and screening in the
      two-dimensional pseudospin-1 dice lattice},}\ }\href@noop {} {\bibfield
      {journal} {\bibinfo  {journal} {Phys. Rev. B}\ }\textbf {\bibinfo {volume}
      {93}},\ \bibinfo {pages} {165433} (\bibinfo {year} {2016})}\BibitemShut
      {NoStop}%
    \bibitem [{\citenamefont {Carbotte}\ \emph {et~al.}(2019)\citenamefont
      {Carbotte}, \citenamefont {Bryenton},\ and\ \citenamefont
      {Nicol}}]{carbotte2019optical}%
      \BibitemOpen
      \bibfield  {author} {\bibinfo {author} {\bibfnamefont {J.}~\bibnamefont
      {Carbotte}}, \bibinfo {author} {\bibfnamefont {K.}~\bibnamefont {Bryenton}},
      \ and\ \bibinfo {author} {\bibfnamefont {E.}~\bibnamefont {Nicol}},\
      }\bibfield  {title} {\enquote {\bibinfo {title} {Optical properties of a
      semi-{D}irac material},}\ }\href@noop {} {\bibfield  {journal} {\bibinfo
      {journal} {Phys. Rev. B}\ }\textbf {\bibinfo {volume} {99}},\ \bibinfo
      {pages} {115406} (\bibinfo {year} {2019})}\BibitemShut {NoStop}%
    \bibitem [{\citenamefont {Mojarro}\ \emph {et~al.}(2020)\citenamefont
      {Mojarro}, \citenamefont {Ibarra-Sierra}, \citenamefont {Sandoval-Santana},
      \citenamefont {Carrillo-Bastos},\ and\ \citenamefont
      {Naumis}}]{mojarro2020electron}%
      \BibitemOpen
      \bibfield  {author} {\bibinfo {author} {\bibfnamefont {M.}~\bibnamefont
      {Mojarro}}, \bibinfo {author} {\bibfnamefont {V.}~\bibnamefont
      {Ibarra-Sierra}}, \bibinfo {author} {\bibfnamefont {J.}~\bibnamefont
      {Sandoval-Santana}}, \bibinfo {author} {\bibfnamefont {R.}~\bibnamefont
      {Carrillo-Bastos}}, \ and\ \bibinfo {author} {\bibfnamefont {G.~G.}\
      \bibnamefont {Naumis}},\ }\bibfield  {title} {\enquote {\bibinfo {title}
      {Electron transitions for {D}irac {H}amiltonians with flat bands under
      electromagnetic radiation: Application to the $\alpha$-${T}_3$ graphene
      model},}\ }\href@noop {} {\bibfield  {journal} {\bibinfo  {journal} {Phys.
      Rev. B}\ }\textbf {\bibinfo {volume} {101}},\ \bibinfo {pages} {165305}
      (\bibinfo {year} {2020})}\BibitemShut {NoStop}%
    \bibitem [{\citenamefont {Han}\ and\ \citenamefont
      {Lai}(2022)}]{han2022optical}%
      \BibitemOpen
      \bibfield  {author} {\bibinfo {author} {\bibfnamefont {C.-D.}\ \bibnamefont
      {Han}}\ and\ \bibinfo {author} {\bibfnamefont {Y.-C.}\ \bibnamefont {Lai}},\
      }\bibfield  {title} {\enquote {\bibinfo {title} {Optical response of
      two-dimensional {D}irac materials with a flat band},}\ }\href@noop {}
      {\bibfield  {journal} {\bibinfo  {journal} {Phys. Rev. B}\ }\textbf {\bibinfo
      {volume} {105}},\ \bibinfo {pages} {155405} (\bibinfo {year}
      {2022})}\BibitemShut {NoStop}%
    \bibitem [{\citenamefont {Iurov}\ \emph {et~al.}(2022)\citenamefont {Iurov},
      \citenamefont {Zhemchuzhna}, \citenamefont {Gumbs}, \citenamefont {Huang},
      \citenamefont {Dahal},\ and\ \citenamefont {Abranyos}}]{iurov2022finite}%
      \BibitemOpen
      \bibfield  {author} {\bibinfo {author} {\bibfnamefont {A.}~\bibnamefont
      {Iurov}}, \bibinfo {author} {\bibfnamefont {L.}~\bibnamefont {Zhemchuzhna}},
      \bibinfo {author} {\bibfnamefont {G.}~\bibnamefont {Gumbs}}, \bibinfo
      {author} {\bibfnamefont {D.}~\bibnamefont {Huang}}, \bibinfo {author}
      {\bibfnamefont {D.}~\bibnamefont {Dahal}}, \ and\ \bibinfo {author}
      {\bibfnamefont {Y.}~\bibnamefont {Abranyos}},\ }\bibfield  {title} {\enquote
      {\bibinfo {title} {Finite-temperature plasmons, damping, and collective
      behavior in the $\alpha$-${T}_3$ model},}\ }\href@noop {} {\bibfield
      {journal} {\bibinfo  {journal} {Phys. Rev. B}\ }\textbf {\bibinfo {volume}
      {105}},\ \bibinfo {pages} {245414} (\bibinfo {year} {2022})}\BibitemShut
      {NoStop}%
    \bibitem [{\citenamefont {Oriekhov}\ and\ \citenamefont
      {Gusynin}(2022)}]{oriekhov2022optical}%
      \BibitemOpen
      \bibfield  {author} {\bibinfo {author} {\bibfnamefont {D.}~\bibnamefont
      {Oriekhov}}\ and\ \bibinfo {author} {\bibfnamefont {V.}~\bibnamefont
      {Gusynin}},\ }\bibfield  {title} {\enquote {\bibinfo {title} {Optical
      conductivity of semi-{D}irac and pseudospin-1 models: Zitterbewegung
      approach},}\ }\href@noop {} {\bibfield  {journal} {\bibinfo  {journal} {Phys.
      Rev. B}\ }\textbf {\bibinfo {volume} {106}},\ \bibinfo {pages} {115143}
      (\bibinfo {year} {2022})}\BibitemShut {NoStop}%
    \bibitem [{\citenamefont {Iurov}\ \emph {et~al.}(2021)\citenamefont {Iurov},
      \citenamefont {Zhemchuzhna}, \citenamefont {Gumbs}, \citenamefont {Huang},
      \citenamefont {Fekete}, \citenamefont {Anwar}, \citenamefont {Dahal},\ and\
      \citenamefont {Weekes}}]{iurov2021tailoring}%
      \BibitemOpen
      \bibfield  {author} {\bibinfo {author} {\bibfnamefont {A.}~\bibnamefont
      {Iurov}}, \bibinfo {author} {\bibfnamefont {L.}~\bibnamefont {Zhemchuzhna}},
      \bibinfo {author} {\bibfnamefont {G.}~\bibnamefont {Gumbs}}, \bibinfo
      {author} {\bibfnamefont {D.}~\bibnamefont {Huang}}, \bibinfo {author}
      {\bibfnamefont {P.}~\bibnamefont {Fekete}}, \bibinfo {author} {\bibfnamefont
      {F.}~\bibnamefont {Anwar}}, \bibinfo {author} {\bibfnamefont
      {D.}~\bibnamefont {Dahal}}, \ and\ \bibinfo {author} {\bibfnamefont
      {N.}~\bibnamefont {Weekes}},\ }\bibfield  {title} {\enquote {\bibinfo {title}
      {Tailoring plasmon excitations in $\alpha$-${T}_3$ armchair nanoribbons},}\
      }\href@noop {} {\bibfield  {journal} {\bibinfo  {journal} {Sci. Rep.}\
      }\textbf {\bibinfo {volume} {11}},\ \bibinfo {pages} {20577} (\bibinfo {year}
      {2021})}\BibitemShut {NoStop}%
    \bibitem [{\citenamefont {Illes}\ and\ \citenamefont
      {Nicol}(2016)}]{illes2016magnetic}%
      \BibitemOpen
      \bibfield  {author} {\bibinfo {author} {\bibfnamefont {E.}~\bibnamefont
      {Illes}}\ and\ \bibinfo {author} {\bibfnamefont {E.}~\bibnamefont {Nicol}},\
      }\bibfield  {title} {\enquote {\bibinfo {title} {Magnetic properties of the
      $\alpha$-${T}_3$ model: Magneto-optical conductivity and the hofstadter
      butterfly},}\ }\href@noop {} {\bibfield  {journal} {\bibinfo  {journal}
      {Phys. Rev. B}\ }\textbf {\bibinfo {volume} {94}},\ \bibinfo {pages} {125435}
      (\bibinfo {year} {2016})}\BibitemShut {NoStop}%
    \bibitem [{\citenamefont {Soni}\ \emph {et~al.}(2020)\citenamefont {Soni},
      \citenamefont {Kaushal}, \citenamefont {Okamoto},\ and\ \citenamefont
      {Dagotto}}]{soni2020flat}%
      \BibitemOpen
      \bibfield  {author} {\bibinfo {author} {\bibfnamefont {R.}~\bibnamefont
      {Soni}}, \bibinfo {author} {\bibfnamefont {N.}~\bibnamefont {Kaushal}},
      \bibinfo {author} {\bibfnamefont {S.}~\bibnamefont {Okamoto}}, \ and\
      \bibinfo {author} {\bibfnamefont {E.}~\bibnamefont {Dagotto}},\ }\bibfield
      {title} {\enquote {\bibinfo {title} {Flat bands and ferrimagnetic order in
      electronically correlated dice-lattice ribbons},}\ }\href@noop {} {\bibfield
      {journal} {\bibinfo  {journal} {Phys. Rev. B}\ }\textbf {\bibinfo {volume}
      {102}},\ \bibinfo {pages} {045105} (\bibinfo {year} {2020})}\BibitemShut
      {NoStop}%
    \bibitem [{\citenamefont {Roslyak}\ \emph {et~al.}(2021)\citenamefont
      {Roslyak}, \citenamefont {Gumbs}, \citenamefont {Balassis},\ and\
      \citenamefont {Elsayed}}]{roslyak2021effect}%
      \BibitemOpen
      \bibfield  {author} {\bibinfo {author} {\bibfnamefont {O.}~\bibnamefont
      {Roslyak}}, \bibinfo {author} {\bibfnamefont {G.}~\bibnamefont {Gumbs}},
      \bibinfo {author} {\bibfnamefont {A.}~\bibnamefont {Balassis}}, \ and\
      \bibinfo {author} {\bibfnamefont {H.}~\bibnamefont {Elsayed}},\ }\bibfield
      {title} {\enquote {\bibinfo {title} {Effect of magnetic field and chemical
      potential on the rkky interaction in the $\alpha$-${T}_3$ lattice},}\
      }\href@noop {} {\bibfield  {journal} {\bibinfo  {journal} {Phys. Rev. B}\
      }\textbf {\bibinfo {volume} {103}},\ \bibinfo {pages} {075418} (\bibinfo
      {year} {2021})}\BibitemShut {NoStop}%
    \bibitem [{\citenamefont {Sun}\ \emph {et~al.}(2022)\citenamefont {Sun},
      \citenamefont {Liu}, \citenamefont {Du},\ and\ \citenamefont
      {Guo}}]{sun2022strain}%
      \BibitemOpen
      \bibfield  {author} {\bibinfo {author} {\bibfnamefont {J.}~\bibnamefont
      {Sun}}, \bibinfo {author} {\bibfnamefont {T.}~\bibnamefont {Liu}}, \bibinfo
      {author} {\bibfnamefont {Y.}~\bibnamefont {Du}}, \ and\ \bibinfo {author}
      {\bibfnamefont {H.}~\bibnamefont {Guo}},\ }\bibfield  {title} {\enquote
      {\bibinfo {title} {Strain-induced pseudo magnetic field in the
      $\alpha$-${T}_3$ lattice},}\ }\href@noop {} {\bibfield  {journal} {\bibinfo
      {journal} {Phys. Rev. B}\ }\textbf {\bibinfo {volume} {106}},\ \bibinfo
      {pages} {155417} (\bibinfo {year} {2022})}\BibitemShut {NoStop}%
    \bibitem [{\citenamefont {Filusch}\ and\ \citenamefont
      {Fehske}(2022)}]{filusch2022tunable}%
      \BibitemOpen
      \bibfield  {author} {\bibinfo {author} {\bibfnamefont {A.}~\bibnamefont
      {Filusch}}\ and\ \bibinfo {author} {\bibfnamefont {H.}~\bibnamefont
      {Fehske}},\ }\bibfield  {title} {\enquote {\bibinfo {title} {Tunable valley
      filtering in dynamically strained $\alpha$-${T}_3$ lattices},}\ }\href@noop
      {} {\bibfield  {journal} {\bibinfo  {journal} {Phys. Rev. B}\ }\textbf
      {\bibinfo {volume} {106}},\ \bibinfo {pages} {245106} (\bibinfo {year}
      {2022})}\BibitemShut {NoStop}%
    \bibitem [{\citenamefont {Li}\ \emph {et~al.}(2023)\citenamefont {Li},
      \citenamefont {Liu},\ and\ \citenamefont {Wang}}]{li2023topological}%
      \BibitemOpen
      \bibfield  {author} {\bibinfo {author} {\bibfnamefont {R.}~\bibnamefont
      {Li}}, \bibinfo {author} {\bibfnamefont {J.-F.}\ \bibnamefont {Liu}}, \ and\
      \bibinfo {author} {\bibfnamefont {J.}~\bibnamefont {Wang}},\ }\bibfield
      {title} {\enquote {\bibinfo {title} {Topological ac charge current and
      continuous invariant in the $\alpha$-${T}_3$ lattice under a periodically
      varying strain},}\ }\href@noop {} {\bibfield  {journal} {\bibinfo  {journal}
      {Phys. Rev. B}\ }\textbf {\bibinfo {volume} {108}},\ \bibinfo {pages}
      {115403} (\bibinfo {year} {2023})}\BibitemShut {NoStop}%
    \bibitem [{\citenamefont {Kov{\'a}cs}\ \emph {et~al.}(2017)\citenamefont
      {Kov{\'a}cs}, \citenamefont {D{\'a}vid}, \citenamefont {D{\'o}ra},\ and\
      \citenamefont {Cserti}}]{kovacs2017frequency}%
      \BibitemOpen
      \bibfield  {author} {\bibinfo {author} {\bibfnamefont {{\'A}.~D.}\
      \bibnamefont {Kov{\'a}cs}}, \bibinfo {author} {\bibfnamefont
      {G.}~\bibnamefont {D{\'a}vid}}, \bibinfo {author} {\bibfnamefont
      {B.}~\bibnamefont {D{\'o}ra}}, \ and\ \bibinfo {author} {\bibfnamefont
      {J.}~\bibnamefont {Cserti}},\ }\bibfield  {title} {\enquote {\bibinfo {title}
      {Frequency-dependent magneto-optical conductivity in the generalized
      $\alpha$-${T}_3$ model},}\ }\href@noop {} {\bibfield  {journal} {\bibinfo
      {journal} {Phys. Rev. B}\ }\textbf {\bibinfo {volume} {95}},\ \bibinfo
      {pages} {035414} (\bibinfo {year} {2017})}\BibitemShut {NoStop}%
    \bibitem [{\citenamefont {Chen}\ \emph
      {et~al.}(2019{\natexlab{a}})\citenamefont {Chen}, \citenamefont {Xu},
      \citenamefont {Wang}, \citenamefont {Liu},\ and\ \citenamefont
      {Ma}}]{chen2019enhanced}%
      \BibitemOpen
      \bibfield  {author} {\bibinfo {author} {\bibfnamefont {Y.-R.}\ \bibnamefont
      {Chen}}, \bibinfo {author} {\bibfnamefont {Y.}~\bibnamefont {Xu}}, \bibinfo
      {author} {\bibfnamefont {J.}~\bibnamefont {Wang}}, \bibinfo {author}
      {\bibfnamefont {J.-F.}\ \bibnamefont {Liu}}, \ and\ \bibinfo {author}
      {\bibfnamefont {Z.}~\bibnamefont {Ma}},\ }\bibfield  {title} {\enquote
      {\bibinfo {title} {Enhanced magneto-optical response due to the flat band in
      nanoribbons made from the $\alpha$-${T}_3$ lattice},}\ }\href@noop {}
      {\bibfield  {journal} {\bibinfo  {journal} {Phys. Rev. B}\ }\textbf {\bibinfo
      {volume} {99}},\ \bibinfo {pages} {045420} (\bibinfo {year}
      {2019}{\natexlab{a}})}\BibitemShut {NoStop}%
    \bibitem [{\citenamefont {Chen}\ \emph
      {et~al.}(2019{\natexlab{b}})\citenamefont {Chen}, \citenamefont {Zuber},
      \citenamefont {Ma},\ and\ \citenamefont {Zhang}}]{chen2019nonlinear}%
      \BibitemOpen
      \bibfield  {author} {\bibinfo {author} {\bibfnamefont {L.}~\bibnamefont
      {Chen}}, \bibinfo {author} {\bibfnamefont {J.}~\bibnamefont {Zuber}},
      \bibinfo {author} {\bibfnamefont {Z.}~\bibnamefont {Ma}}, \ and\ \bibinfo
      {author} {\bibfnamefont {C.}~\bibnamefont {Zhang}},\ }\bibfield  {title}
      {\enquote {\bibinfo {title} {Nonlinear optical response of the
      $\alpha$-${T}_3$ model due to the nontrivial topology of the band
      dispersion},}\ }\href@noop {} {\bibfield  {journal} {\bibinfo  {journal}
      {Phys. Rev. B}\ }\textbf {\bibinfo {volume} {100}},\ \bibinfo {pages}
      {035440} (\bibinfo {year} {2019}{\natexlab{b}})}\BibitemShut {NoStop}%
    \bibitem [{\citenamefont {Balassis}\ \emph {et~al.}(2020)\citenamefont
      {Balassis}, \citenamefont {Dahal}, \citenamefont {Gumbs}, \citenamefont
      {Iurov}, \citenamefont {Huang},\ and\ \citenamefont
      {Roslyak}}]{balassis2020magnetoplasmons}%
      \BibitemOpen
      \bibfield  {author} {\bibinfo {author} {\bibfnamefont {A.}~\bibnamefont
      {Balassis}}, \bibinfo {author} {\bibfnamefont {D.}~\bibnamefont {Dahal}},
      \bibinfo {author} {\bibfnamefont {G.}~\bibnamefont {Gumbs}}, \bibinfo
      {author} {\bibfnamefont {A.}~\bibnamefont {Iurov}}, \bibinfo {author}
      {\bibfnamefont {D.}~\bibnamefont {Huang}}, \ and\ \bibinfo {author}
      {\bibfnamefont {O.}~\bibnamefont {Roslyak}},\ }\bibfield  {title} {\enquote
      {\bibinfo {title} {Magnetoplasmons for the $\alpha$-${T}_3$ model with filled
      landau levels},}\ }\href@noop {} {\bibfield  {journal} {\bibinfo  {journal}
      {J. Phys.: Condens. Matter}\ }\textbf {\bibinfo {volume} {32}},\ \bibinfo
      {pages} {485301} (\bibinfo {year} {2020})}\BibitemShut {NoStop}%
    \bibitem [{\citenamefont {Vigh}\ \emph {et~al.}(2013)\citenamefont {Vigh},
      \citenamefont {Oroszl{\'a}ny}, \citenamefont {Vajna}, \citenamefont
      {San-Jose}, \citenamefont {D{\'a}vid}, \citenamefont {Cserti},\ and\
      \citenamefont {D{\'o}ra}}]{vigh2013diverging}%
      \BibitemOpen
      \bibfield  {author} {\bibinfo {author} {\bibfnamefont {M.}~\bibnamefont
      {Vigh}}, \bibinfo {author} {\bibfnamefont {L.}~\bibnamefont {Oroszl{\'a}ny}},
      \bibinfo {author} {\bibfnamefont {S.}~\bibnamefont {Vajna}}, \bibinfo
      {author} {\bibfnamefont {P.}~\bibnamefont {San-Jose}}, \bibinfo {author}
      {\bibfnamefont {G.}~\bibnamefont {D{\'a}vid}}, \bibinfo {author}
      {\bibfnamefont {J.}~\bibnamefont {Cserti}}, \ and\ \bibinfo {author}
      {\bibfnamefont {B.}~\bibnamefont {D{\'o}ra}},\ }\bibfield  {title} {\enquote
      {\bibinfo {title} {Diverging dc conductivity due to a flat band in a
      disordered system of pseudospin-1 {D}irac-{W}eyl fermions},}\ }\href@noop {}
      {\bibfield  {journal} {\bibinfo  {journal} {Phys. Rev. B}\ }\textbf {\bibinfo
      {volume} {88}},\ \bibinfo {pages} {161413} (\bibinfo {year}
      {2013})}\BibitemShut {NoStop}%
    \bibitem [{\citenamefont {Louvet}\ \emph {et~al.}(2015)\citenamefont {Louvet},
      \citenamefont {Delplace}, \citenamefont {Fedorenko},\ and\ \citenamefont
      {Carpentier}}]{louvet2015origin}%
      \BibitemOpen
      \bibfield  {author} {\bibinfo {author} {\bibfnamefont {T.}~\bibnamefont
      {Louvet}}, \bibinfo {author} {\bibfnamefont {P.}~\bibnamefont {Delplace}},
      \bibinfo {author} {\bibfnamefont {A.~A.}\ \bibnamefont {Fedorenko}}, \ and\
      \bibinfo {author} {\bibfnamefont {D.}~\bibnamefont {Carpentier}},\ }\bibfield
       {title} {\enquote {\bibinfo {title} {On the origin of minimal conductivity
      at a band crossing},}\ }\href@noop {} {\bibfield  {journal} {\bibinfo
      {journal} {Phys. Rev. B}\ }\textbf {\bibinfo {volume} {92}},\ \bibinfo
      {pages} {155116} (\bibinfo {year} {2015})}\BibitemShut {NoStop}%
    \bibitem [{\citenamefont {Wang}\ \emph {et~al.}(2020)\citenamefont {Wang},
      \citenamefont {Liu}, \citenamefont {Wang},\ and\ \citenamefont
      {Liu}}]{wang2020integer}%
      \BibitemOpen
      \bibfield  {author} {\bibinfo {author} {\bibfnamefont {J.~J.}\ \bibnamefont
      {Wang}}, \bibinfo {author} {\bibfnamefont {S.}~\bibnamefont {Liu}}, \bibinfo
      {author} {\bibfnamefont {J.}~\bibnamefont {Wang}}, \ and\ \bibinfo {author}
      {\bibfnamefont {J.-F.}\ \bibnamefont {Liu}},\ }\bibfield  {title} {\enquote
      {\bibinfo {title} {Integer quantum {H}all effect of the $\alpha$-${T}_3$
      model with a broken flat band},}\ }\href@noop {} {\bibfield  {journal}
      {\bibinfo  {journal} {Phys. Rev. B}\ }\textbf {\bibinfo {volume} {102}},\
      \bibinfo {pages} {235414} (\bibinfo {year} {2020})}\BibitemShut {NoStop}%
    \bibitem [{\citenamefont {Wang}\ \emph
      {et~al.}(2021{\natexlab{b}})\citenamefont {Wang}, \citenamefont {Wang},
      \citenamefont {Wang},\ and\ \citenamefont {Liu}}]{wang2021flat}%
      \BibitemOpen
      \bibfield  {author} {\bibinfo {author} {\bibfnamefont {X.-H.}\ \bibnamefont
      {Wang}}, \bibinfo {author} {\bibfnamefont {J.~J.}\ \bibnamefont {Wang}},
      \bibinfo {author} {\bibfnamefont {J.}~\bibnamefont {Wang}}, \ and\ \bibinfo
      {author} {\bibfnamefont {J.-F.}\ \bibnamefont {Liu}},\ }\bibfield  {title}
      {\enquote {\bibinfo {title} {Flat band assisted topological charge pump in
      the dice lattice},}\ }\href@noop {} {\bibfield  {journal} {\bibinfo
      {journal} {Phys. Rev. B}\ }\textbf {\bibinfo {volume} {103}},\ \bibinfo
      {pages} {195442} (\bibinfo {year} {2021}{\natexlab{b}})}\BibitemShut
      {NoStop}%
    \bibitem [{\citenamefont {Zhou}(2021)}]{zhou2021andreev}%
      \BibitemOpen
      \bibfield  {author} {\bibinfo {author} {\bibfnamefont {X.}~\bibnamefont
      {Zhou}},\ }\bibfield  {title} {\enquote {\bibinfo {title} {Andreev reflection
      and {J}osephson effect in the $\alpha$-${T}_3$ lattice},}\ }\href@noop {}
      {\bibfield  {journal} {\bibinfo  {journal} {Phys. Rev. B}\ }\textbf {\bibinfo
      {volume} {104}},\ \bibinfo {pages} {125441} (\bibinfo {year}
      {2021})}\BibitemShut {NoStop}%
    \bibitem [{\citenamefont {Biswas}\ and\ \citenamefont
      {Ghosh}(2016)}]{biswas2016magnetotransport}%
      \BibitemOpen
      \bibfield  {author} {\bibinfo {author} {\bibfnamefont {T.}~\bibnamefont
      {Biswas}}\ and\ \bibinfo {author} {\bibfnamefont {T.~K.}\ \bibnamefont
      {Ghosh}},\ }\bibfield  {title} {\enquote {\bibinfo {title} {Magnetotransport
      properties of the $\alpha$-${T}_3$ model},}\ }\href@noop {} {\bibfield
      {journal} {\bibinfo  {journal} {J. Phys.: Condens. Matter}\ }\textbf
      {\bibinfo {volume} {28}},\ \bibinfo {pages} {495302} (\bibinfo {year}
      {2016})}\BibitemShut {NoStop}%
    \bibitem [{\citenamefont {Islam}\ and\ \citenamefont
      {Dutta}(2017)}]{islam2017valley}%
      \BibitemOpen
      \bibfield  {author} {\bibinfo {author} {\bibfnamefont {S.~F.}\ \bibnamefont
      {Islam}}\ and\ \bibinfo {author} {\bibfnamefont {P.}~\bibnamefont {Dutta}},\
      }\bibfield  {title} {\enquote {\bibinfo {title} {Valley-polarized
      magnetoconductivity and particle-hole symmetry breaking in a periodically
      modulated $\alpha$-${T}_3$ lattice},}\ }\href@noop {} {\bibfield  {journal}
      {\bibinfo  {journal} {Phys. Rev. B}\ }\textbf {\bibinfo {volume} {96}},\
      \bibinfo {pages} {045418} (\bibinfo {year} {2017})}\BibitemShut {NoStop}%
    \bibitem [{\citenamefont {Duan}(2023)}]{duan2023seebeck}%
      \BibitemOpen
      \bibfield  {author} {\bibinfo {author} {\bibfnamefont {W.}~\bibnamefont
      {Duan}},\ }\bibfield  {title} {\enquote {\bibinfo {title} {Seebeck and
      {N}ernst effects of pseudospin-1 fermions in the $\alpha$-${T}_3$ model under
      magnetic fields},}\ }\href@noop {} {\bibfield  {journal} {\bibinfo  {journal}
      {Phys. Rev. B}\ }\textbf {\bibinfo {volume} {108}},\ \bibinfo {pages}
      {155428} (\bibinfo {year} {2023})}\BibitemShut {NoStop}%
    \bibitem [{\citenamefont {Huang}\ \emph {et~al.}(2019)\citenamefont {Huang},
      \citenamefont {Iurov}, \citenamefont {Xu}, \citenamefont {Lai},\ and\
      \citenamefont {Gumbs}}]{huang2019interplay}%
      \BibitemOpen
      \bibfield  {author} {\bibinfo {author} {\bibfnamefont {D.}~\bibnamefont
      {Huang}}, \bibinfo {author} {\bibfnamefont {A.}~\bibnamefont {Iurov}},
      \bibinfo {author} {\bibfnamefont {H.-Y.}\ \bibnamefont {Xu}}, \bibinfo
      {author} {\bibfnamefont {Y.-C.}\ \bibnamefont {Lai}}, \ and\ \bibinfo
      {author} {\bibfnamefont {G.}~\bibnamefont {Gumbs}},\ }\bibfield  {title}
      {\enquote {\bibinfo {title} {Interplay of {L}orentz-{B}erry forces in
      position-momentum spaces for valley-dependent impurity scattering in
      $\alpha$-${T}_3$ lattices},}\ }\href@noop {} {\bibfield  {journal} {\bibinfo
      {journal} {Phys. Rev. B}\ }\textbf {\bibinfo {volume} {99}},\ \bibinfo
      {pages} {245412} (\bibinfo {year} {2019})}\BibitemShut {NoStop}%
    \bibitem [{\citenamefont {Wang}\ and\ \citenamefont
      {Ran}(2011)}]{wang2011nearly}%
      \BibitemOpen
      \bibfield  {author} {\bibinfo {author} {\bibfnamefont {F.}~\bibnamefont
      {Wang}}\ and\ \bibinfo {author} {\bibfnamefont {Y.}~\bibnamefont {Ran}},\
      }\bibfield  {title} {\enquote {\bibinfo {title} {Nearly flat band with chern
      number ${C}=2$ on the dice lattice},}\ }\href@noop {} {\bibfield  {journal}
      {\bibinfo  {journal} {Phys. Rev. B}\ }\textbf {\bibinfo {volume} {84}},\
      \bibinfo {pages} {241103} (\bibinfo {year} {2011})}\BibitemShut {NoStop}%
    \bibitem [{\citenamefont {K{\"o}ksal}\ \emph {et~al.}(2023)\citenamefont
      {K{\"o}ksal}, \citenamefont {Li},\ and\ \citenamefont
      {Pentcheva}}]{koksal2023high}%
      \BibitemOpen
      \bibfield  {author} {\bibinfo {author} {\bibfnamefont {O.}~\bibnamefont
      {K{\"o}ksal}}, \bibinfo {author} {\bibfnamefont {L.}~\bibnamefont {Li}}, \
      and\ \bibinfo {author} {\bibfnamefont {R.}~\bibnamefont {Pentcheva}},\
      }\bibfield  {title} {\enquote {\bibinfo {title} {High chern numbers in a
      perovskite-derived dice lattice $\mathrm{({L}a {X} {O}_3)_3}$/$\mathrm{({L}a
      {A}l {O}_3)_3(111)}$ with {X}= {T}i, {M}n and {C}o},}\ }\href@noop {}
      {\bibfield  {journal} {\bibinfo  {journal} {Sci. Rep.}\ }\textbf {\bibinfo
      {volume} {13}},\ \bibinfo {pages} {10615} (\bibinfo {year}
      {2023})}\BibitemShut {NoStop}%
    \bibitem [{\citenamefont {Zhu}\ \emph {et~al.}(2016)\citenamefont {Zhu},
      \citenamefont {Wang}, \citenamefont {Guan}, \citenamefont {Liu},
      \citenamefont {Zhang}, \citenamefont {Chen},\ and\ \citenamefont
      {Yang}}]{zhu2016blue}%
      \BibitemOpen
      \bibfield  {author} {\bibinfo {author} {\bibfnamefont {L.}~\bibnamefont
      {Zhu}}, \bibinfo {author} {\bibfnamefont {S.-S.}\ \bibnamefont {Wang}},
      \bibinfo {author} {\bibfnamefont {S.}~\bibnamefont {Guan}}, \bibinfo {author}
      {\bibfnamefont {Y.}~\bibnamefont {Liu}}, \bibinfo {author} {\bibfnamefont
      {T.}~\bibnamefont {Zhang}}, \bibinfo {author} {\bibfnamefont
      {G.}~\bibnamefont {Chen}}, \ and\ \bibinfo {author} {\bibfnamefont {S.~A.}\
      \bibnamefont {Yang}},\ }\bibfield  {title} {\enquote {\bibinfo {title} {Blue
      phosphorene oxide: strain-tunable quantum phase transitions and novel 2d
      emergent fermions},}\ }\href@noop {} {\bibfield  {journal} {\bibinfo
      {journal} {Nano Lett.}\ }\textbf {\bibinfo {volume} {16}},\ \bibinfo {pages}
      {6548} (\bibinfo {year} {2016})}\BibitemShut {NoStop}%
    \bibitem [{\citenamefont {Malcolm}\ and\ \citenamefont
      {Nicol}(2015)}]{malcolm2015magneto}%
      \BibitemOpen
      \bibfield  {author} {\bibinfo {author} {\bibfnamefont {J.~D.}\ \bibnamefont
      {Malcolm}}\ and\ \bibinfo {author} {\bibfnamefont {E.~J.}\ \bibnamefont
      {Nicol}},\ }\bibfield  {title} {\enquote {\bibinfo {title} {Magneto-optics of
      massless {K}ane fermions: Role of the flat band and unusual {B}erry phase},}\
      }\href@noop {} {\bibfield  {journal} {\bibinfo  {journal} {Phys. Rev. B}\
      }\textbf {\bibinfo {volume} {92}},\ \bibinfo {pages} {035118} (\bibinfo
      {year} {2015})}\BibitemShut {NoStop}%
    \bibitem [{\citenamefont {Orlita}\ \emph {et~al.}(2014)\citenamefont {Orlita},
      \citenamefont {Basko}, \citenamefont {Zholudev}, \citenamefont {Teppe},
      \citenamefont {Knap}, \citenamefont {Gavrilenko}, \citenamefont {Mikhailov},
      \citenamefont {Dvoretskii}, \citenamefont {Neugebauer}, \citenamefont
      {Faugeras} \emph {et~al.}}]{orlita2014observation}%
      \BibitemOpen
      \bibfield  {author} {\bibinfo {author} {\bibfnamefont {M.}~\bibnamefont
      {Orlita}}, \bibinfo {author} {\bibfnamefont {D.}~\bibnamefont {Basko}},
      \bibinfo {author} {\bibfnamefont {M.}~\bibnamefont {Zholudev}}, \bibinfo
      {author} {\bibfnamefont {F.}~\bibnamefont {Teppe}}, \bibinfo {author}
      {\bibfnamefont {W.}~\bibnamefont {Knap}}, \bibinfo {author} {\bibfnamefont
      {V.}~\bibnamefont {Gavrilenko}}, \bibinfo {author} {\bibfnamefont
      {N.}~\bibnamefont {Mikhailov}}, \bibinfo {author} {\bibfnamefont
      {S.}~\bibnamefont {Dvoretskii}}, \bibinfo {author} {\bibfnamefont
      {P.}~\bibnamefont {Neugebauer}}, \bibinfo {author} {\bibfnamefont
      {C.}~\bibnamefont {Faugeras}},  \emph {et~al.},\ }\bibfield  {title}
      {\enquote {\bibinfo {title} {Observation of three-dimensional massless kane
      fermions in a zinc-blende crystal},}\ }\href@noop {} {\bibfield  {journal}
      {\bibinfo  {journal} {Nat. Phys.}\ }\textbf {\bibinfo {volume} {10}},\
      \bibinfo {pages} {233} (\bibinfo {year} {2014})}\BibitemShut {NoStop}%
    \bibitem [{\citenamefont {Andrijauskas}\ \emph {et~al.}(2015)\citenamefont
      {Andrijauskas}, \citenamefont {Anisimovas}, \citenamefont
      {Ra{\v{c}}i{\=u}nas}, \citenamefont {Mekys}, \citenamefont {Kudria{\v{s}}ov},
      \citenamefont {Spielman},\ and\ \citenamefont
      {Juzeli{\=u}nas}}]{andrijauskas2015three}%
      \BibitemOpen
      \bibfield  {author} {\bibinfo {author} {\bibfnamefont {T.}~\bibnamefont
      {Andrijauskas}}, \bibinfo {author} {\bibfnamefont {E.}~\bibnamefont
      {Anisimovas}}, \bibinfo {author} {\bibfnamefont {M.}~\bibnamefont
      {Ra{\v{c}}i{\=u}nas}}, \bibinfo {author} {\bibfnamefont {A.}~\bibnamefont
      {Mekys}}, \bibinfo {author} {\bibfnamefont {V.}~\bibnamefont
      {Kudria{\v{s}}ov}}, \bibinfo {author} {\bibfnamefont {I.}~\bibnamefont
      {Spielman}}, \ and\ \bibinfo {author} {\bibfnamefont {G.}~\bibnamefont
      {Juzeli{\=u}nas}},\ }\bibfield  {title} {\enquote {\bibinfo {title}
      {Three-level {H}aldane-like model on a dice optical lattice},}\ }\href@noop
      {} {\bibfield  {journal} {\bibinfo  {journal} {Phys. Rev. A}\ }\textbf
      {\bibinfo {volume} {92}},\ \bibinfo {pages} {033617} (\bibinfo {year}
      {2015})}\BibitemShut {NoStop}%
    \bibitem [{\citenamefont {Dey}\ \emph {et~al.}(2020)\citenamefont {Dey},
      \citenamefont {Kapri}, \citenamefont {Pal},\ and\ \citenamefont
      {Ghosh}}]{dey2020unconventional}%
      \BibitemOpen
      \bibfield  {author} {\bibinfo {author} {\bibfnamefont {B.}~\bibnamefont
      {Dey}}, \bibinfo {author} {\bibfnamefont {P.}~\bibnamefont {Kapri}}, \bibinfo
      {author} {\bibfnamefont {O.}~\bibnamefont {Pal}}, \ and\ \bibinfo {author}
      {\bibfnamefont {T.~K.}\ \bibnamefont {Ghosh}},\ }\bibfield  {title} {\enquote
      {\bibinfo {title} {Unconventional phases in a {H}aldane model of dice
      lattice},}\ }\href@noop {} {\bibfield  {journal} {\bibinfo  {journal} {Phys.
      Rev. B}\ }\textbf {\bibinfo {volume} {101}},\ \bibinfo {pages} {235406}
      (\bibinfo {year} {2020})}\BibitemShut {NoStop}%
    \bibitem [{\citenamefont {Haldane}(1988)}]{haldane1988model}%
      \BibitemOpen
      \bibfield  {author} {\bibinfo {author} {\bibfnamefont {F.~D.~M.}\
      \bibnamefont {Haldane}},\ }\bibfield  {title} {\enquote {\bibinfo {title}
      {Model for a quantum {H}all effect without {L}andau levels: Condensed-matter
      realization of the" parity anomaly"},}\ }\href@noop {} {\bibfield  {journal}
      {\bibinfo  {journal} {Phys. Rev. Lett.}\ }\textbf {\bibinfo {volume} {61}},\
      \bibinfo {pages} {2015} (\bibinfo {year} {1988})}\BibitemShut {NoStop}%
    \bibitem [{\citenamefont {Wang}\ and\ \citenamefont
      {Liu}(2021)}]{wang2021quantum}%
      \BibitemOpen
      \bibfield  {author} {\bibinfo {author} {\bibfnamefont {J.}~\bibnamefont
      {Wang}}\ and\ \bibinfo {author} {\bibfnamefont {J.-F.}\ \bibnamefont {Liu}},\
      }\bibfield  {title} {\enquote {\bibinfo {title} {Quantum spin {H}all phase
      transition in the $\alpha$-${T}_3$ lattice},}\ }\href@noop {} {\bibfield
      {journal} {\bibinfo  {journal} {Phys. Rev. B}\ }\textbf {\bibinfo {volume}
      {103}},\ \bibinfo {pages} {075419} (\bibinfo {year} {2021})}\BibitemShut
      {NoStop}%
    \bibitem [{\citenamefont {Hao}(2022)}]{hao2022zigzag}%
      \BibitemOpen
      \bibfield  {author} {\bibinfo {author} {\bibfnamefont {L.}~\bibnamefont
      {Hao}},\ }\bibfield  {title} {\enquote {\bibinfo {title} {Zigzag dice lattice
      ribbons: Distinct edge morphologies and structure-spectrum
      correspondences},}\ }\href@noop {} {\bibfield  {journal} {\bibinfo  {journal}
      {Phys. Rev. Mater.}\ }\textbf {\bibinfo {volume} {6}},\ \bibinfo {pages}
      {034002} (\bibinfo {year} {2022})}\BibitemShut {NoStop}%
    \bibitem [{\citenamefont {Real}\ \emph {et~al.}(2017)\citenamefont {Real},
      \citenamefont {Cantillano}, \citenamefont {L{\'o}pez-Gonz{\'a}lez},
      \citenamefont {Szameit}, \citenamefont {Aono}, \citenamefont {Naruse},
      \citenamefont {Kim}, \citenamefont {Wang},\ and\ \citenamefont
      {Vicencio}}]{real2017flat}%
      \BibitemOpen
      \bibfield  {author} {\bibinfo {author} {\bibfnamefont {B.}~\bibnamefont
      {Real}}, \bibinfo {author} {\bibfnamefont {C.}~\bibnamefont {Cantillano}},
      \bibinfo {author} {\bibfnamefont {D.}~\bibnamefont {L{\'o}pez-Gonz{\'a}lez}},
      \bibinfo {author} {\bibfnamefont {A.}~\bibnamefont {Szameit}}, \bibinfo
      {author} {\bibfnamefont {M.}~\bibnamefont {Aono}}, \bibinfo {author}
      {\bibfnamefont {M.}~\bibnamefont {Naruse}}, \bibinfo {author} {\bibfnamefont
      {S.-J.}\ \bibnamefont {Kim}}, \bibinfo {author} {\bibfnamefont
      {K.}~\bibnamefont {Wang}}, \ and\ \bibinfo {author} {\bibfnamefont {R.~A.}\
      \bibnamefont {Vicencio}},\ }\bibfield  {title} {\enquote {\bibinfo {title}
      {Flat-band light dynamics in stub photonic lattices},}\ }\href@noop {}
      {\bibfield  {journal} {\bibinfo  {journal} {Sci. Rep.}\ }\textbf {\bibinfo
      {volume} {7}},\ \bibinfo {pages} {15085} (\bibinfo {year}
      {2017})}\BibitemShut {NoStop}%
    \bibitem [{\citenamefont {Bartlett}\ \emph {et~al.}(2021)\citenamefont
      {Bartlett}, \citenamefont {Hu},\ and\ \citenamefont
      {Zhao}}]{bartlett2021illuminating}%
      \BibitemOpen
      \bibfield  {author} {\bibinfo {author} {\bibfnamefont {J.}~\bibnamefont
      {Bartlett}}, \bibinfo {author} {\bibfnamefont {H.}~\bibnamefont {Hu}}, \ and\
      \bibinfo {author} {\bibfnamefont {E.}~\bibnamefont {Zhao}},\ }\bibfield
      {title} {\enquote {\bibinfo {title} {Illuminating the bulk-boundary
      correspondence of a non-hermitian stub lattice with {M}ajorana stars},}\
      }\href@noop {} {\bibfield  {journal} {\bibinfo  {journal} {Phys. Rev. B}\
      }\textbf {\bibinfo {volume} {104}},\ \bibinfo {pages} {195131} (\bibinfo
      {year} {2021})}\BibitemShut {NoStop}%
    \bibitem [{\citenamefont {C{\'a}ceres-Aravena}\ \emph
      {et~al.}(2022)\citenamefont {C{\'a}ceres-Aravena}, \citenamefont {Real},
      \citenamefont {Guzm{\'a}n-Silva}, \citenamefont {Amo}, \citenamefont
      {Torres},\ and\ \citenamefont {Vicencio}}]{caceres2022experimental}%
      \BibitemOpen
      \bibfield  {author} {\bibinfo {author} {\bibfnamefont {G.}~\bibnamefont
      {C{\'a}ceres-Aravena}}, \bibinfo {author} {\bibfnamefont {B.}~\bibnamefont
      {Real}}, \bibinfo {author} {\bibfnamefont {D.}~\bibnamefont
      {Guzm{\'a}n-Silva}}, \bibinfo {author} {\bibfnamefont {A.}~\bibnamefont
      {Amo}}, \bibinfo {author} {\bibfnamefont {L.~E.~F.}\ \bibnamefont {Torres}},
      \ and\ \bibinfo {author} {\bibfnamefont {R.~A.}\ \bibnamefont {Vicencio}},\
      }\bibfield  {title} {\enquote {\bibinfo {title} {Experimental observation of
      edge states in {S}{S}{H}-stub photonic lattices},}\ }\href@noop {} {\bibfield
       {journal} {\bibinfo  {journal} {Phys. Rev. Res.}\ }\textbf {\bibinfo
      {volume} {4}},\ \bibinfo {pages} {013185} (\bibinfo {year}
      {2022})}\BibitemShut {NoStop}%
    \bibitem [{\citenamefont {Majorana}(1932)}]{majorana1932atomi}%
      \BibitemOpen
      \bibfield  {author} {\bibinfo {author} {\bibfnamefont {E.}~\bibnamefont
      {Majorana}},\ }\bibfield  {title} {\enquote {\bibinfo {title} {Atomi
      orientati in campo magnetico variabile},}\ }\href@noop {} {\bibfield
      {journal} {\bibinfo  {journal} {Il Nuovo Cimento (1924-1942)}\ }\textbf
      {\bibinfo {volume} {9}},\ \bibinfo {pages} {43} (\bibinfo {year}
      {1932})}\BibitemShut {NoStop}%
    \bibitem [{\citenamefont {Hannay}(1998)}]{hannay1998berry}%
      \BibitemOpen
      \bibfield  {author} {\bibinfo {author} {\bibfnamefont {J.}~\bibnamefont
      {Hannay}},\ }\bibfield  {title} {\enquote {\bibinfo {title} {The {B}erry
      phase for spin in the {M}ajorana representation},}\ }\href@noop {} {\bibfield
       {journal} {\bibinfo  {journal} {J. Phys. A: Math. Gen.}\ }\textbf {\bibinfo
      {volume} {31}},\ \bibinfo {pages} {L53} (\bibinfo {year} {1998})}\BibitemShut
      {NoStop}%
    \bibitem [{\citenamefont {Liu}\ and\ \citenamefont
      {Fu}(2014)}]{liu2014representation}%
      \BibitemOpen
      \bibfield  {author} {\bibinfo {author} {\bibfnamefont {H.}~\bibnamefont
      {Liu}}\ and\ \bibinfo {author} {\bibfnamefont {L.}~\bibnamefont {Fu}},\
      }\bibfield  {title} {\enquote {\bibinfo {title} {Representation of {B}erry
      phase by the trajectories of {M}ajorana stars},}\ }\href@noop {} {\bibfield
      {journal} {\bibinfo  {journal} {Phys. Rev. Lett.}\ }\textbf {\bibinfo
      {volume} {113}},\ \bibinfo {pages} {240403} (\bibinfo {year}
      {2014})}\BibitemShut {NoStop}%
    \bibitem [{\citenamefont {Pratama}\ and\ \citenamefont
      {Nakanishi}(2024)}]{pratama24-letter}%
      \BibitemOpen
      \bibfield  {author} {\bibinfo {author} {\bibfnamefont {F.~R.}\ \bibnamefont
      {Pratama}}\ and\ \bibinfo {author} {\bibfnamefont {T.}~\bibnamefont
      {Nakanishi}},\ }\bibfield  {title} {\enquote {\bibinfo {title} {Topological
      edge state of massless fermion with non-quantized and zero {B}erry phases},}\
      }\href@noop {} {\bibfield  {journal} {\bibinfo  {journal} {Submitted}\ }
      (\bibinfo {year} {2024})}\BibitemShut {NoStop}%
    \bibitem [{Note1()}]{Note1}%
      \BibitemOpen
      \bibinfo {note} {The positions of all sites in each unit cell are regarded as
      identical. This treatment is equivalent to fixing a gauge in the Hamiltonian
      such that the intracell hopping is real~\cite
      {delplace2011,ezawa2014}.}\BibitemShut {Stop}%
    \bibitem [{Note2()}]{Note2}%
      \BibitemOpen
      \bibinfo {note} {$Z=\protect \mathrm {Arg}(X+iY)=\protect \mathrm
      {atan2}(Y,X)$, where $\protect \mathrm {atan2}$ is two-argument arctangent
      function. Particularly, ${Z}=0$ for ${Y}=0,~{X}>0$ and ${Z}=\pi $ for
      ${Y}=0,~{X}<0$.}\BibitemShut {Stop}%
    \bibitem [{Note3()}]{Note3}%
      \BibitemOpen
      \bibinfo {note} {For $V_A^\prime <0$, the BZ is defined for $k\in [0,2\pi
      /a_0]$, because $F(k)=V_A-|V_A^\prime | e^{-ika_0} = V_A+|V_A^\prime |
      e^{-i(k+\pi /a_0)a_0} $}\BibitemShut {NoStop}%
    \bibitem [{\citenamefont {Lieb}(1989)}]{lieb1989two}%
      \BibitemOpen
      \bibfield  {author} {\bibinfo {author} {\bibfnamefont {E.~H.}\ \bibnamefont
      {Lieb}},\ }\bibfield  {title} {\enquote {\bibinfo {title} {Two theorems on
      the hubbard model},}\ }\href@noop {} {\bibfield  {journal} {\bibinfo
      {journal} {Phys. Rev. Lett.}\ }\textbf {\bibinfo {volume} {62}},\ \bibinfo
      {pages} {1201} (\bibinfo {year} {1989})}\BibitemShut {NoStop}%
    \bibitem [{\citenamefont {Wakabayashi}\ \emph {et~al.}(2010)\citenamefont
      {Wakabayashi}, \citenamefont {Sasaki}, \citenamefont {Nakanishi},\ and\
      \citenamefont {Enoki}}]{wakabayashi2010electronic}%
      \BibitemOpen
      \bibfield  {author} {\bibinfo {author} {\bibfnamefont {K.}~\bibnamefont
      {Wakabayashi}}, \bibinfo {author} {\bibfnamefont {K.-i.}\ \bibnamefont
      {Sasaki}}, \bibinfo {author} {\bibfnamefont {T.}~\bibnamefont {Nakanishi}}, \
      and\ \bibinfo {author} {\bibfnamefont {T.}~\bibnamefont {Enoki}},\ }\bibfield
       {title} {\enquote {\bibinfo {title} {Electronic states of graphene
      nanoribbons and analytical solutions},}\ }\href@noop {} {\bibfield  {journal}
      {\bibinfo  {journal} {Sci. Technol. Adv. Mater.}\ }\textbf {\bibinfo {volume}
      {11}},\ \bibinfo {pages} {054504} (\bibinfo {year} {2010})}\BibitemShut
      {NoStop}%
    \bibitem [{\citenamefont {Schwinger}(1952)}]{schwinger1952}%
      \BibitemOpen
      \bibfield  {author} {\bibinfo {author} {\bibfnamefont {J.}~\bibnamefont
      {Schwinger}},\ }\bibfield  {title} {\enquote {\bibinfo {title} {On angular
      momentum},}\ }\href@noop {} {\bibfield  {journal} {\bibinfo  {journal} {US
      Atomic Energy Commission, Report No. NYO-3071}\ } (\bibinfo {year}
      {1952})}\BibitemShut {NoStop}%
    \bibitem [{\citenamefont {Ezawa}\ \emph {et~al.}(2013)\citenamefont {Ezawa},
      \citenamefont {Tanaka},\ and\ \citenamefont
      {Nagaosa}}]{ezawa2013topological}%
      \BibitemOpen
      \bibfield  {author} {\bibinfo {author} {\bibfnamefont {M.}~\bibnamefont
      {Ezawa}}, \bibinfo {author} {\bibfnamefont {Y.}~\bibnamefont {Tanaka}}, \
      and\ \bibinfo {author} {\bibfnamefont {N.}~\bibnamefont {Nagaosa}},\
      }\bibfield  {title} {\enquote {\bibinfo {title} {Topological phase transition
      without gap closing},}\ }\href@noop {} {\bibfield  {journal} {\bibinfo
      {journal} {Sci. Rep.}\ }\textbf {\bibinfo {volume} {3}},\ \bibinfo {pages}
      {2790} (\bibinfo {year} {2013})}\BibitemShut {NoStop}%
    \bibitem [{\citenamefont {Longhi}(2014)}]{longhi2014aharonov}%
      \BibitemOpen
      \bibfield  {author} {\bibinfo {author} {\bibfnamefont {S.}~\bibnamefont
      {Longhi}},\ }\bibfield  {title} {\enquote {\bibinfo {title} {Aharonov-{B}ohm
      photonic cages in waveguide and coupled resonator lattices by synthetic
      magnetic fields},}\ }\href@noop {} {\bibfield  {journal} {\bibinfo  {journal}
      {Opt. Lett.}\ }\textbf {\bibinfo {volume} {39}},\ \bibinfo {pages} {5892}
      (\bibinfo {year} {2014})}\BibitemShut {NoStop}%
    \bibitem [{\citenamefont {Mukherjee}\ and\ \citenamefont
      {Thomson}(2015)}]{mukherjee2015observation}%
      \BibitemOpen
      \bibfield  {author} {\bibinfo {author} {\bibfnamefont {S.}~\bibnamefont
      {Mukherjee}}\ and\ \bibinfo {author} {\bibfnamefont {R.~R.}\ \bibnamefont
      {Thomson}},\ }\bibfield  {title} {\enquote {\bibinfo {title} {Observation of
      localized flat-band modes in a quasi-one-dimensional photonic rhombic
      lattice},}\ }\href@noop {} {\bibfield  {journal} {\bibinfo  {journal} {Opt.
      Lett.}\ }\textbf {\bibinfo {volume} {40}},\ \bibinfo {pages} {5443} (\bibinfo
      {year} {2015})}\BibitemShut {NoStop}%
    \bibitem [{\citenamefont {Pelegr{\'\i}}\ \emph {et~al.}(2019)\citenamefont
      {Pelegr{\'\i}}, \citenamefont {Marques}, \citenamefont {Dias}, \citenamefont
      {Daley}, \citenamefont {Ahufinger},\ and\ \citenamefont
      {Mompart}}]{pelegri2019topological}%
      \BibitemOpen
      \bibfield  {author} {\bibinfo {author} {\bibfnamefont {G.}~\bibnamefont
      {Pelegr{\'\i}}}, \bibinfo {author} {\bibfnamefont {A.}~\bibnamefont
      {Marques}}, \bibinfo {author} {\bibfnamefont {R.}~\bibnamefont {Dias}},
      \bibinfo {author} {\bibfnamefont {A.}~\bibnamefont {Daley}}, \bibinfo
      {author} {\bibfnamefont {V.}~\bibnamefont {Ahufinger}}, \ and\ \bibinfo
      {author} {\bibfnamefont {J.}~\bibnamefont {Mompart}},\ }\bibfield  {title}
      {\enquote {\bibinfo {title} {Topological edge states with ultracold atoms
      carrying orbital angular momentum in a diamond chain},}\ }\href@noop {}
      {\bibfield  {journal} {\bibinfo  {journal} {Phys. Rev. A}\ }\textbf {\bibinfo
      {volume} {99}},\ \bibinfo {pages} {023612} (\bibinfo {year}
      {2019})}\BibitemShut {NoStop}%
    \bibitem [{\citenamefont {Oriekhov}\ \emph {et~al.}(2018)\citenamefont
      {Oriekhov}, \citenamefont {Gorbar},\ and\ \citenamefont
      {Gusynin}}]{oriekhov2018electronic}%
      \BibitemOpen
      \bibfield  {author} {\bibinfo {author} {\bibfnamefont {D.}~\bibnamefont
      {Oriekhov}}, \bibinfo {author} {\bibfnamefont {E.}~\bibnamefont {Gorbar}}, \
      and\ \bibinfo {author} {\bibfnamefont {V.}~\bibnamefont {Gusynin}},\
      }\bibfield  {title} {\enquote {\bibinfo {title} {Electronic states of
      pseudospin-1 fermions in dice lattice ribbon},}\ }\href@noop {} {\bibfield
      {journal} {\bibinfo  {journal} {Low Temp. Phys.}\ }\textbf {\bibinfo {volume}
      {44}},\ \bibinfo {pages} {1313} (\bibinfo {year} {2018})}\BibitemShut
      {NoStop}%
    \end{thebibliography}

\end{document}